\documentclass[aps,superscriptaddress,preprint,amsmath,amssymb,nofootinbib,]{revtex4-1}

\usepackage{graphicx}% Include figure files
\usepackage{dcolumn}% Align table columns on decimal point
\usepackage{bm}% bold math

\newcommand{\beq}{\begin{equation}}
\newcommand{\eeq}{\end{equation}}

\newcommand{\del}{\partial}

\newcommand{\3}{4.9 cm}
\newcommand{\2}{7.0 cm}
\newcommand{\1}{12.5 cm}
\newcommand{\5}{6.5 cm}

\begin{document}

\setcounter{tocdepth}{1} 

\preprint{CERN-PH-TH/2012-251} 

\title{\boldmath 
Precision studies of the Higgs boson decay channel $H\to ZZ \to 4\ell$
with {\sc MEKD}}

\author{Paul Avery}
\author{Dimitri Bourilkov}
\author{Mingshui Chen}
\author{Tongguang Cheng}
\author{Alexey Drozdetskiy}
\author{James S. Gainer}
\email{Corresponding author: gainer@phys.ufl.edu}
\author{Andrey Korytov}
\author{Konstantin T. Matchev}  
\author{Predrag Milenovic}
\author{Guenakh Mitselmakher}
\affiliation{Physics Department, University of Florida, Gainesville, FL 32611, USA.}
\author{Myeonghun Park}
\affiliation{CERN Physics Department, Theory Division, CH-1211 Geneva 23, Switzerland.}
\author{Aurelijus Rinkevicius}
\author{Matthew Snowball}
\affiliation{Physics Department, University of Florida, Gainesville, FL 32611, USA.}

\date{December 24, 2012} 

\begin{abstract}
The importance of the $H \to ZZ \to 4\ell$ ``golden''
channel was shown by its major role 
in the discovery, by the ATLAS and CMS collaborations, of a
Higgs-like boson with mass near $125$ GeV.   
We analyze the discrimination power of the matrix element method
both for separating the signal from the irreducible ZZ background 
and for distinguishing various spin and parity hypotheses
describing a signal in this channel. We show that the proper treatment of
interference effects associated with permutations of identical leptons
in the $4e$ and $4\mu$ final states plays an important role in
achieving the best sensitivity in measuring the properties of the newly
discovered boson. 
We provide a code, {\sc MEKD}, that calculates kinematic 
discriminants based on the full leading order matrix elements and
which will aid experimentalists and phenomenologists in their continuing 
studies of the $H \to ZZ \to 4\ell$ channel. 
\end{abstract}

\maketitle

\tableofcontents

\setcounter{tocdepth}{1} 

\clearpage
\pagebreak[4]

\section{Introduction}
\label{sec:introduction}
%===============================================================================

The CERN LHC collaborations recently reported the observation of a new
bosonic particle with mass $m\sim 125$~GeV~\cite{Aad:2012gk,
Chatrchyan:2012gu}. 
The production rates in the main discovery channels are consistent
with the expectations for the Higgs boson, $H$, of the Standard Model
(SM).

The most useful channels for the discovery of the Higgs-like boson 
and early measurements of its properties are 
$H\to \gamma\gamma$~\cite{ATLASHGG,CMSHGG,ATLASHGGlatest}
and $H\to ZZ \to 4\ell$~\cite{ATLASHZZ,CMSHZZ,CMSHZZlatest, ATLASHZZlatest}.
Each of these discovery channels has strengths and weaknesses.
For example, observation of the appropriate excess in diphoton events
immediately implies that the discovered resonance is not spin one
\cite{Landau:1948kw,Yang:1950rg}. 
At the same time, because of the much larger background, measuring the
exact properties of the object in $\gamma\gamma$ events will be quite
challenging~\cite{Cottingham:1995wy,Kumar:2011yta,
Alves:2011kc,Ellis:2012wg,Bolognesi:2012mm,Alves:2012fb,Choi:2012yg}. 

In contrast, the ``golden channel" 
$H\to ZZ \to 4\ell$\footnote{
Throughout this paper we shall use the convention that $Z$ stands for
both on-shell and off-shell $Z$ bosons, as well as $\gamma^\ast$ when
allowed, while $Z^\ast$ is used for either an off-shell $Z$ or for
$\gamma^\ast$.  We never make the approximation that a $Z$ is
on-shell.
}
offers the
opportunity for clean measurements of the mass, spin, parity, etc.~of
the new resonance in a controlled environment with low backgrounds.
Furthermore, the four lepton final state allows
experiments to probe the polarization of the intermediate $Z$ bosons
through angular correlations. Thus, most of previous theoretical work on
spin and parity discrimination has concentrated on this channel
\cite{Nelson:1986ki,Soni:1993jc,Chang:1993jy,Arens:1994wd,Cottingham:1995wy,Choi:2002jk,Buszello:2002uu,Schalla:2004aa,
Godbole:2007cn,Kovalchuk:2008zz,Kovalchuk:2009zz,Cao:2009ah,Gao:2010qx,DeRujula:2010ys,Englert:2010ud,DeSanctis:2011yc,
Bolognesi:2012mm,Boughezal:2012tz,Stolarski:2012ps}; see also
\cite{Englert:2012ct,Moffat:2012pb,Coleppa:2012eh,Cea:2012ud,Kumar:2012ba}.
In fact, CMS and ATLAS have already shown that the interpretation of this 
resonance as CP odd is strongly disfavored, using information from this 
channel~\cite{CMSHZZlatest, ATLASHZZlatest}.

At the same time, relatively little theoretical effort has gone into
using the kinematics of four leptons, besides their invariant mass,
for discriminating between the Higgs signal and the SM background in
this channel; see,
however~\cite{Gao:2010qx,DeRujula:2010ys,Gainer:2011xz,ATLASHZZ,CMSHZZ}.
Discriminating between signal and background is an important issue, 
though, whether for determining the significance of the signal in this
channel or for accurately measuring the spin and 
parity properties of the resonance.  This is
because for an SM Higgs boson with mass near $125$ GeV,
the rate of background events in comparison with that of the signal
is not negligible; in fact, the signal to background ratio for a four lepton mass window 
of $\pm 2 \sigma_{m_{4\ell}}$ is roughly 2:1~\cite{CMSHZZlatest, ATLASHZZlatest}.
(Here, $\sigma_{m_{4\ell}}$ stands for the experimental four lepton
mass resolution.)

Another motivation for revisiting the previous work on this channel is
that much of the previous literature is limited to the case of a heavy
Higgs boson, with mass above the $ZZ$ threshold, where both $Z$ bosons
produced in the Higgs decay will be on-shell.
Most previous studies of this channel in the low Higgs
boson mass range have lacked diagrams involving permutations of
identical leptons, and hence the associated interference, 
in the $4e$ and $4\mu$ final states.
Such permutations are relatively unimportant when both $Z$ bosons are
mostly on-shell. 
However, as we show in this paper, the inclusion of these interference
effects plays an important role in achieving the best possible
sensitivities in the low mass range and is crucially important for
separating spin zero and spin two resonance hypotheses.

One of the great advantages of the golden channel is that the final
state is fully reconstructed and well-measured.
However, the presence of four leptons in the final state means
that, at leading order, there are eight independent observed degrees
of freedom (in the Higgs CM frame), not counting the irrelevant
azimuthal orientation of the event.   
The existence of eight meaningful kinematic variables strongly
motivates the use of a Multivariate Analysis (MVA)~\cite{Bhat:2010zz}
technique, and in particular the Matrix Element Method
(MEM)~\cite{Gao:2010qx,DeRujula:2010ys,Gainer:2011xz,Campbell:2012cz,Campbell:2012ct,Bolognesi:2012mm,Chen:2012jy},
thereby allowing all of the information in each event to be used in either
distinguishing signal from background or in distinguishing between different
signal spin/parity hypotheses. Both the MEM and other MVA approaches
presumably achieve an approximately identical level of discrimination.
However, the MEM has several
advantages over other MVAs.  The quantity used in the MEM analysis is the
(squared) matrix element, which is (up to approximations used) 
uniquely defined from first principles, while MVA-based discriminants
require an ad hoc training on very large Monte Carlo samples.
Since the matrix element has a clear, well-understood physical
meaning, one is able to make a direct connection between the features
of a statistical analysis and the underlying physics. This is
particularly true in the case of the golden channel, as the
distributions of kinematic variables can be determined from the
amplitudes for producing $Z$ bosons with given helicities
\cite{Hagiwara:1986vm,Duncan:1985vj}.  

In contrast with even the recent past, there are now a number of
commonly used and well-tested programs which can be utilized to
calculate matrix elements automatically~\cite{Ask:2012sm}.
These include tools to generate model files from an arbitrary
Lagrangian, such as {\sc FeynRules} \cite{Christensen:2008py} and
{\sc LanHEP}~\cite{Semenov:2008jy}, as well as tools to calculate the matrix
elements using these model files, such as 
{\sc MadGraph}~\cite{Stelzer:1994ta,Alwall:2011uj}, 
{\sc CalcHEP} \cite{Belyaev:2012qa}, and
{\sc CompHEP}~\cite{Pukhov:1999gg,Boos:2009un}.
Taken together, these tools allow for relatively automatic
implementation of new models and automatic generation of matrix
elements, as explained at the TASI-2011 summer
school~\cite{Kong:2012vg}  and the MC4BSM-2012
workshop~\cite{MC4BSM6,Ask:2012sm}.
The existence of such tools considerably simplifies calculations of
leading order matrix elements for any desired process and strongly
motivates the use of the MEM where possible.
It is this approach that we take in the presented paper.

In Sec.~\ref{sec:MEM}, we review the MEM and
describe the associated variable, the kinematic discriminant $KD$,
which quantifies the extent to which a particular event is described
by one hypothesis as opposed to another.  We also show the superiority
of the MEM over analyses involving fewer variables.
Next, we calculate $KD$ for signal and
background events using different tools: {\sc MadGraph}
\cite{Alwall:2011uj} in Sec.~\ref{sec:MG}, and {\sc CalcHEP} \cite{Belyaev:2012qa} 
and {\sc NLOME} \cite{Campbell:2010ff,NLOME} 
in Appendix~\ref{sec:CH}. 
In Sec.~\ref{sec:performance} we compare the results 
for the kinematic discriminant computed with these three tools
and show that they are in agreement.
We also analyze the relevance of interference in the
$4e$ and $4\mu$ final states in the context of SM Higgs boson
signal vs.~background separation.
In Sec.~\ref{sec:IS}, we discuss the added benefit from
incorporating experimentally-known information
about the initial state into the analysis.
Sec.~\ref{sec:jcpVsBackground} quantifies the effect of the spin
and CP of the resonance on the ability to separate signal from
background.

In Sec.~\ref{sec:JCP} we demonstrate the application of the MEM
to discrimination between different spin/parity signal hypotheses.
We show that properly taking into account the above-mentioned
interference effects in the $4e$ and $4\mu$ final states significantly
enhances the separation power between alternative signal hypotheses.
Sec.~\ref{sec:conclusions} summarizes our findings and 
outlines directions for future work.
The notation for the
relevant kinematic variables in the Higgs golden channel
are given in Appendix~\ref{sec:notation}. 

In order to facilitate future MEM-based studies of the properties of
the newly discovered four lepton resonance by the experimental and
theoretical communities, we are making public one of the two new $KD$
codes used in this paper (the {\sc MadGraph}-based code, which we call
{\sc MEKD}). 
The instructions on how to install and run the code are given in
Appendix~\ref{sec:MEKD}.
This code calculates matrix elements and kinematic discriminants,
natively including diagrams with swapped identical leptons (and the
associated interference) in the $4e$ and $4\mu$ final states,
background diagrams with $\gamma^\ast$ propagators for the 
``doubly resonant'' $q \bar q \to ZZ \to 4\ell$ process and 
singly resonant $q \bar q \to Z \to 4\ell$ production~\cite{CMS:2012bw}. 
It does not use a narrow width approximation for either the $Z$ bosons
or the signal resonance.

\section{Preliminaries}
\label{sec:MEM}

We now define the MEM and note some modifications to the general
procedure which are useful in studying the golden channel.  We then
describe a practical method  for displaying the sensitivity of
different kinematic discriminants. 
Finally, we demonstrate the increase in discrimination power that may be
obtained using the MEM, in comparison with analyses which use fewer variables.

\subsection{The matrix element method and its kinematic discriminant $KD$}
\label{MEM and variations}

For a four lepton event described by kinematic information
$\mathbf{p}$, the best possible discriminant between two production hypotheses 
{\it A} (e.g. signal) and {\it B} (e.g. background),
is the ratio of the probabilities to observe such an event, given the alternative
production hypotheses:
\begin{equation}
D( A; B) =
\frac
{P( \mathbf{p} \, | \, A )}
{P( \mathbf{p} \, | \, B )}.
\end{equation}
Here, $P( \mathbf{p} \, | \, A )$ and $P( \mathbf{p} \, | \, B )$ are the
probability density functions ($pdf$s) for observing the event 
in the cases of hypotheses {\it A} and {\it B}, respectively.
Note that the alternative definition
of the discriminant $\tilde D$:
\begin{equation}
\tilde D(A; B) =
\frac
{\alpha \, P( \mathbf{p} \, | A ) + \beta \, P(\mathbf{p} \, | B )}
{P( \mathbf{p} \, | \, B)},
\end{equation}
may seem to be more suited for the case when the alternative hypotheses are \textit{signal+background} 
or \textit{background-only}. However,  $\tilde D$ is a 
monotonic function of $D$ and, hence, considering $\tilde D$ rather
than $D$ does not add or subtract any information.

In particle physics, probabilities $P( \mathbf{p})$ can be represented by the
differential cross section of a signal or background process with
respect to the variables considered, $d\sigma/d\mathbf{p}$, 
scaled by the instrumental efficiencies $\epsilon(\mathbf{p})$, 
and normalized by the appropriate total cross section within the instrumental acceptance:
\begin{equation}
P( \mathbf{p}) =
\frac
{  d\sigma / d\mathbf{p} \,\, \epsilon(\mathbf{p}) }
{ \int  d\sigma / d\mathbf{p} \,\, \epsilon(\mathbf{p}) \,\, d\mathbf{p} }.
\end{equation}
Following the same logic of comparing discriminants $D$ and $\tilde
D$, one can see that the normalization constants in the definitions of $P(
\mathbf{p})$ do not matter and can be dropped; the instrumental
efficiencies appearing  in the numerator and denominator cancel out;
and one arrives at the following definition of the kinematic
discriminant between two chosen processes A and B,  e.g. signal and background:
\begin{equation}
KD(A; \, B) = \ln 
\left(
\frac
{ d\sigma_A / d\mathbf{p} }
{ d\sigma_B / d\mathbf{p} } 
\right).
\end{equation}
We use log of the ratio only for technical convenience; this is 
dictated by the large dynamic range in the ratio of cross sections involved.
If all signal and background processes involve effectively massless
initial state partons (as is often the case), the phase space factors
are identical for all processes and hence cancel in the ratio.  
Thus, we find that 
\begin{equation}\label{D final}
KD(A; B) = \ln 
\left(
\frac
{\sum\limits_{a,b}   \, f_{a}(x_1)  \cdot f_{b}(x_2)  \cdot \left| {\cal M}_A (a + b   \to 4\ell ) \right|^2}
{\sum\limits_{a',b'} \, f_{a'}(x_1) \cdot f_{b'}(x_2) \cdot \left| {\cal M}_B (a' + b' \to 4\ell ) \right|^2}
\right).
\end{equation}
In this equation, $a$ and $b$ ($a'$ and $b'$) stand for different types of partons
which can produce the four lepton final state via process A (B) with
corresponding matrix elements $\mathcal{M}_A$ ( $\mathcal{M}_B$).
The momenta of the partons are fully defined by the final state and
are denoted by $x_1$ and $x_2$ in units of the initial proton energy.
The functions $f(x)$ are the parton distribution functions (PDFs), with the
subscripts indicating which  types of partons they correspond to.
 
The products of PDFs in the numerator and denominator of Eq.~(\ref{D final})
in general do not cancel. In particular they do not cancel if 
hypothesis {\it A} is $gg \to H \to ZZ \to 4\ell$ and 
hypothesis {\it B} is $q \bar{q} \to Z Z^\ast \to 4\ell$.
Note that Eq.~(\ref{D final}) is invariant under any changes of variables, as any resulting
Jacobian would be common to both the signal and background
expressions and hence would cancel in the ratio.

One may wish to ignore the information related to the momentum distribution
of the initial state partons inside the proton and effectively consider only the information
encoded in the decay itself. Then the kinematic discriminant (\ref{D final})
simplifies to become:
\begin{equation}\label{D simple}
KD(A; B) = \ln 
\left(
\frac
{ \left| {\cal M}_A (a + b   \to 4\ell ) \right|^2}
{ \left| {\cal M}_B (a' + b' \to 4\ell ) \right|^2}
\right).
\end{equation}
In Sec.~\ref{sec:IS}, we explicitly analyze whether this simplification results in
a substantial loss in our ability to discriminate signal from background.

There is an experimental complication that we have ignored so
far, namely that the momenta of the final state leptons are not perfectly measured. 
Generally, one can take this into account by including an integration
over transfer functions for the final state momenta in both the numerator and
denominator of Eqs.~(\ref{D final}) or~(\ref{D simple});
these integrals do not cancel in the ratio.  The transfer functions would
need to include parameterizations of momentum mismeasurements and
reconstruction efficiencies in the entire phase space relevant for the analysis.
Since leptons are well-measured, one can use delta function transfer
functions for many applications involving leptons in the final state.
However, in the case in which the signal process is the production of a
narrow resonance, the value of the signal matrix element will be
very sensitive to small perturbations in the measured invariant mass
when the instrumental mass resolution is similar to or greater than
the resonance width.  
This is the case for the channel we consider; the width of a
$125$ GeV SM Higgs boson is only about 4~MeV~\cite{Denner:2011mq},
while the instrumental four lepton mass resolution is in neighborhood
of 1~GeV~\cite{ATLASHZZ, CMSHZZ, CMSHZZlatest, ATLASHZZlatest}.  
To avoid complications associated with using transfer functions, we
employ the approach used in the CMS $H \to ZZ^\ast \to 4\ell$
analyses~\cite{CMSHZZ,CMSHZZlatest}. 
In this approach, the four lepton mass information is factorized from the
rest of the kinematic information by calculating the signal matrix element
for $m_H = m^{obs}_{4\ell}$, where $m^{obs}_{4\ell}$ is the observed
four lepton mass of a given event:
\begin{equation}
\left| {\cal M}_{H} (a + b   \to 4\ell ) \right|^2 =
\left| {\cal M}_{H} ( gg \to H  \to 4\ell \, | \, m_H = m^{obs}_{4\ell} ) \right|^2 ,
\end{equation}
This matrix element can be used both in Eqs.~(\ref{D final}) and~(\ref{D simple}). 
The instrumental four lepton mass resolution is then encoded in 2D-$pdf$'s:
\begin{itemize}
\item for signal:     $pdf_H(m_{4\ell}, KD \, | \, m_H) = pdf_H(m_{4\ell} \, | \, m_H) \cdot pdf_H(KD \, | \,  m_{4\ell})$;
\item for background: $pdf_{ZZ}(m_{4\ell}, KD )             = pdf_{ZZ}(m_{4\ell} )           \cdot pdf_{ZZ}(KD \, | \, m_{4\ell})$,
\end{itemize}
which are ultimately used for the construction of the test statistic itself.

\subsection{Event generation} 
\label{sec:leptoncuts}

For the analyses described below, we use parton-level events generated
with {\sc MadGraph} \cite{Alwall:2011uj, Alwall:2007st,
  Alwall:2008pm}.
Both signal and background events were generated within the 
four lepton mass window $120 < m_{4\ell} < 130$ GeV.
When generating the signal, we use a nominal Higgs boson mass of
$m_H=125$ GeV.
For background $ZZ$ events, only processes initiated by the first two
generation quarks were generated.  
The lepton acceptance cuts were~\cite{CMSHZZlatest}
\begin{itemize}
\item $p_T > 7$ GeV and $|\eta| < 2.5$ for electrons,
\item $p_T > 5$ GeV and $|\eta| < 2.4$ for muons.
\end{itemize}
Events are generated for each possible choice of final state flavors,
namely $e^-e^+\mu^-\mu^+$ (``different favor'', DF), and $e^-e^+e^-e^+$ and
$\mu^-\mu^+\mu^-\mu^+$ (``same flavor'', SF).  
In the case of DF events, 
opposite-charge lepton pairs are formed by flavor
$(e^-e^+)(\mu^-\mu^+)$. 
In the SF case, two different pairings of opposite-charge leptons are possible,
e.g. $(e_1^-e_2^+)(e_3^-e_4^+)$ and $(e_1^-e_4^+)(e_3^-e_2^+)$. 
The invariant mass 
of a pair closest to the mass of Z boson is $m_{Z1}$ ($m_{12}$) in CMS (ATLAS) notations, 
while the invariant mass of the other pair is $m_{Z2}$ ($m_{34})$. 
We accept the event if at least one pairing of leptons passes 
the invariant mass cuts of 40 and 12~GeV on $m_{Z1}$ ($m_{12}$) and $m_{Z2}$ ($m_{34})$, respectively.   
We do not perform any
detector simulation as our primary concern is with the comparison of the 
different analyses as opposed to the absolute performance of any given analysis.

\subsection{ROC curves}
\label{sec:ROC}

We wish to characterize the extent to which our $KD$ is useful
in enhancing the sensitivity of an analysis.  One can
generate pseudoexperiments using the two-dimensional $pdf$s
(involving KD and invariant mass $m_{4\ell}$) described above and
perform a statistical analysis on how well two hypotheses
can be separated. In fact, we perform such a prototypical 
analysis in Sec.~\ref{sec:JCP}.  
However, a simpler and more intuitive means of displaying the extent 
to which signal and background may be separated is provided by 
``Receiver Operating Characteristic'' (ROC) curves \cite{ROC}.

Using generated events (see previous subsection), we build
distributions of $KD(A;B)$ for processes A and B that we wish to compare. 
Then, we calculate cumulative probabilities $\epsilon_A( \, KD \ge KD_{\textsc {CUT}} \, )$
and $\epsilon_B( \, KD \ge KD_{\textsc {CUT}} \, )$. The plot of 
$\epsilon_A( \, KD \ge KD_{\textsc {CUT}} )$
versus $\epsilon_B( \, KD \ge KD_{\textsc {CUT}} )$, as one smoothly increases 
the cut value $KD_{\textsc {CUT}}$, is the ROC curve.
The curve gives the fraction of
process A events that pass a given cut on $KD$ in terms of the fraction
of process B events passing the same given cut.
The further the curve is from the diagonal ($\epsilon_A = \epsilon_B$), 
the better the cut has separated the two processes.

\subsection{Comparison with single variable analyses}

The most sensitive kinematic variable for distinguishing signal and
background is, of course, the invariant mass of the event.  
However, if we want
to move beyond this and obtain greater sensitivity, a natural question
is whether we truly need to use all of the kinematic variables in our
analysis, or whether using one or two particularly sensitive variables
would suffice. We therefore investigate the angular variables 
$\theta^\ast$ and $\Phi$, in the
convention of Refs.~\cite{Gao:2010qx,Bolognesi:2012mm}, as well as
$m_{Z2}$, the mass of the less massive opposite-charge dilepton pair~\cite{mZ2}.  
We choose $m_{Z2}$ in particular, as due to the $Z\gamma^\ast$ contribution 
to the background, the signal and background shapes are significantly different.
The $ZZ$ background has a substantial fraction of events with 
$m_{Z2}$ values close to the cut value of 12 GeV, while for a Higgs boson
decaying to actual $Z$ bosons at tree level, the low values of $m_{Z2}$ are
substantially less likely (due to the loop suppression, 
the $H\to Z\gamma^\ast$ decay can be neglected). 
The motivation for considering the angle $\theta^\ast$, which is the
$Z$ boson polar angle in the center of mass frame of the event, is
that the background $\theta^\ast$ distribution is much more peaked in the forward 
and backward directions (in comparison with signal). This is due to the
$t$-channel nature of the $q \bar q \to ZZ$ processes.
The angle $\Phi$, the angle between the two dilepton decay planes, 
has often been considered in the literature for distinguishing between 
spin and parity hypotheses (see, e.g.,  Refs.~\cite{Nelson:1986ki,Choi:2002jk}).  

\begin{figure}[t]
\centering
\includegraphics[width= \1]{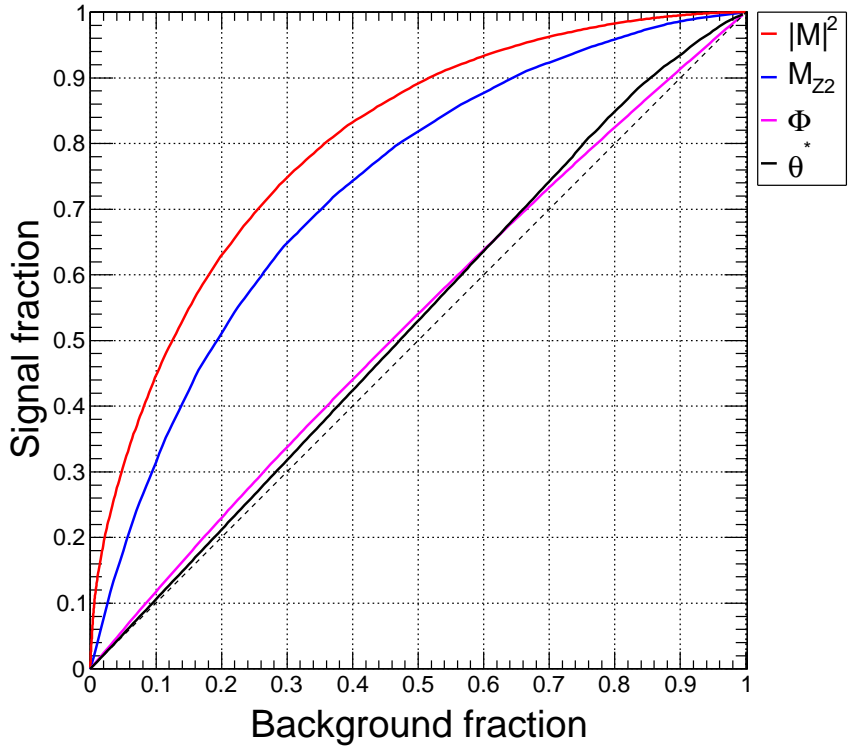}
\caption{\label{fig:ROC} Comparison of different ROC curves based on: 
  the value of $KD$ obtained from matrix elements (red curve), 
  $m_{Z2}$ (blue curve), $\Phi$ (magenta curve), and
  $\theta^\ast$ (black curve).  The black dashed diagonal line is the ROC
  curve obtained from cutting events indiscriminately (e.g., by
  flipping a fair coin or by only considering some fraction of the data
  set).}
\end{figure}

Using simulated SM Higgs boson and $ZZ$ background events, 
we construct numerical $pdf$s $P_s(x)$ and $P_b(x)$ for each of these variables
(here $x$ stands for either $m_{Z2}$ or $\theta^\ast$ or $\Phi$).
We then take the ratio of these probabilities to construct
discriminants and produce corresponding ROC curves.  
These three ROC curves are shown in Fig.~\ref{fig:ROC}, 
along with the ROC curve constructed using the
MEM $KD(H;ZZ)$ calculated using the {\sc MadGraph} matrix element 
(see Sec.~\ref{sec:MG} for details).  
One can see that the analysis utilizing $KD(H;ZZ)$ 
based on the complete matrix element is the most sensitive. We
also note that $m_{Z2}$ is very sensitive in its own right. This observation
will become important in Sec.~\ref{sec:jcpVsBackground}, where we
discuss the signal vs. background separation when assuming different signal
hypotheses.

\subsection{Overview of available tools for calculating $KD$}
\label{sec:calculation}

Since an event generator performs a Monte Carlo integration of
a matrix element squared over the relevant phase space, in principle
{\em any} event generator can be used to calculate the matrix element,
given the final state kinematics. Unfortunately, not all event generators have 
implemented this functionality. 

In this paper, we shall use
{\sc MadGraph}~\cite{Stelzer:1994ta,Alwall:2011uj}, 
and {\sc CalcHEP} \cite{Belyaev:2012qa}, 
which can easily provide
analytical expressions for the matrix element for the relevant $2\to
4$ scattering processes, e.g, $gg\to H\to ZZ^\ast\to 4\ell$ for signal and $q\bar{q}\to ZZ^\ast\to 4\ell$ for the background. 
{\sc MadGraph} and {\sc CalcHEP} are on a relatively equal footing --- 
they are both highly automated and very flexible, and thus provide a 
useful cross-check on each other (this test is performed in Appendix~\ref{sec:CH}).
Using the built-in flexibility of these generators, we construct
a higher-level package, {\sc MEKD} (Sec.~\ref{sec:MG}), that 
calculates complete leading order matrix elements for $gg \to X \to 4\ell$
and $q \bar q \to 4\ell$ for any given four lepton final state. This
package was used along with the {\sc MELA} package, to be described below, 
by CMS in their most recent update on the observation of the newly
discovered boson
decaying to $ZZ \to 4\ell$ and the measurements of its properties
in this channel~\cite{CMSHZZlatest}.

As an alternative to the automated tools, one could utilize a
specialized package, perhaps
because it has the expressions for the relevant matrix elements 
readily available or because it offers extra functionality not accessible in
the automated tools.
Two examples of this type include {\sc NLOME}\footnote{We thank J.~Campbell and C.~Williams for making the beta version of 
the {\sc NLOME} code available to us.}~\cite{NLOME}, which is based on the
{\sc MCFM} event generator \cite{Campbell:2010ff}, 
and {\sc JHUgen} \cite{JHUgen}. 
The crucial advantage of {\sc NLOME} over all other competitors is
its next-to-leading order (NLO) functionality \cite{Campbell:2012cz,Campbell:2012ct}.
On the other hand, the $H \to ZZ \to 4\ell$ channel 
is built around the matrix element for the DF $e^-e^+\mu^-\mu^+$
final state. Therefore, when applying it to events with the SF
$4e$ or $4\mu$ final states, the interference effects are missed. 
Also, signals of alternative spins and parities are not available in this package.  
{\sc JHUgen} allows one to extract the leading order matrix element for a broad spectrum of possible
signal hypotheses, but does not calculate the matrix element for backgrounds. 

There is yet another option: one may wish to code one's own software package using expressions 
for matrix elements found in the literature, such as those in Ref.~\cite{Gao:2010qx,Bolognesi:2012mm} 
for signal and Ref.~\cite{Gainer:2011xz,Chen:2012jy} for background.
This approach was used by both CMS and ATLAS. Although the corresponding 
software packages used by the two experiments are not publicly available, 
for completeness we describe their main features, 
derived from the references the codes are based on. 

In their new boson discovery paper \cite{CMSHZZ}, the CMS collaboration used 
the {\sc MELA} package which is largely based on 
the work in \cite{Gao:2010qx,Bolognesi:2012mm}.
For an event with four lepton mass $m_{4\ell}$ above the on-shell ZZ threshold, 
the signal and background probabilities for the event, 
$P_{\textsc{H}}( \, \mathbf{\Omega} \, | \, m_{\textsc{H}} = m_{4\ell} \, )$ and 
$P_{\textsc{ZZ}}( \, \mathbf{\Omega} \, | \, m_{4\ell} \, )$, were calculated
analytically using equations from Refs.~\cite{Gao:2010qx, Gainer:2011xz}. 
Here, the vector $\mathbf{\Omega}$ denotes five angles describing the four lepton
system in its center of mass frame (see Appendix~\ref{sec:notation}).
For an event with four lepton mass below the on-shell ZZ threshold,
the dilepton masses $m_{Z1}$ and $m_{Z2}$ must be included in the matrix element calculations. 
The signal probability 
$P_{\textsc{H}}( \, m_{Z1}, m_{Z2}, \mathbf{\Omega} \, | \, m_{\textsc{H}} = m_{4\ell} \, )$
was calculated
analytically~\cite{Bolognesi:2012mm},
while the background probability was
tabulated from template distributions for $m_{Z1}$, $m_{Z2}$, and the five angles, 
obtained with the {\sc POWHEG} simulation at generator level \cite{CMSHZZ,MELAtalk}.
The analytic computations of the matrix elements 
did not include same-type lepton permutations and
associated interferences effects in the SF four lepton final states.

In the recently released $H \to ZZ \to 4\ell$ analysis update by
CMS~\cite{CMSHZZlatest}, 
the {\sc MELA} package was upgraded and now includes the background probability 
$P_{\textsc{ZZ}}( \, m_{Z1}, m_{Z2}, \mathbf{\Omega} \, | \, m_{4\ell} \, )$
for the low mass range calculated analytically using parameterizations
from Ref.~\cite{Chen:2012jy}.
The permutation/interference effects for the SF four lepton
final states are still ignored. In the same Ref.~\cite{CMSHZZlatest},
CMS also used the {\sc MEKD} package,
which has complete leading order matrix elements for signal
and background, including the proper treatment of lepton permutations
and the associated interference  present in SF four lepton final states.

In their recent update on the $H \to ZZ \to 4\ell$ channel~\cite{ATLASHZZlatest},
ATLAS also used the matrix element method for testing various spin-parity
hypotheses for the newly discovered boson. In these analyses, the kinematic
discriminants are built solely from signal matrix elements based on
Ref.~\cite{Bolognesi:2012mm}. Hence, the constructed kinematic
discriminants do not account for permutations of identical leptons
and the associated interference. Ignoring the presence of background
in these MEM-based analyses also makes them even less optimal. 

Given that the permutation/interference effects have not always 
been consistently implemented in the past, 
a large part of the results and discussion presented below 
are focused on the investigation of 
the effects from interference on the physics results.
We show that in some cases the interference effects play a significant role and should not
be neglected in separating alternative hypotheses describing the observed excess of
events near the mass of 125~GeV.

\section{Matrix Element Kinematic Discriminant (MEKD)}
\label{sec:MG}
%===============================================================================
The {\sc MEKD} code is a publicly available package~\cite{KDcode};
the instructions for download and usage are provided in Appendix~\ref{sec:MEKD}.
This code uses {\sc MadGraph} to calculate complete LO matrix elements for signal $gg \to X \to ZZ \to 4\ell$
and background process $q \bar q \to ZZ \to 4\ell$ and also builds the kinematic discriminants 
according to Eqs.~(\ref{D final}) and~(\ref{D simple}). 
The SM Lagrangian and the non-SM additions are implemented automatically 
into {\sc MadGraph} via {\sc FeynRules}~\cite{Christensen:2008py}. 

The {\sc MEKD} calculations are validated twice: first, by comparing matrix elements
calculated by {\sc MadGraph} and {\sc CalcHEP}, both embedded into
the same {\sc MEKD} framework; and, second, by comparing the matrix elements calculated within
the {\sc MEKD} framework with the LO matrix elements calculated by the standalone {\sc NLOME} package.
The {\sc MadGraph} vs. {\sc CalcHEP} validation is performed for all three final states,
$4e$, $4\mu$, $2e2\mu$, separately, while {\sc MadGraph} vs. {\sc NLOME} validation 
is performed only for the $2e2\mu$ final state (the version of {\sc NLOME} available to us 
did not include interference for the $4e$ and $4\mu$ final states). The results 
are found to be identical for all practical purposes; 
the details can be found in Appendix~\ref{sec:CH}.

In the current version of {\sc MEKD}, the SM Lagrangian is extended to allow
for non-SM couplings of a spin zero or spin two resonance to gluons and to
$Z$ bosons. To be specific, following \cite{Gao:2010qx,Bolognesi:2012mm}, 
we implement the following terms in the Lagrangian for a
spin zero resonance:
\begin{eqnarray}
\mathcal{L}_{HZZ} &\ni&  -\frac{g_{1z} }{2} H Z_\mu Z^\mu -\frac{g_{2z}}{4}  H Z_{\mu\nu} Z^{\mu \nu}
 -\frac{g_{3z}}{2} Z_{\mu\alpha}Z^{\mu\beta}
\left(\partial_\beta\, \partial^\alpha H\right) -\frac{g_{4z}}{4}  H Z_{\mu \nu} \tilde{Z}^{\mu \nu} , \\
\mathcal{L}_{Hgg} &\ni&  -\frac{g_{1g} }{2} H F_\mu^a F^{a,\mu} -\frac{g_{2g}}{4}  H F_{\mu\nu}^a F^{a, \mu \nu}
 -\frac{g_{3g}}{2} F_{\mu\alpha}^a F^{a,\mu\beta}
\left(\partial_\beta\, \partial^\alpha H \right) -\frac{g_{4g}}{4}  H F_{\mu \nu}^a \tilde{F}^{a, \mu \nu},~~~~
\end{eqnarray}
where
 $H$ is generic scalar field, $Z(F)^\mu$ is the vector potential for
 the $Z$ boson (gluon), and $Z(F)^{\mu\nu}$ is the field strength
 tensor for the $Z$ boson (gluon). For the specific case of the SM
 Higgs boson, one has
 \begin{eqnarray}
&& \left(g_{1z}, g_{2z}, g_{3z}, g_{4z} \right) = \left(\frac{2 m_Z^2} {<v> }, 0,0,0 \right) ,  \\
&& \left(g_{1g}, g_{2g}, g_{3g}, g_{4g} \right) = \left(0, g_{\textrm{eff} ,0,0} \right).
\end{eqnarray}
Another hypothesis which we will consider below is a CP odd spin zero
resonance, for which we have
 \begin{eqnarray}
&& \left(g_{1z}, g_{2z}, g_{3z}, g_{4z} \right) = \left(0, 0,0,g^\prime_{\textrm{eff,Z}} \right) ,  \\
&& \left(g_{1g}, g_{2g}, g_{3g}, g_{4g} \right) = \left(0, 0 ,0,g^\prime_{\textrm{eff,g}} \right),
\end{eqnarray}
where the product of $g^\prime_{\textrm{eff,Z}}$ and
  $g^\prime_{\textrm{eff,g}}$ is chosen so that we obtain the same
  total cross section times branching ratio as in the case of the SM
  Higgs boson. 
  Of course, for a spin zero resonance, no angular distributions depend
  on which particular couplings with gluons are non-vanishing, at least at
  leading order.

In the case of a general spin two resonance, we have
\begin{eqnarray} 
\mathcal{L}_{GZZ}~\ni   &&k_{1z} h_{\mu \nu} Z^{\mu \alpha} Z^{\nu}_\alpha + k_{2z} \left(\partial_\alpha \partial_\beta h_{\mu \nu} \right )Z^{\mu \alpha} Z^{\nu \beta} + k_{3z} h_{\beta \nu}\left( \partial^\alpha Z^{\mu \nu} \partial^{\beta} Z_{\mu \alpha} + \del^{\alpha} Z_{\mu \alpha}\del^{\beta} Z^{\mu \nu} \right)  \nonumber\\
 &&+ k_{4z} h_{\mu \nu} \del^{\mu} Z^{\alpha \beta} \del^{\nu} Z_{\alpha \beta}+k_{5z} h_{\mu \nu} Z^{\mu} Z^{\nu}
 +k_{6z} (\del_{\alpha} h_{\mu \nu}) Z^{\nu} \del^{\mu} Z^{\alpha} + k_{7z} h_{\mu \nu} \del^{\mu} Z^{\alpha} \del^{\nu} Z_{\alpha} \nonumber \\
 &&+ k_{8z} h_{\mu \nu} \del^{\mu} Z^{\alpha \beta} \del^{\nu} \tilde{Z}_{\alpha \beta}+k_{9z} ( \del_{\alpha} h_{\mu \alpha} ) \epsilon^{\mu \nu \rho \sigma} 
 \left( \del^{\alpha} Z_{\nu}\right) Z_{\rho}  \nonumber \\
 &&+ k_{10z} \left(\del_{\beta} \del_{\rho} h_{\mu \alpha} \right) \epsilon^{\mu \nu \rho \sigma}
 \left(\del^{\alpha} Z_{\nu} \del_{\sigma} Z^{\beta} +\del^{\alpha} Z^{\beta}\del_{\sigma} Z_{\nu}\right),
 \end{eqnarray}
and an analogous effective Lagrangian ${\mathcal{L}}_{Ggg}$ describing
the coupling of an arbitrary spin two resonance to gluons.

In what follows we will consider the hypothesis of a
massive graviton~\cite{Randall:1999ee}, 
for which the values of couplings are
 \begin{eqnarray}
&& \left(k_{1z}, k_{2z}, k_{3z}, k_{4z}, k_{5z}, k_{6z}, k_{7z},
  k_{8z}, k_{9z}, k_{10z} \right) = \left(k_{1z}, 0,0,0,  - m_Z^2 k_{1z},0,0,0,0,0 \right) ,  \\
&& \left(k_{1g}, k_{2g}, k_{3g}, k_{4g}, k_{5g}, k_{6g}, k_{7g},
  k_{8g}, k_{9g}, k_{10g} \right) = \left(k_{1g}, 0 ,0,0,0,0,0,0,0,0 \right),
\end{eqnarray}
(see e.g. Ref.~\cite{Han:1998sg}), with the $k_{1z}$ and $k_{1g}$ set to give
the same total cross section as in the SM Higgs boson case.

The user has complete freedom to adjust any of the couplings in the
spin zero and spin two
Lagrangians described above. The case of a spin one boson and other extended functionality
are coming shortly. Hence, the {\sc MEKD} code can be used for
general studies of the Higgs-like resonance, and in particular for the
measurement of these couplings~\cite{newUFpaper}.

\section{Separation of the standard model Higgs boson and background}
\label{sec:performance}

\begin{figure}[t]
\centering
\includegraphics[width=\2]{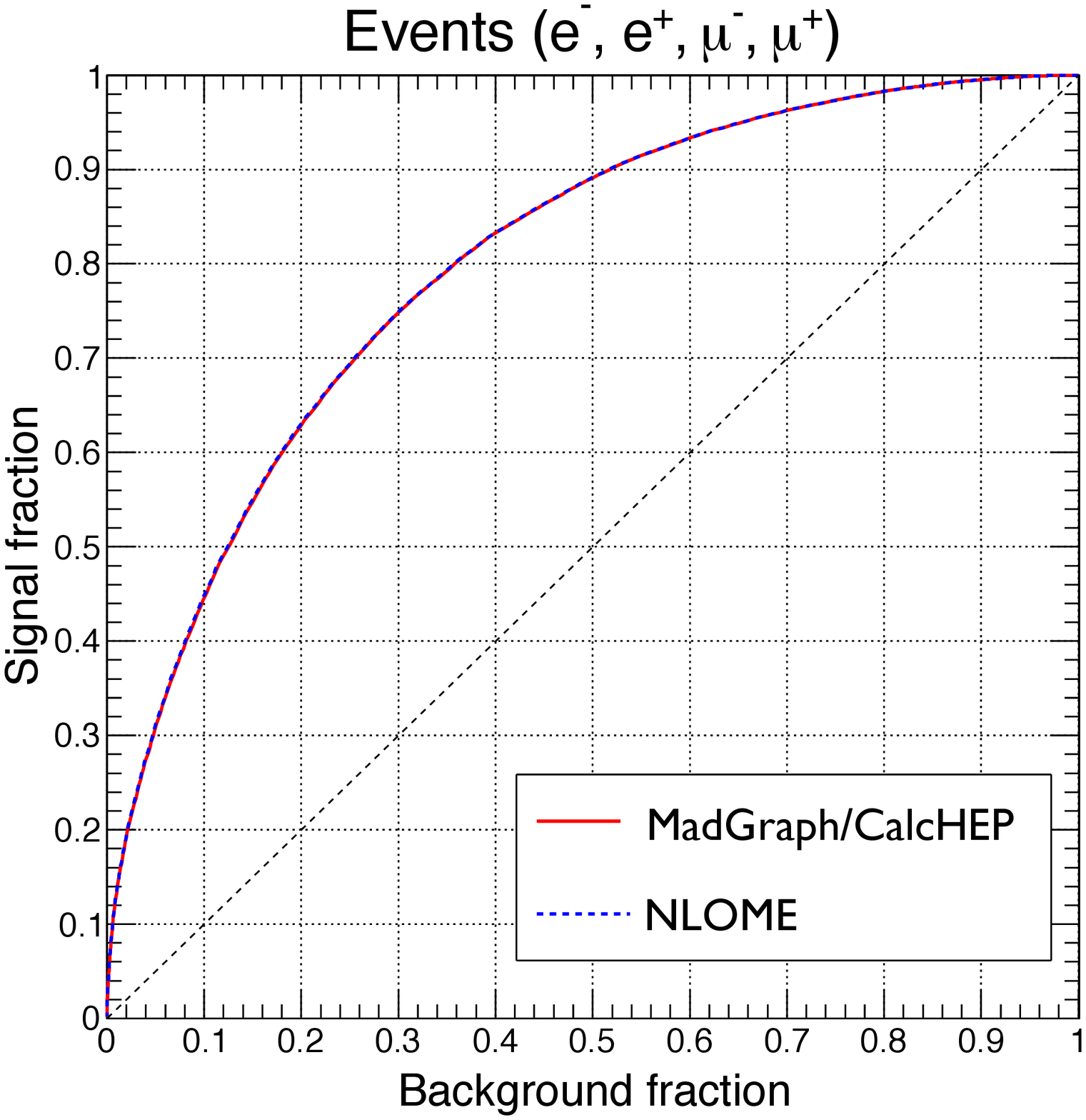}
\includegraphics[width=\2]{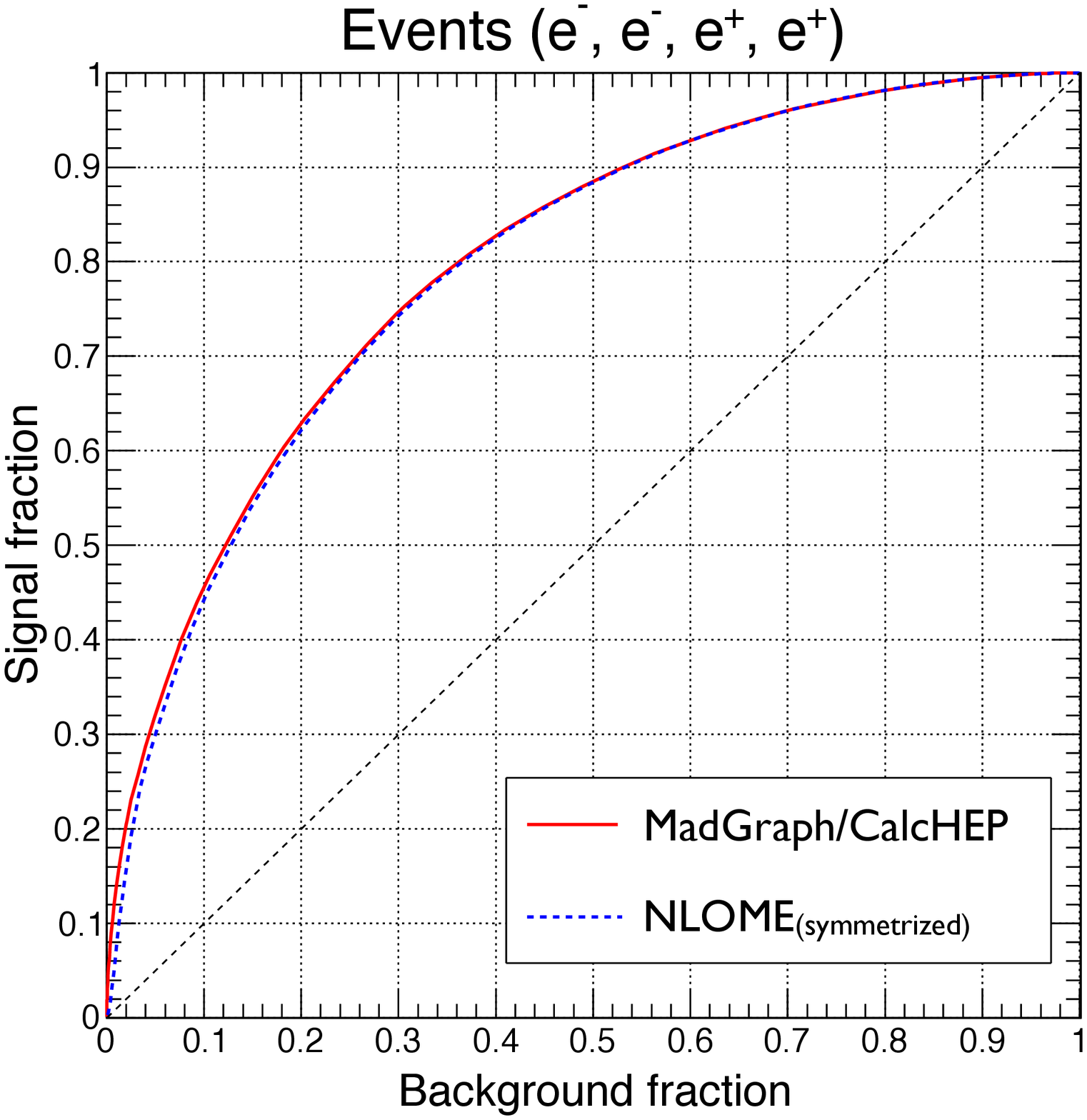}
\caption{\label{fig:ROC4methods} Comparison of ROC curves obtained from
the different implementations of the kinematic discriminant:
from {\sc MadGraph/CalcHEP} (red solid lines) or from {\sc NLOME} (blue dotted lines),
for DF (left) and SF (right) $4\ell$ events.}
\end{figure}

As discussed in Sec.~\ref{sec:ROC}, the 
effectiveness of a kinematic discriminant can be assessed using the
corresponding ROC curve.
Figure~\ref{fig:ROC4methods} shows ROC curves characterizing 
the performance of the kinematic discriminant $KD(H;ZZ)$ that is built to sort events
based on whether they are more likely to be produced via the SM Higgs
boson or by $ZZ$ background processes. 
As shown in Appendix~\ref{sec:CH}, {\sc MadGraph} and {\sc CalcHEP} give
identical results, so for the purposes of this figure we refer to them as {\sc MadGraph/CalcHEP}.

Fig.~\ref{fig:ROC4methods} (left) shows that the ROC curves for $KD(H;ZZ)$ 
as obtained with {\sc NLOME} and {\sc MadGraph/CalcHEP} for $2e2\mu$ events
are {\em identical}, as expected from the comparisons in Appendix~\ref{sec:CH}.
One can see that if one were to use a simple cut on $KD$ as a part
of the event selection in a SM Higgs boson search analysis, then
one could suppress background by about a factor of two, while keeping 90\%
of signal events. 

Fig.~\ref{fig:ROC4methods} (right) shows that the 
power of the kinematic discriminant
to separate signal from background in the SF 
four lepton final states is approximately the same.  However, 
a closer look shows a very small difference, $\mathcal{O}(1\%)$, between {\sc MadGraph/CalcHEP}
and {\sc NLOME(MCFM)} results. This is due to the fact that {\sc NLOME(MCFM)} 
is missing permutations of identical leptons and the associated interference
present for SF four lepton final states. 
To alleviate the problem somewhat, we recycle the DF
matrix element
$$
\left|{\cal M}_{DF} \left( (e^-e^+)(\mu^-\mu^+) \right) \right|^2
$$
to make a ``symmetrized'' matrix element for the SF
four lepton final states:
\beq
\left| {\cal M}_{SF}\left( e_1^-e_2^+e_3^-e_4^+ \right) \right|^2_{\mathrm{symmetrized}}
\approx
\left| {\cal M}_{DF}\left( (e_1^-e_2^+)(e_3^-e_4^+) \right) \right|^2+
\left| {\cal M}_{DF}\left( (e_1^-e_4^+)(e_3^-e_2^+) \right) \right|^2,
\label{patch1}
\eeq
which now includes permutations of identical leptons,
but, of course, still misses the interference terms.

\begin{figure}[t]
\centering
\includegraphics[width=\3]{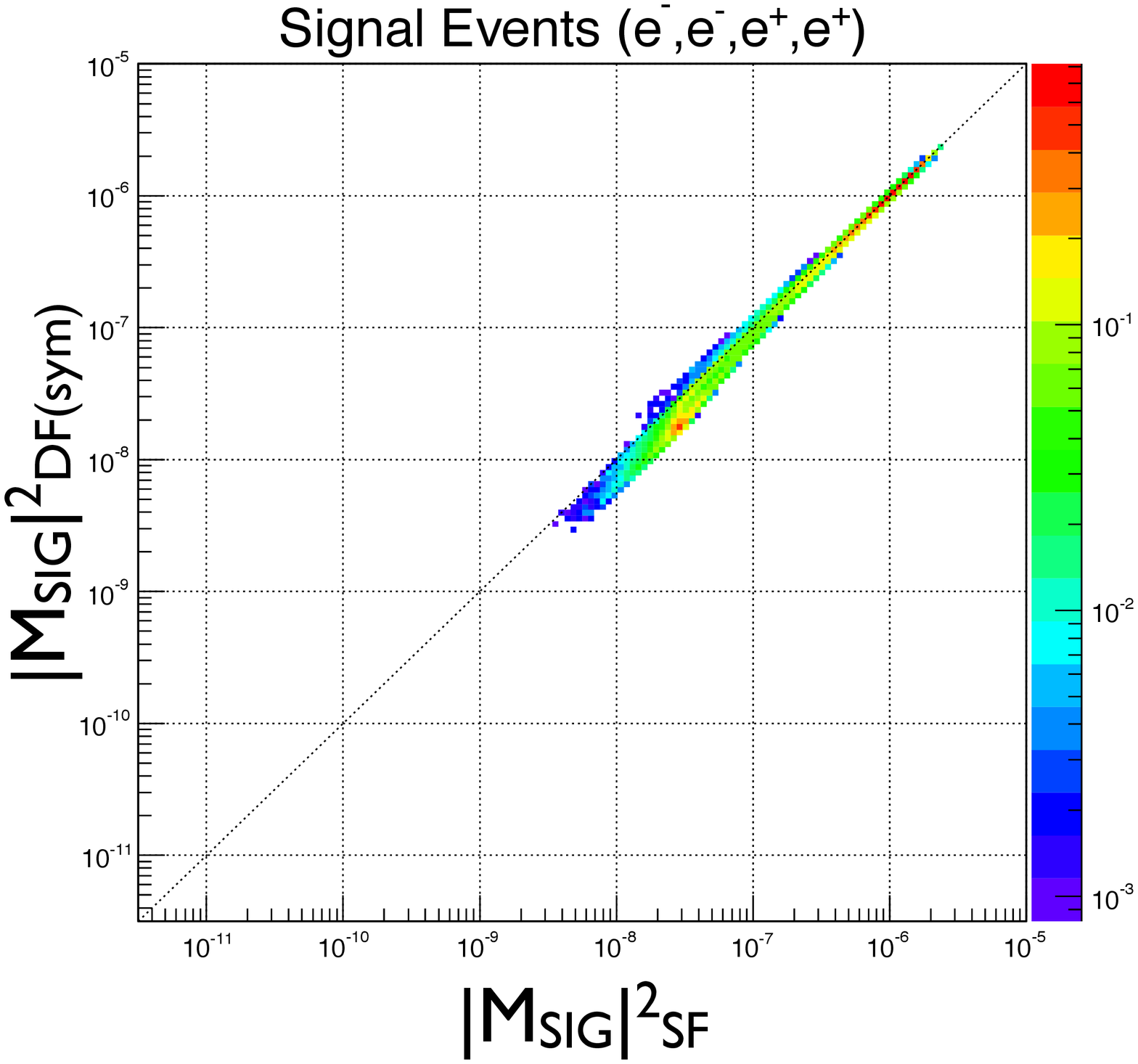}
\includegraphics[width=\3]{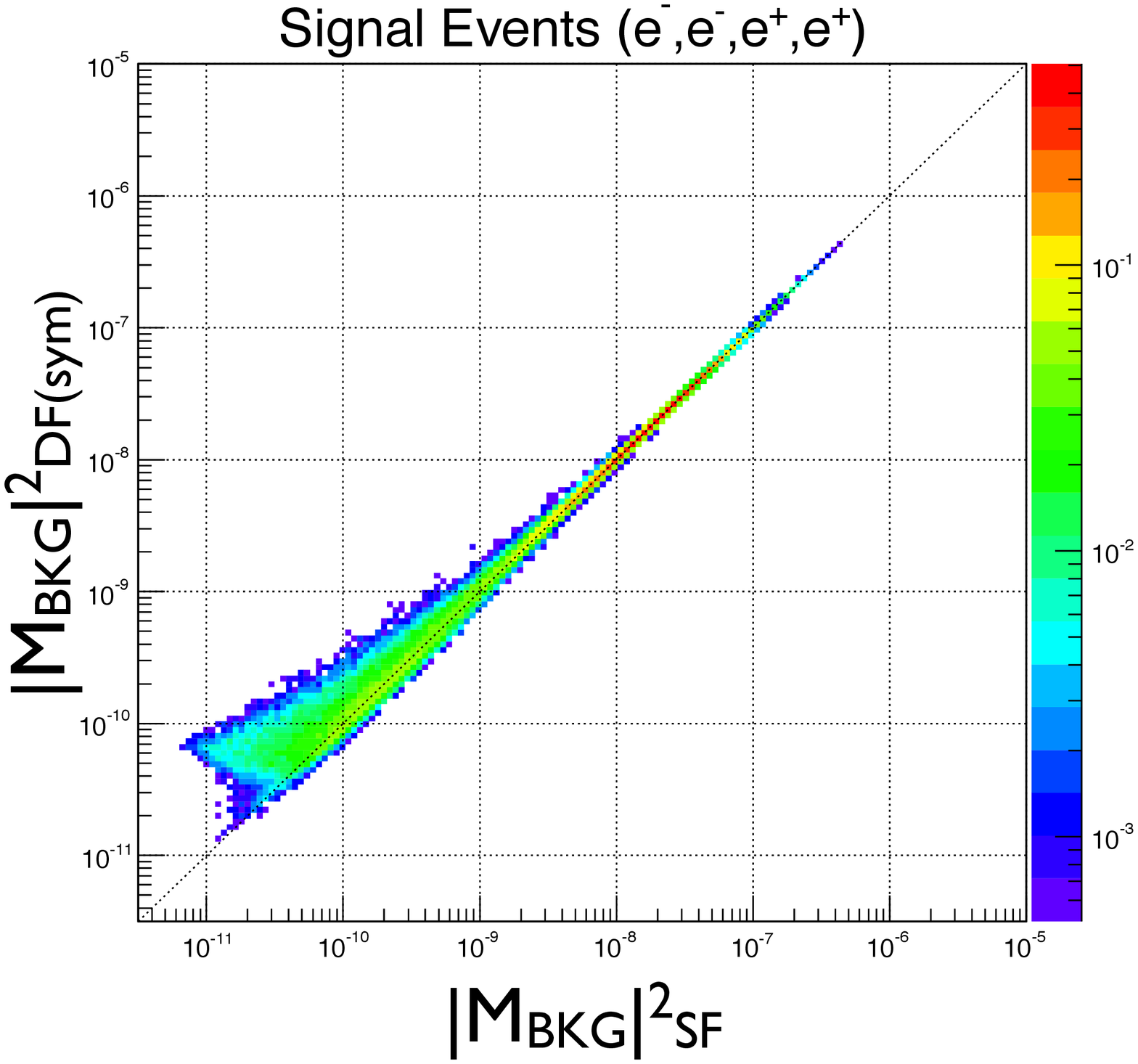}
\includegraphics[width=\3]{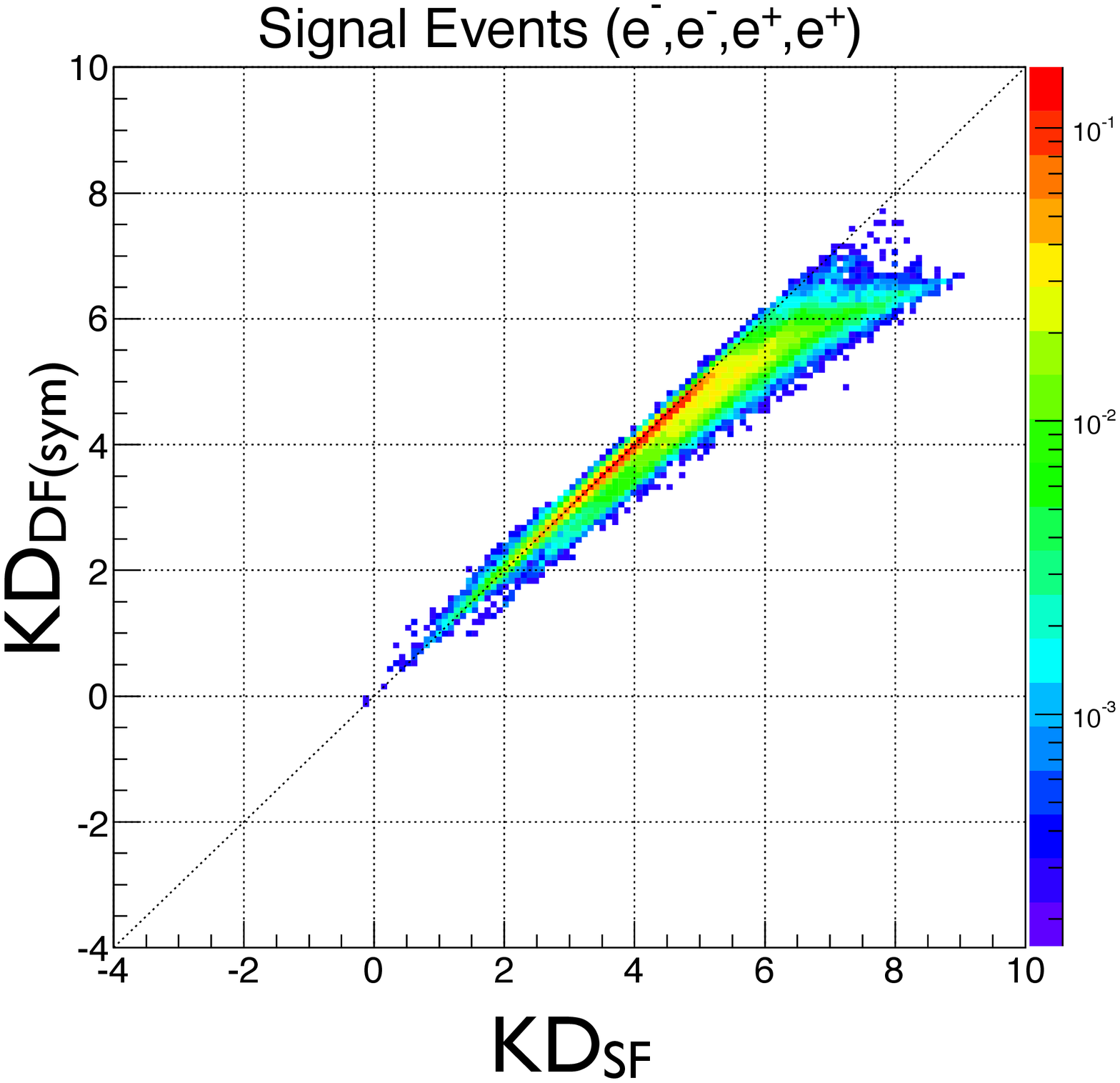}\\
\includegraphics[width=\3]{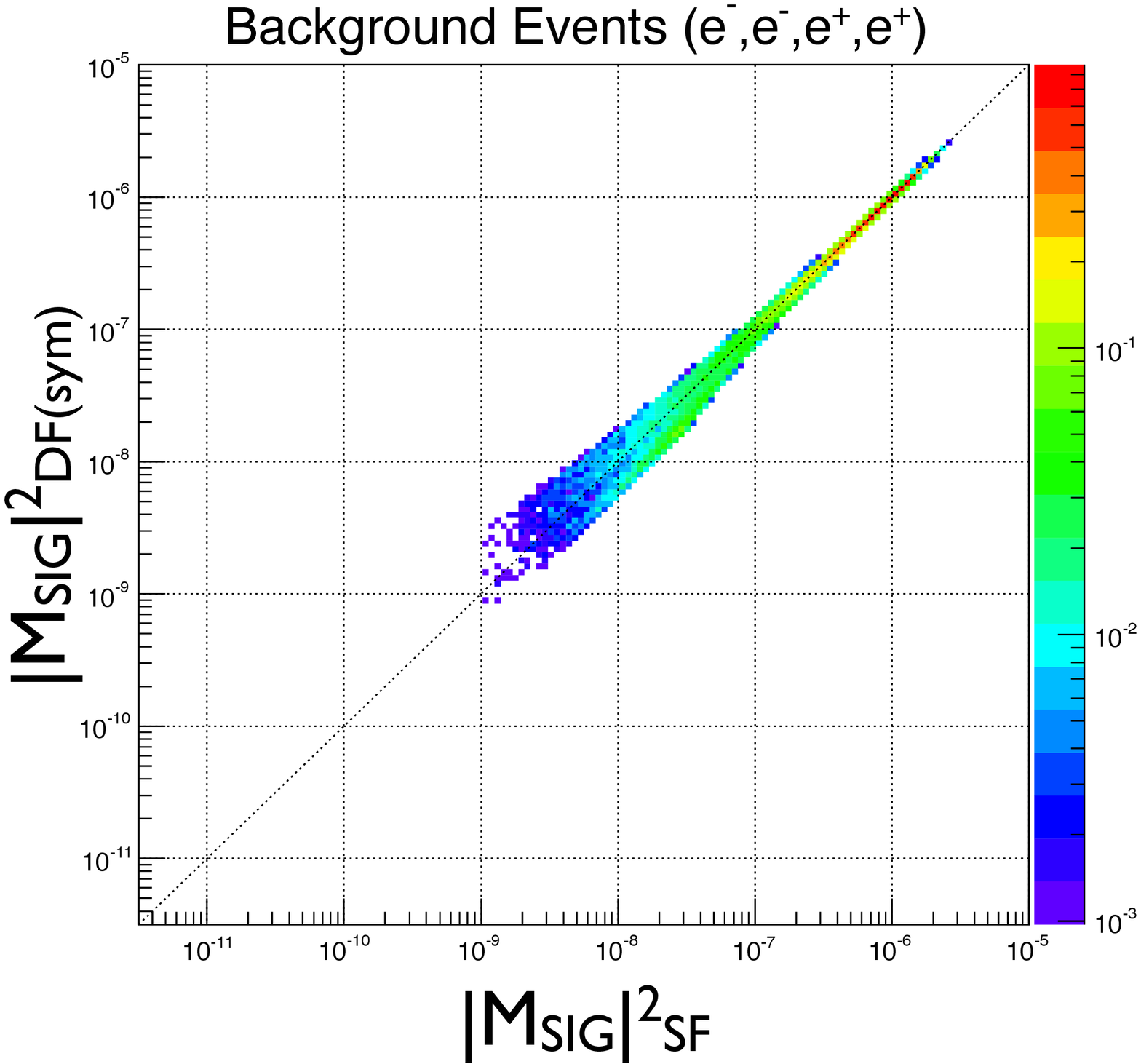}
\includegraphics[width=\3]{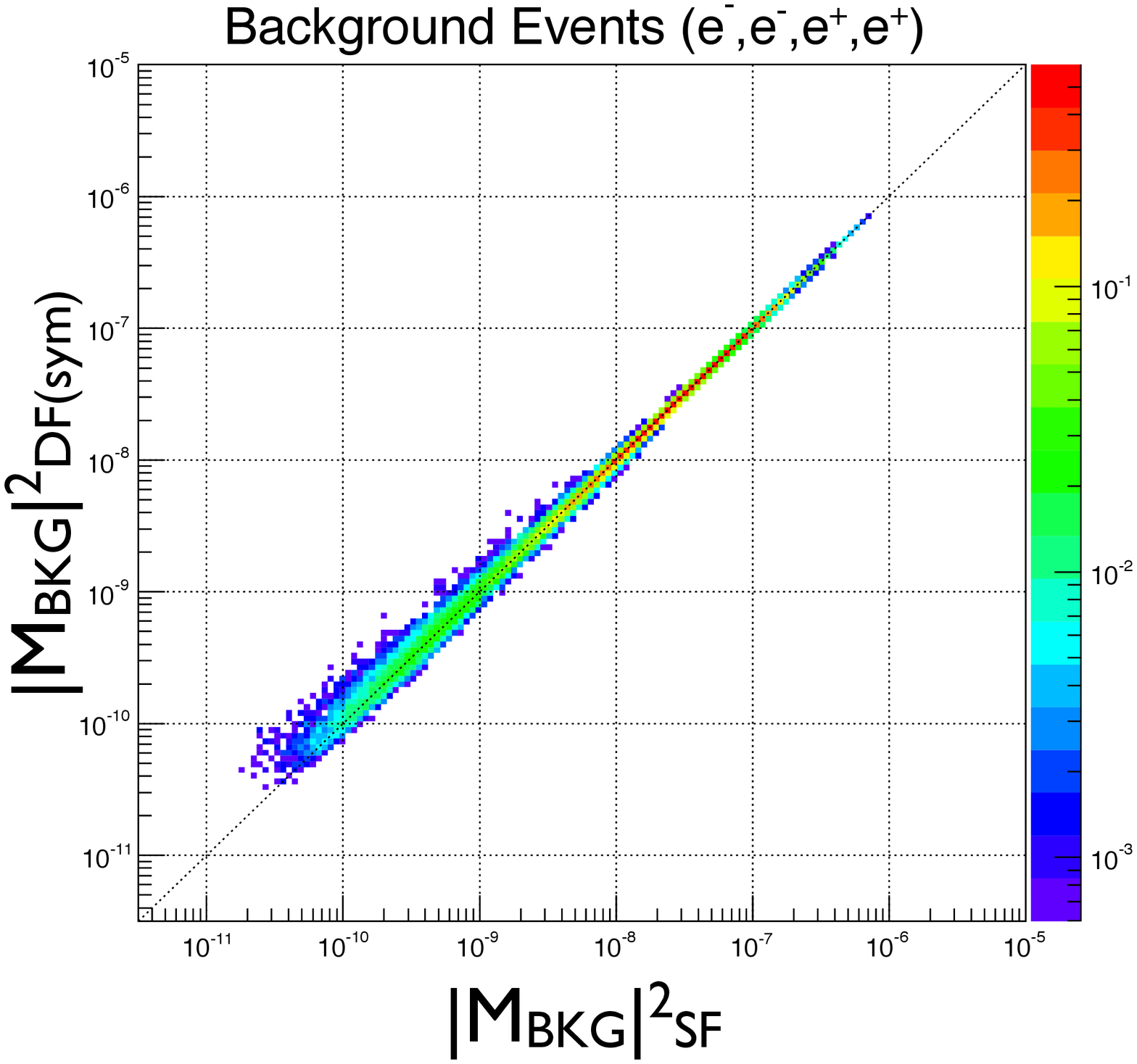}
\includegraphics[width=\3]{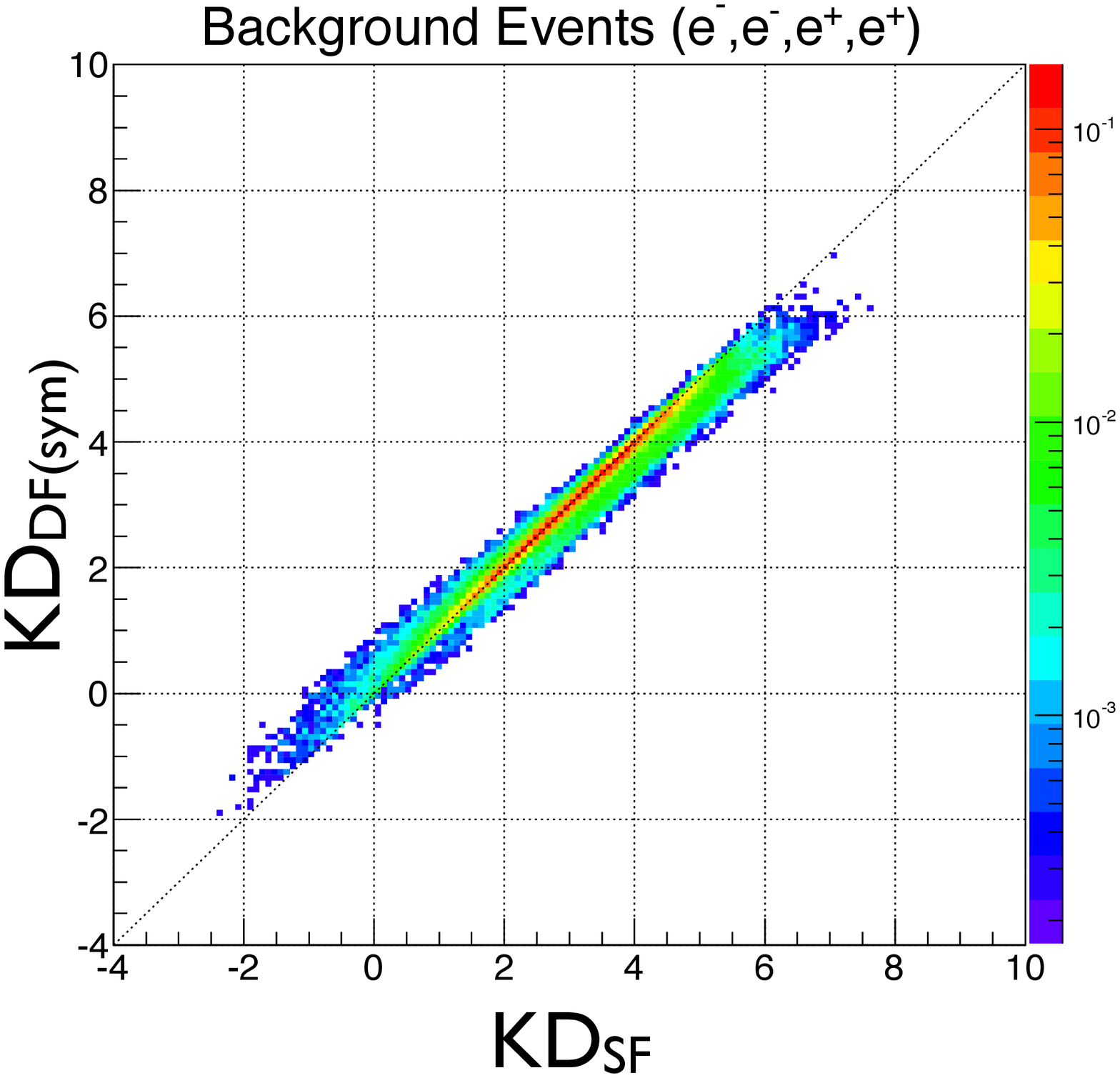}
\caption{\label{fig:SF} Comparison of the signal (left column) and
  background (middle column) matrix elements, as well as the
  kinematic discriminant $KD_{\textsc{MAD}}$ (right column), for signal events
  (top row) and background events (bottom row) as calculated with the
  full matrix element for SF events $|{\cal M}|^2_{SF}$ (including the interference) 
versus the approximation $|{\cal M}|^2_{DF(sym)}$ obtained by the patch given in Eq.~(\ref{patch1}).
All quantities are calculated with {\sc MadGraph}.
}
\end{figure}

The effect of neglecting interference is more visible
in Fig.~\ref{fig:SF}, where we compare three quantities:
$ \left| \mathcal{M}_{\textsc{H}} \right|^2 $ (left column),
$ \left| \mathcal{M}_{\textsc{ZZ}} \right|^2 $ (middle column),
and $KD(H;ZZ)$ (right column), 
all calculated in two different ways using {\sc MadGraph}. 
The $x$ axis shows the full results for the SF final state.
The $y$ axis shows the results obtained using the patch shown in Eq.~(\ref{patch1}).
Since we use the exact same code, the only difference between $x$ and $y$ values
is whether or not the interference contributions are taken into
account in the corresponding code.
Fig.~\ref{fig:SF} shows a fair amount of scatter for the matrix
elements themselves as well as for their ratio; this explicitly shows
the role of interference.

For the matrix elements, the scatter becomes noticeably worse 
for low values of the squared matrix element. 
This can be easily understood: for SF events, 
there are two amplitudes corresponding to the two ways of pairing the
final state leptons.
Normally, one pairing has one $Z$ on-shell and the other $Z$ off-shell,
while the other pairing has both $Z$s off-shell. The result is then
dominated by the on-shell amplitude squared and the interference terms
are negligible.
In order for the interference terms to become noticeable, both pairings should have {\em two} 
off-shell $Z$s in which case the overall matrix element squared will be small.
We can easily observe this effect, especially for the background
matrix element (the middle plots in the figure), 
for which the $Z\gamma^\ast$ amplitude is large, which more easily allows the
$Z$s to be off-shell.
The larger spread at low $|{\cal M}_{BKG}|^2$ appears at
high $KD$ (right column plots), since the $KD$ ratio is inversely
proportional to $|{\cal M}_{BKG}|^2$.

The reason why the obvious scatter seen
in Fig.~\ref{fig:SF} has relatively little effect on the ROC curve
in Fig.~\ref{fig:ROC4methods} (right) is easy to understand. The observed scatter 
(up to $\pm$0.5 units of $KD$) is substantially smaller
than the typical range of $KD$ values (about 4 units).
Therefore, using the ``incorrect'' $KD$ does not significantly
broaden the signal and background $KD$ distributions, and the
signal vs. background separation power is not affected too much.
The similarity of the left and right plots for DF and
SF four lepton events in Fig.~\ref{fig:ROC4methods} is yet another sign that
the relative role of interference effects is not expected
to be large in separating signal and background. 
As we discuss in Sec.~\ref{sec:JCP}, this is not the case when separating
alternative signal hypotheses.

\section{
Including information about the initial state}
\label{sec:IS}
%===================================================================================
In the Higgs golden channel, the final state is fully reconstructed
and thus the momenta of the initial state partons may be inferred.
In our discussion thus far, we have ignored this additional information,
so a very relevant question is how much does one gain from considering it.
\begin{figure}[t]
\centering
\includegraphics[width=\1]{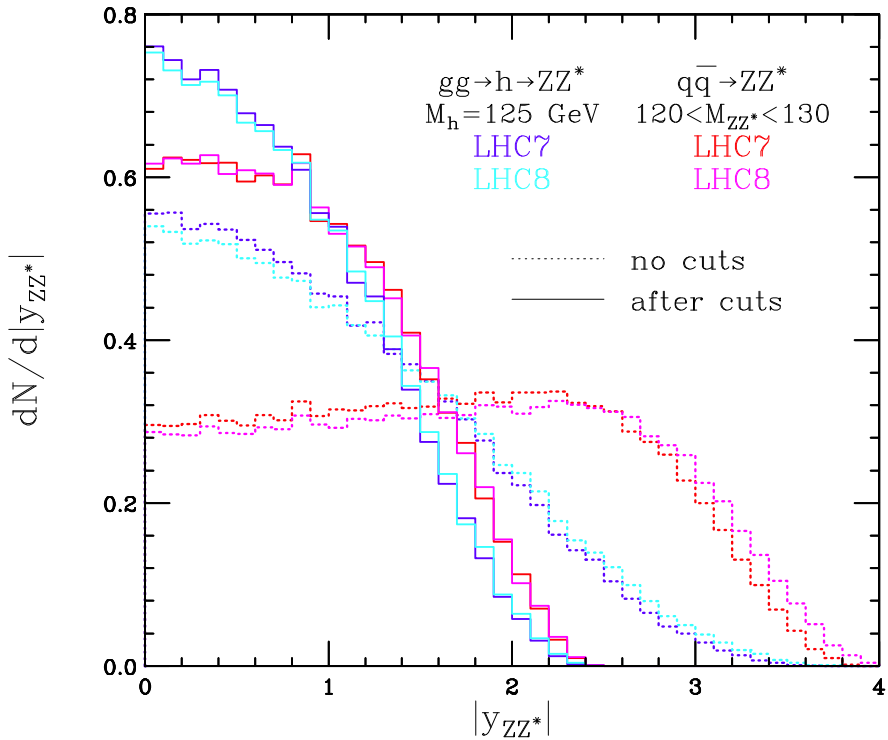}
\caption{\label{fig:rapidity} Unit-normalized rapidity distributions of $ZZ^\ast$ events for
signal (bluish colors) and background (reddish colors). Results are shown for two 
different LHC energies (7 and 8 TeV), before cuts (dashed lines) and
after cuts (solid lines).}
\end{figure} 
To illustrate the idea, in Fig.~\ref{fig:rapidity}
we plot the rapidity distributions for signal and background events
generated with {\sc PYTHIA} \cite{Sjostrand:2007gs}
at the two relevant LHC energies.
The dashed histograms represent parton-level distributions before cuts.
There is a substantial difference in the shapes of the signal and the background ---
the $Z$ bosons in the background events are much more forward, 
because they are produced from an asymmetric $q\bar{q}$ state,
where typically the quark carries a larger momentum fraction.
In contrast, the signal distributions are much more central, since the 
Higgs boson is produced from a symmetric $gg$ initial state,
where the gluons are likely to share the momentum more evenly.

However, these large differences in the event rapidity distributions 
wash out in the presence of cuts
(solid histograms in Fig.~\ref{fig:rapidity}). The leptons produced in the decays
of $Z$s at high rapidity are much more likely to fail the lepton acceptance cuts,
which shaves off the high rapidity tail in the background distributions.
As a result, the signal and background rapidity distributions become similar, 
but not identical. It is therefore worth asking how much one can gain by utilizing the 
residual rapidity differences observed in the figure.

\begin{figure}[t]
\centering
\includegraphics[width=\1]{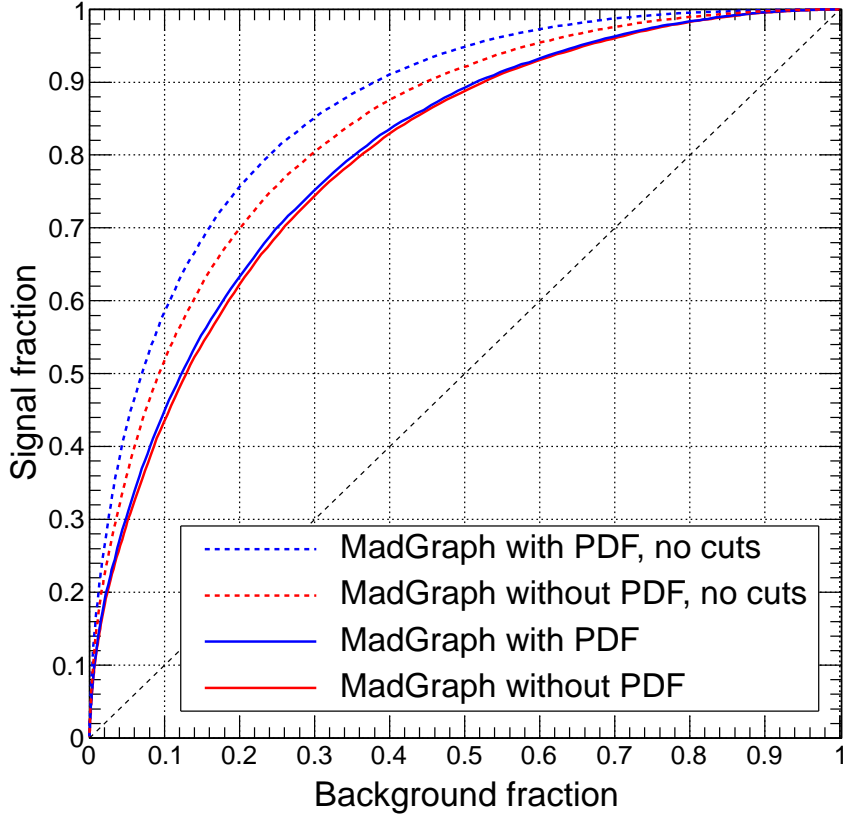}
\caption{\label{fig:ROCIS} Impact of parton distribution functions on ROC curves. 
The red curves are based on $KD(H;ZZ)$ from Eq.~(\ref{D simple}) and do not account for the 
longitudinal boost of the event, while the blue curves are based on
$KD(H;ZZ)$ from Eq.~(\ref{D final}) and include the effect from the parton distribution functions.
Solid (dashed) lines are obtained from event samples with (without) the lepton acceptance cuts.}
\end{figure}

To this end, in Fig.~\ref{fig:ROCIS} we compare the ROC curves obtained from
$KD$ with (blue lines) and without (red lines) PDF information included.
In the absence of the lepton acceptance cuts (dashed lines), the
difference between the ROC curves for the two processes is quite
significant, as one might have guessed from Fig.~\ref{fig:rapidity}.
However, after the cuts (solid lines), the curves become quite similar
and the advantage of using $KD$ with PDFs (Eq.~(\ref{D final})) with respect to using 
$KD$ without PDFs (Eq.~(\ref{D simple})) can be quantified as being at the
percent level.

\section{Signal-background separation for non-SM signals}
\label{sec:jcpVsBackground}
%==============================================================

\begin{figure}[t]
\centering
\includegraphics[width=\3]{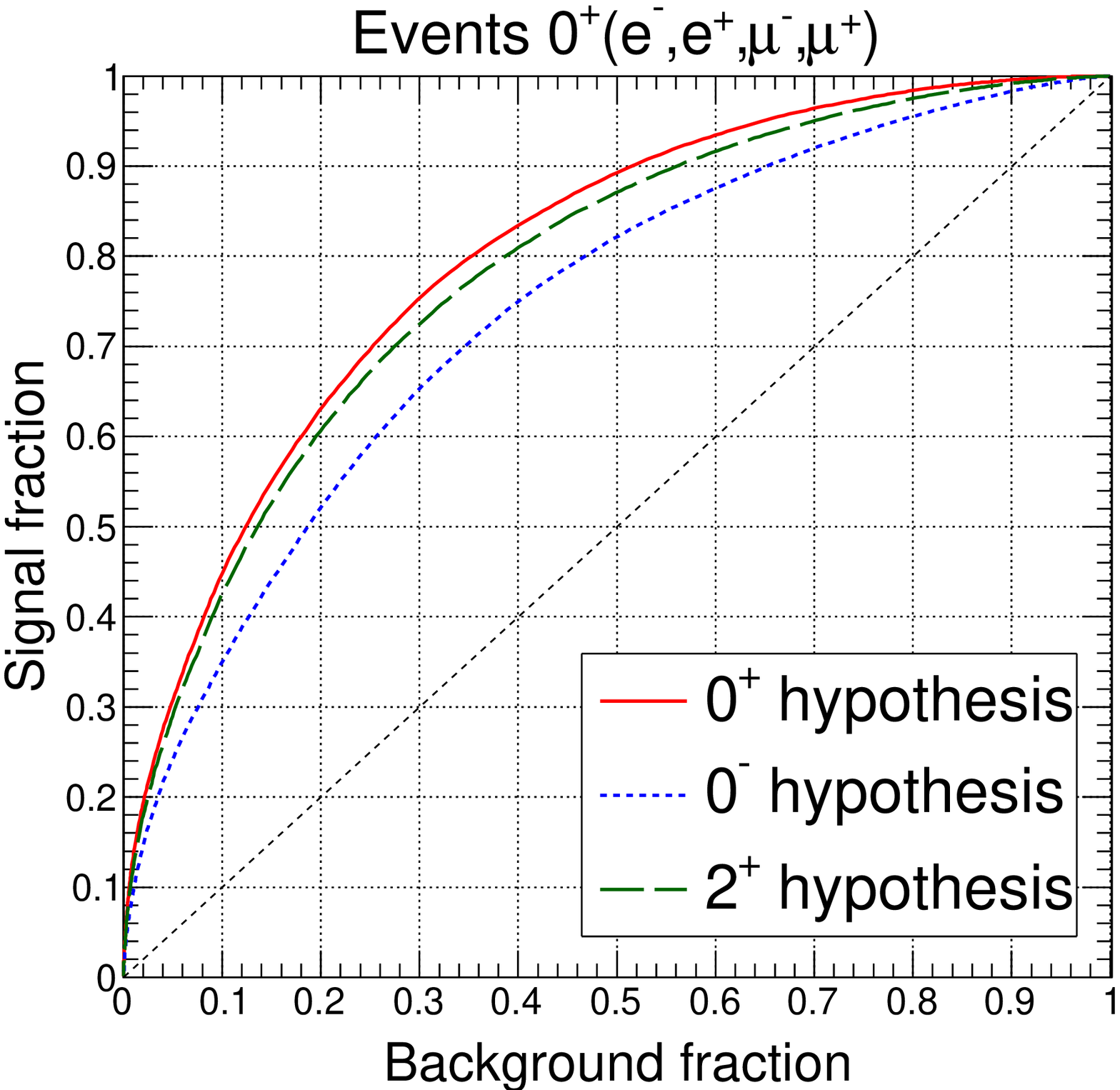}
\includegraphics[width=\3]{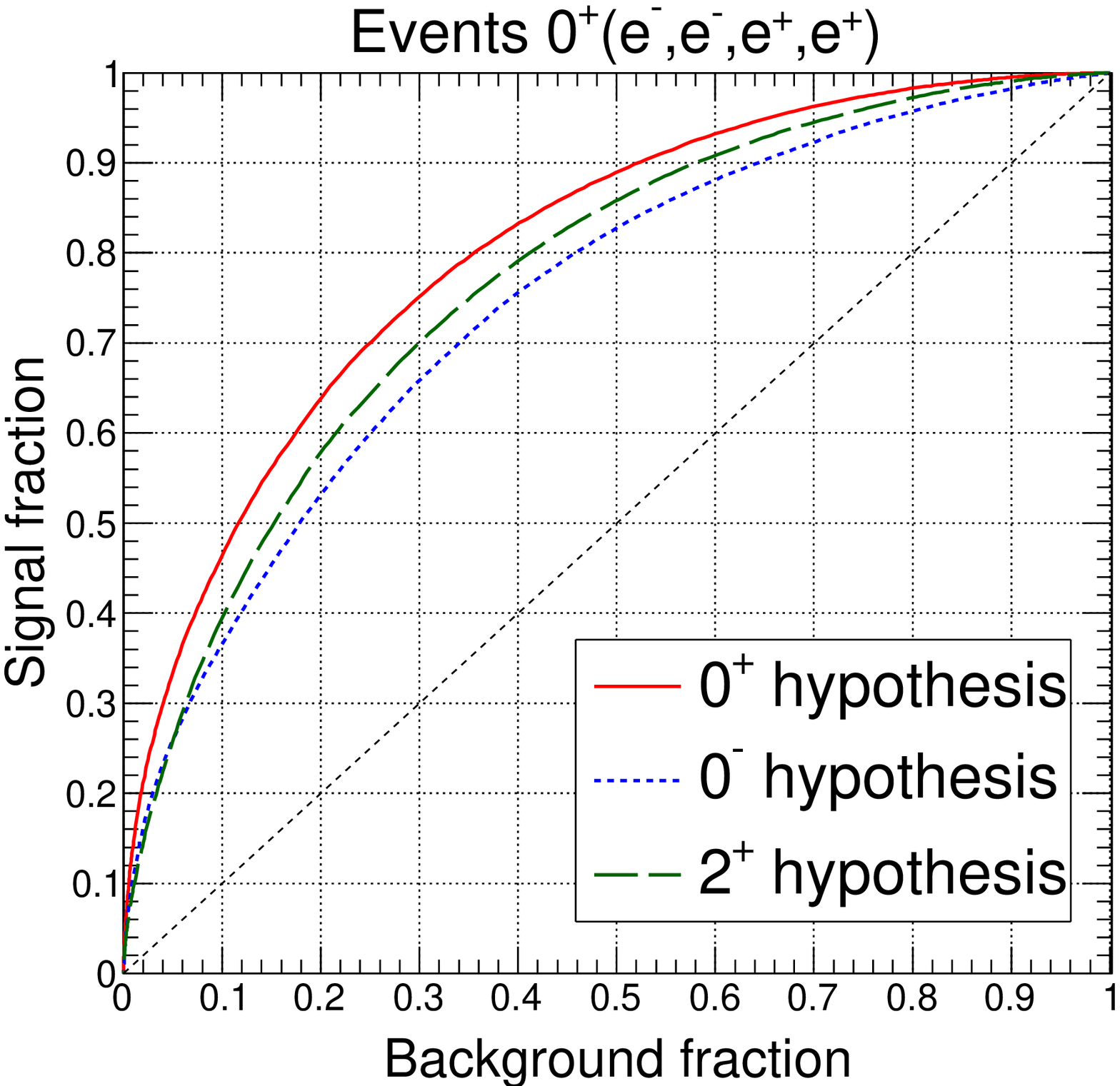}
\includegraphics[width=\3]{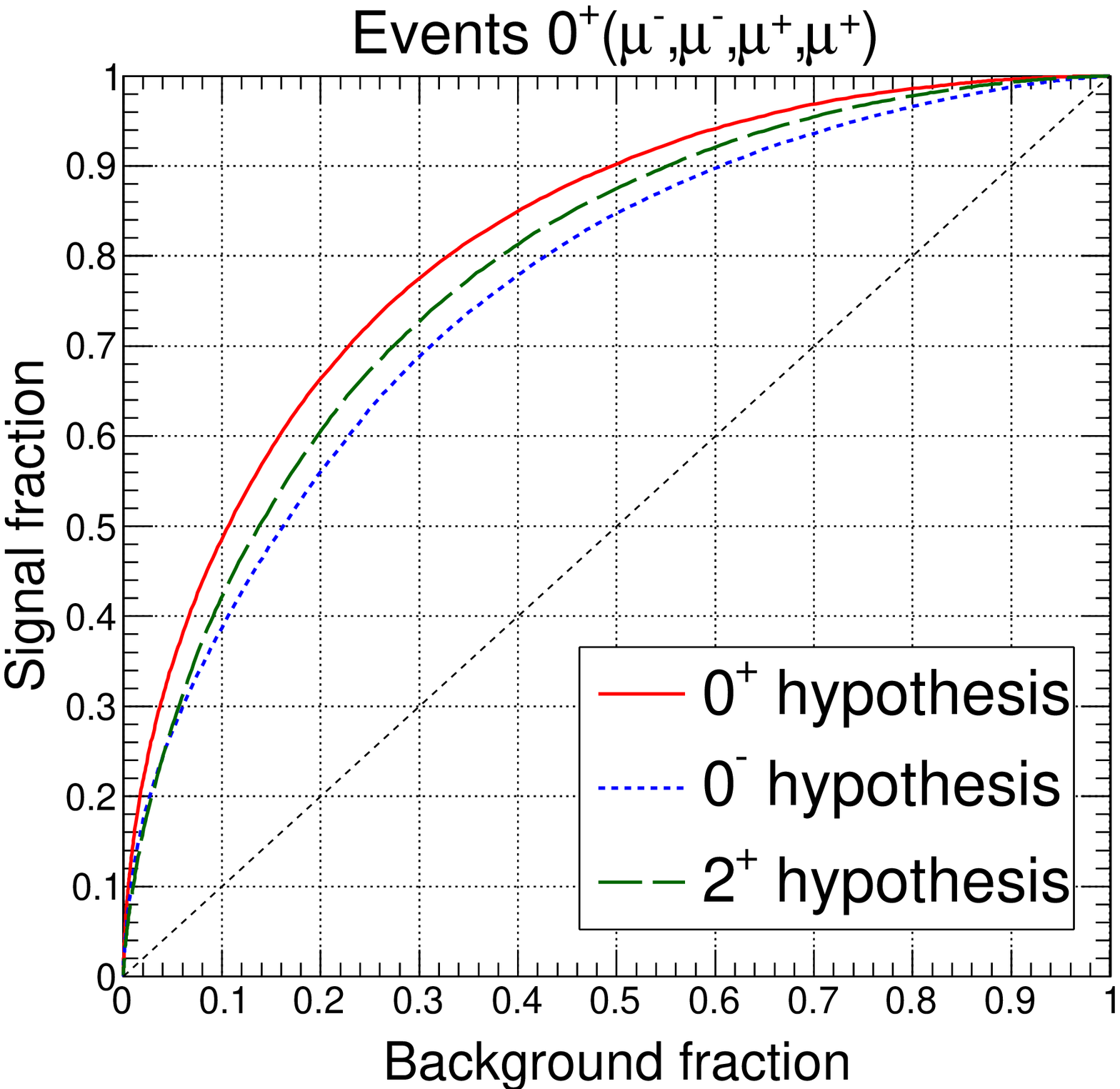} \\
\includegraphics[width=\3]{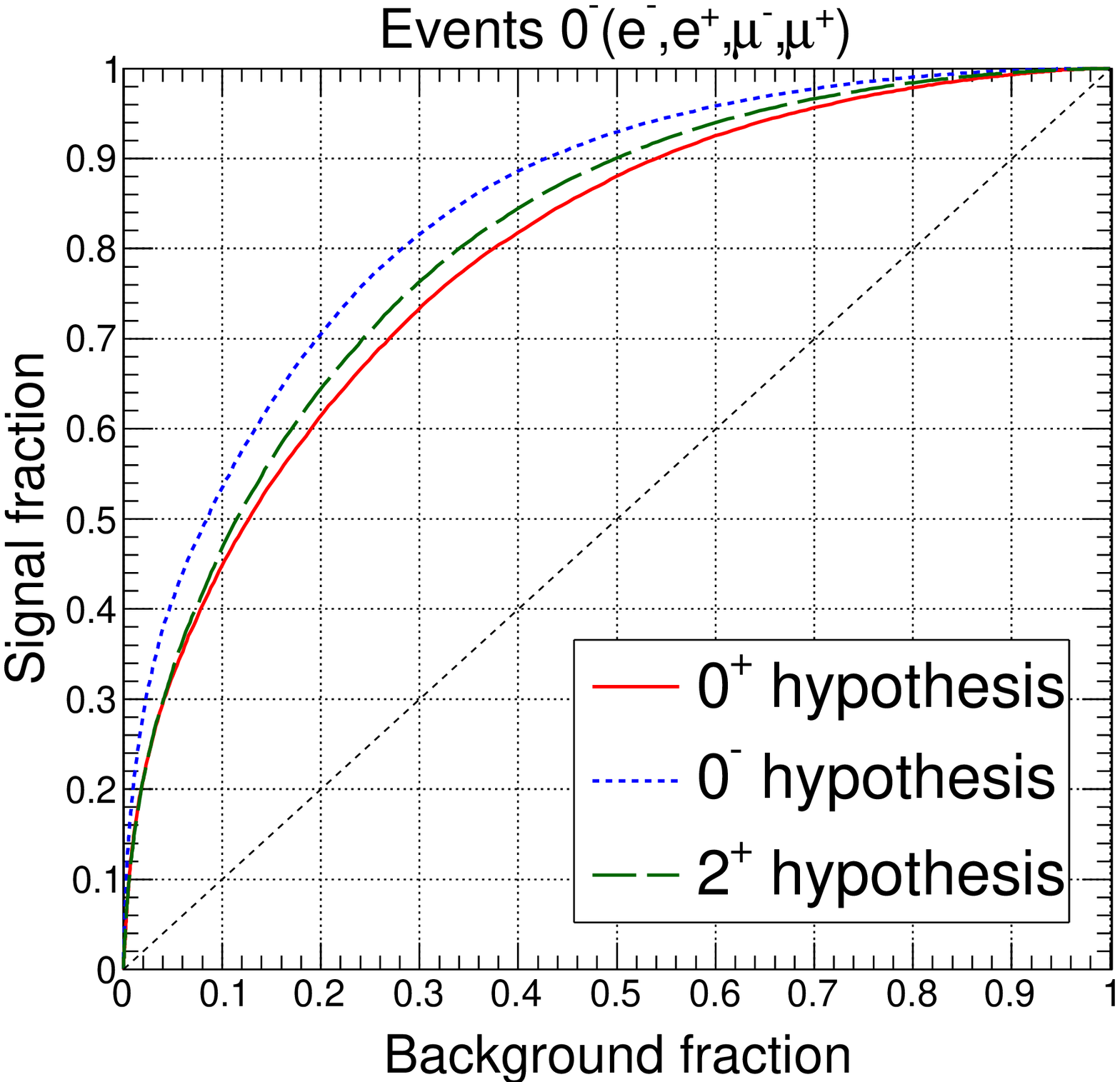}
\includegraphics[width=\3]{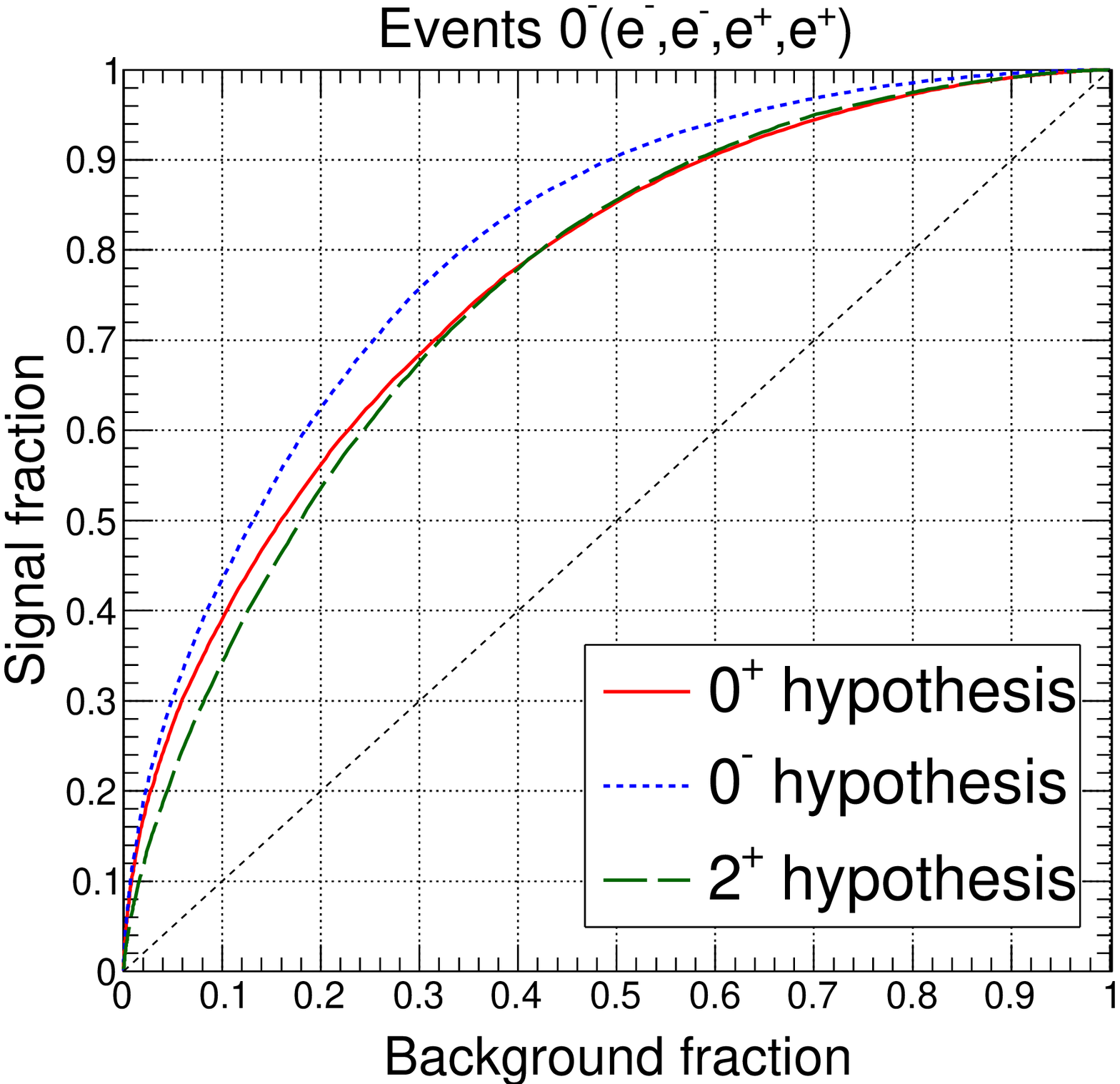}
\includegraphics[width=\3]{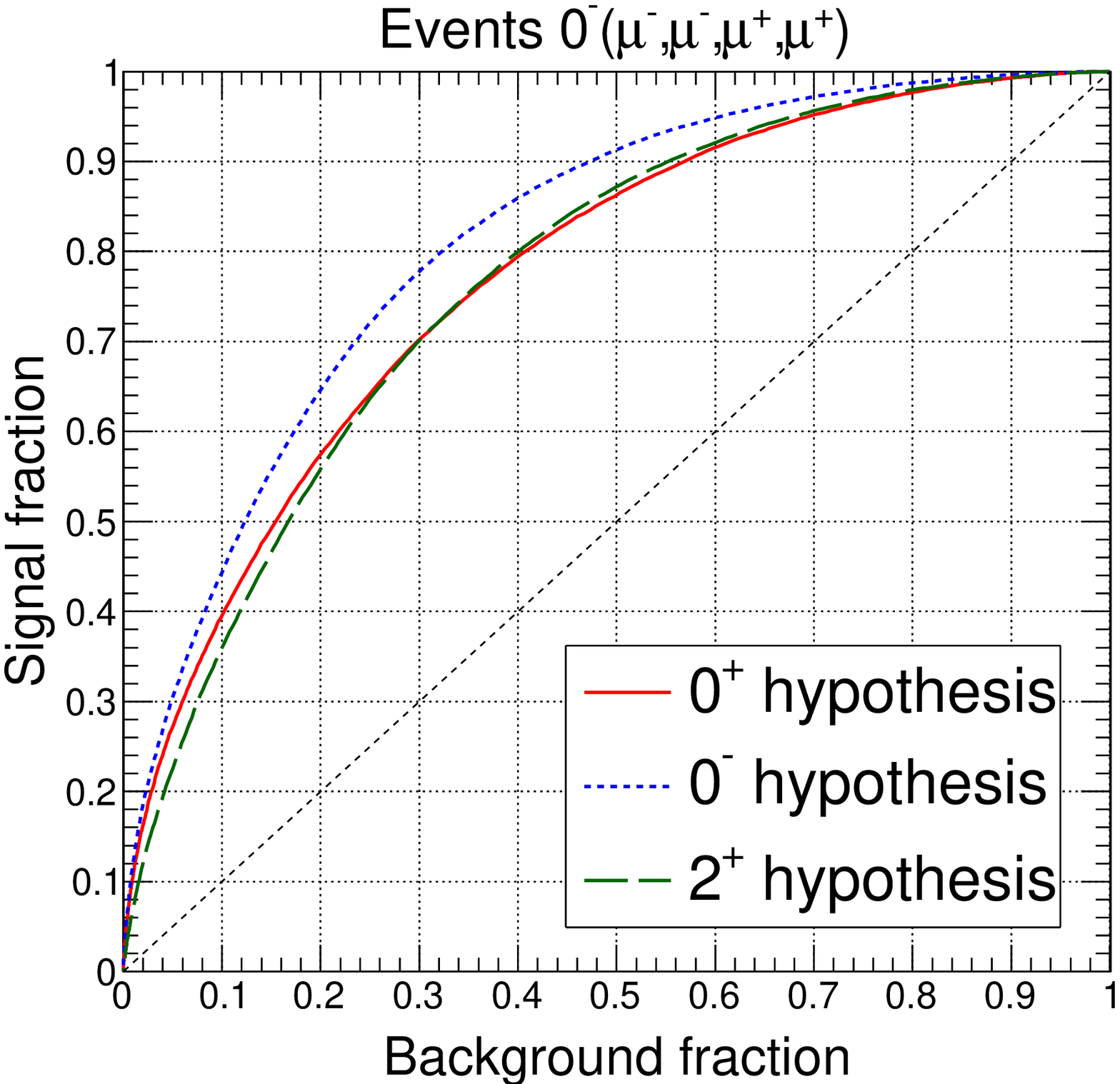} \\
\includegraphics[width=\3]{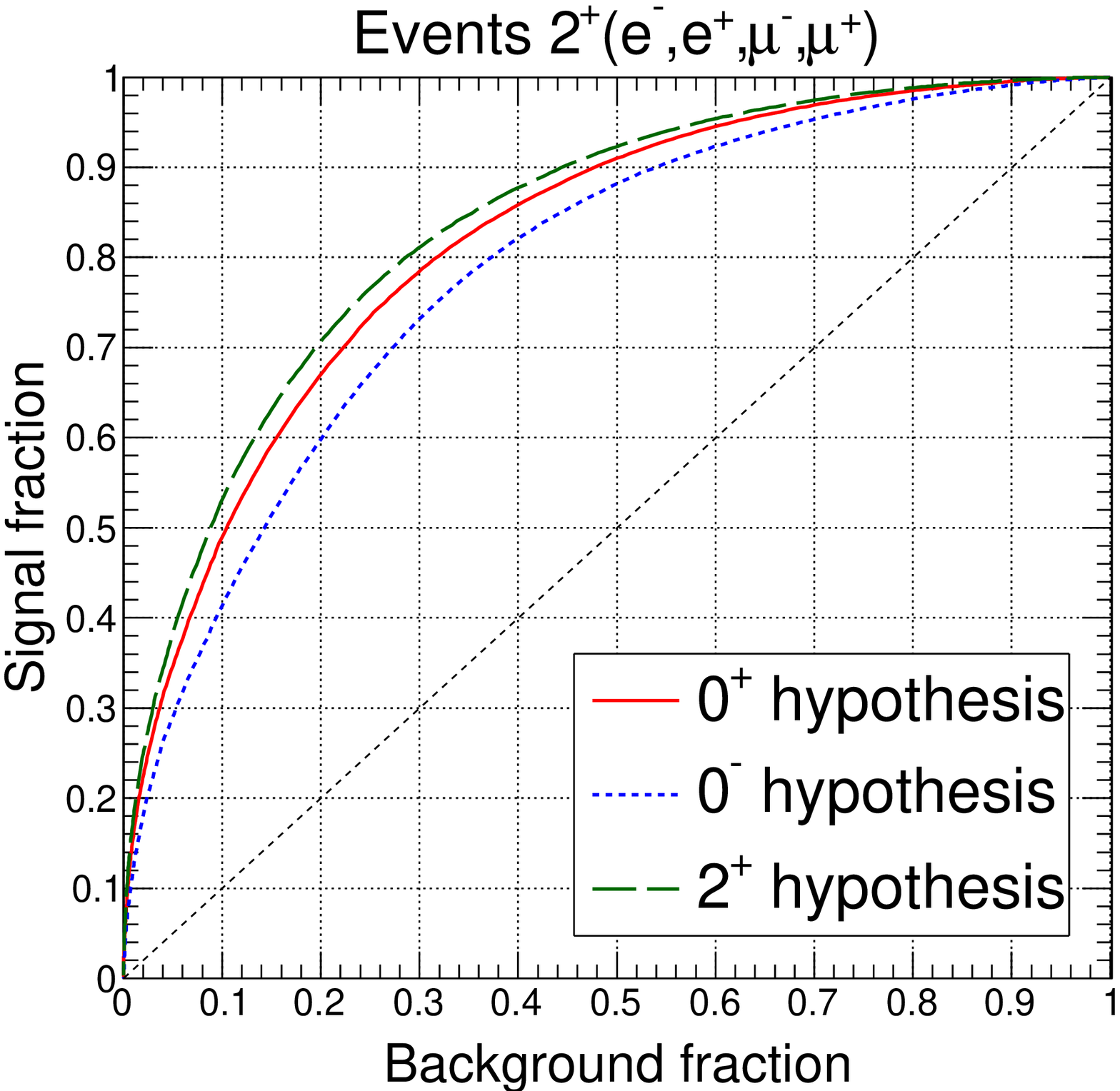}
\includegraphics[width=\3]{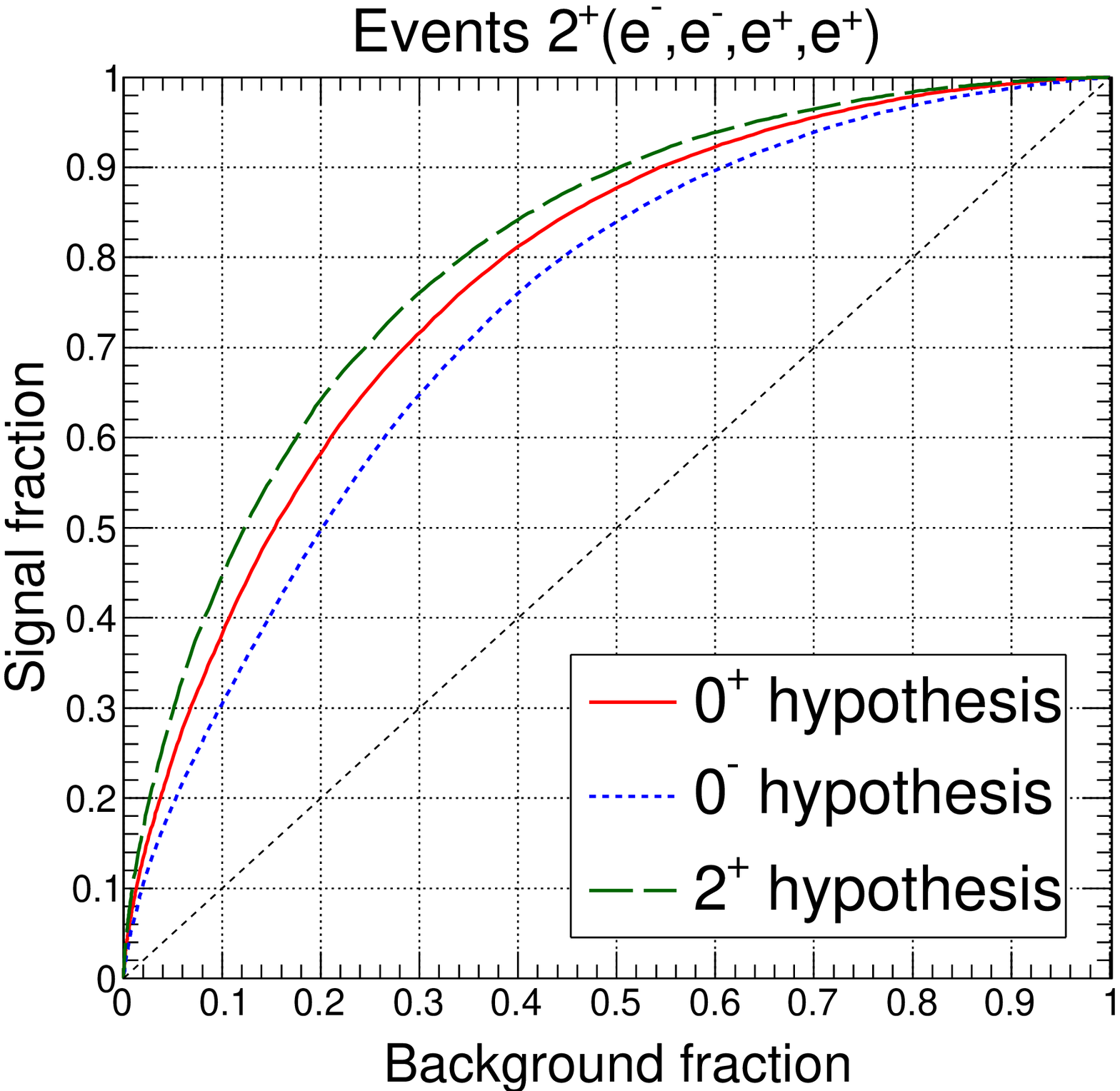}
\includegraphics[width=\3]{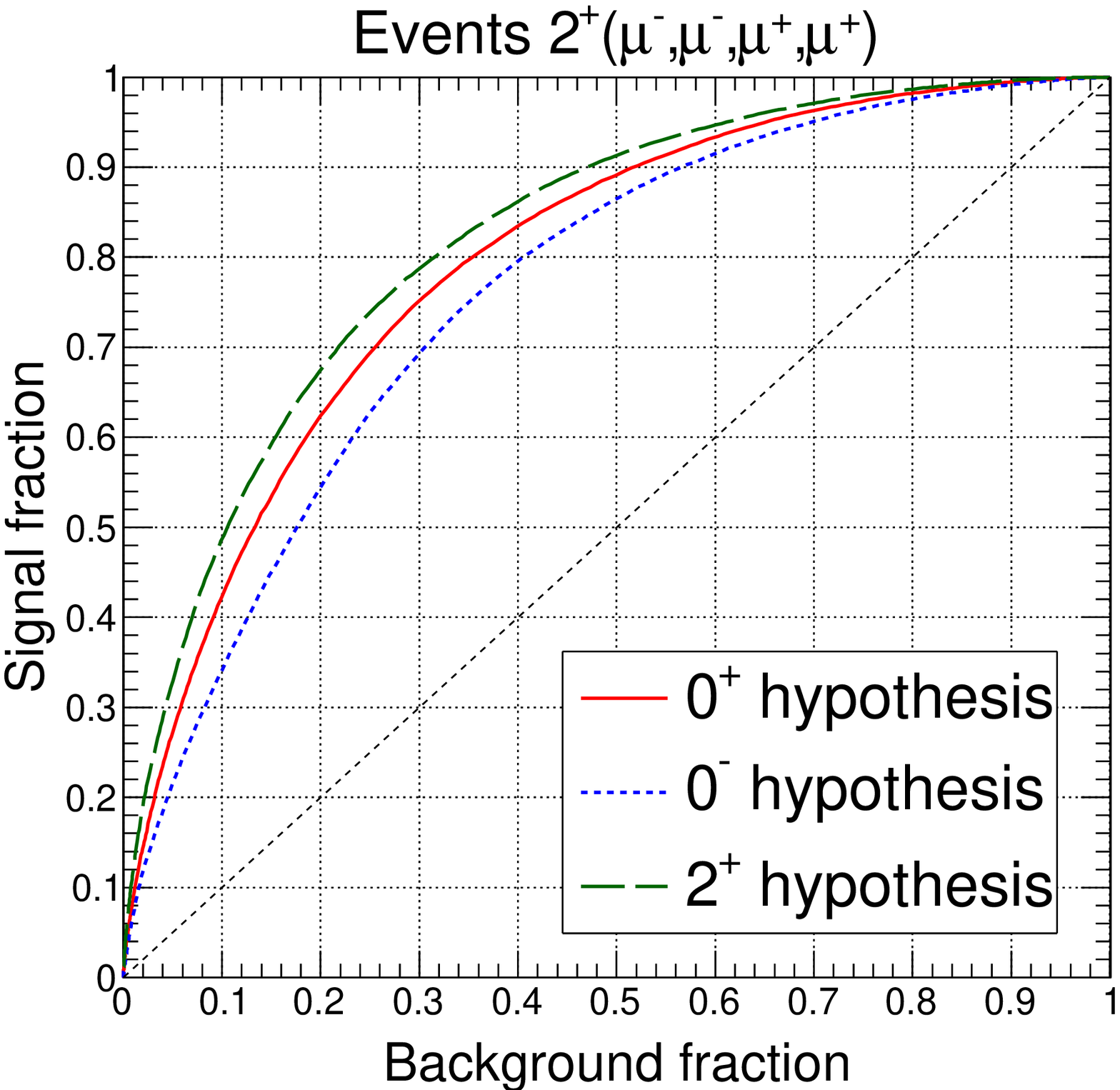} 
\caption{\label{fig:ROC_JCP} ROC curves (as defined in
  Subsection~\ref{sec:ROC}) describing the separation between signal
  and background using {\sc MEKD} where the signal is SM Higgs (top row), CP
  odd Higgs (middle row) or graviton (bottom row).  The final states
  considered are $e^+e^-\mu^+\mu^-$ (left column), $e^+e^-e^+e^-$
  (middle column), and $\mu^+\mu^-\mu^+\mu^-$ (right column).
}
\end{figure}

We showed the utility of the MEM, and in particular a tool like {\sc MEKD},
for distinguishing the SM Higgs from the SM irreducible background.
However, in general, the efficacy of the MEM in separating an arbitrary
resonance from the background may depend on the spin and CP
properties of the new state one is searching for.  
To illustrate this and to quantify 
the importance of this effect for several cases of interest,  in
Fig.~\ref{fig:ROC_JCP} we show ROC curves indicating the ability of the MEM $KD$ to separate
signal from background for three potential signals: 
SM Higgs boson ($\mathrm{J^{CP}}=0^+$), 
CP-odd scalar boson ($\mathrm{J^{CP}}=0^-$), 
and a massive graviton ($\mathrm{J^{CP}}=2^+_m$) 
in each of the three possible flavor states. We construct kinematic discriminants
$KD(0^+; ZZ)$, $KD(0^-; ZZ)$, and $KD(2^+_m; ZZ)$ and apply
all of them for each of the signal spin/parity samples. 

There are three important lessons that one can derive from Fig.~\ref{fig:ROC_JCP}:

\begin{itemize}
\item The optimal separation between signal and background is achieved
only when we use a kinematic discriminant $KD$ constructed with the 
correct signal hypothesis.
\item However, the difference between the three discriminants is not
dramatic. This is easy to understand since all three signal models assume, by construction,
that the new resonances decay to $Z$ bosons, 
while the background has a fair amount of the $Z \gamma^\ast$ and 
even some $\gamma^\ast \gamma^\ast$ intermediate states. 
As was discussed in Sec.~\ref{sec:MEM} (Fig.~\ref{fig:ROC}),
the $m_{Z2}$ observable carries a lot of weight in the overall
discrimination power of $KD(X;ZZ)$, which makes all three signals fairly
alike as far as their separation from the background is concerned.
\item There are differences between the ROC curves 
for the $e^+e^-\mu^+\mu^-$ final state and the SF final states
for $0^-$ and $2^+_m$ signals. This suggests that, 
compared with the SM Higgs boson, 
the interference effects may play a more
important role when the $0^-$ and $2^+_m$ hypotheses are considered.
This is discussed further in the next section.
\end{itemize}

\section{Spin and parity discrimination}
\label{sec:JCP}
\begin{figure}[t]
\centering
\includegraphics[width=\2]{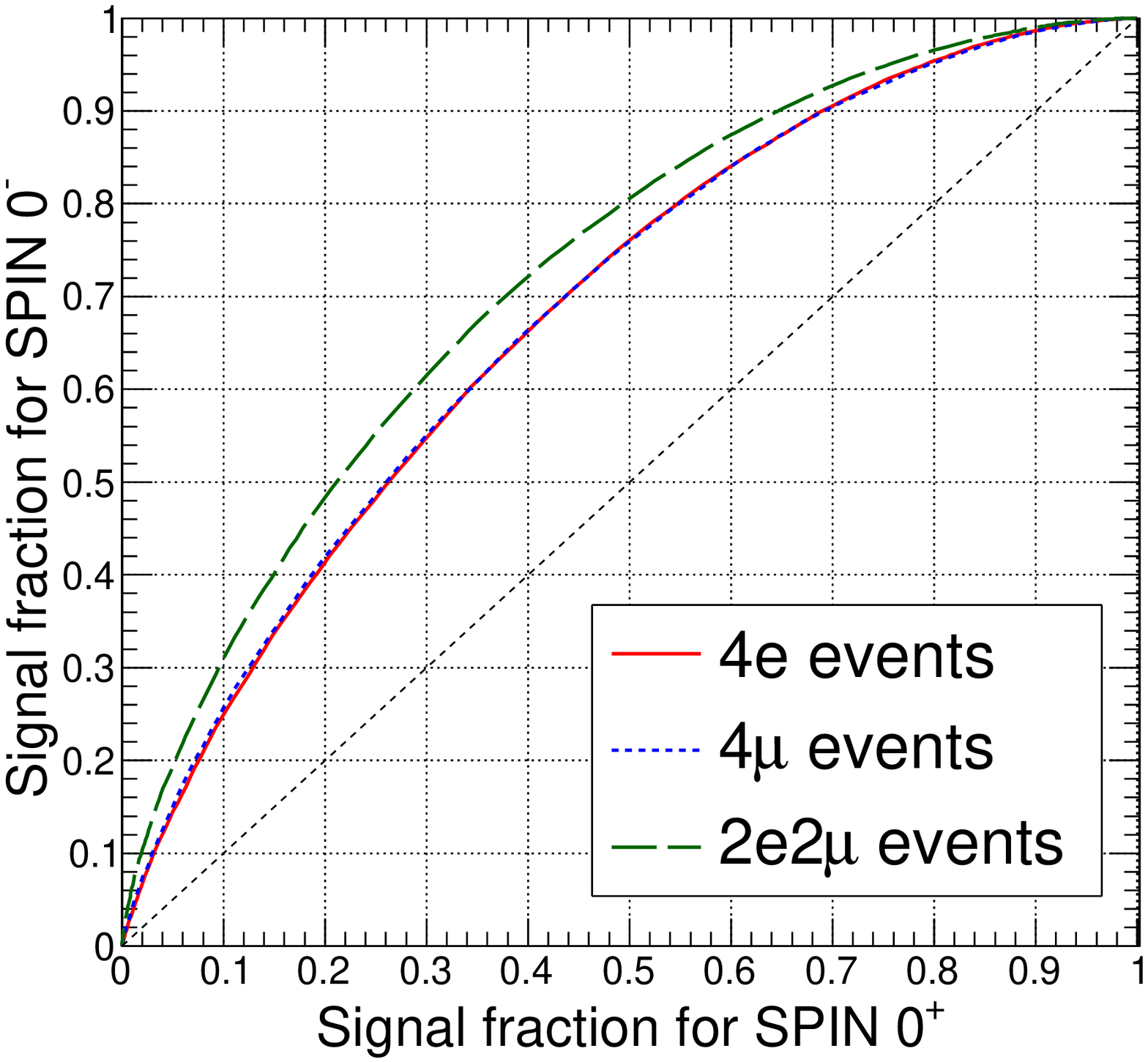}
\includegraphics[width=\2]{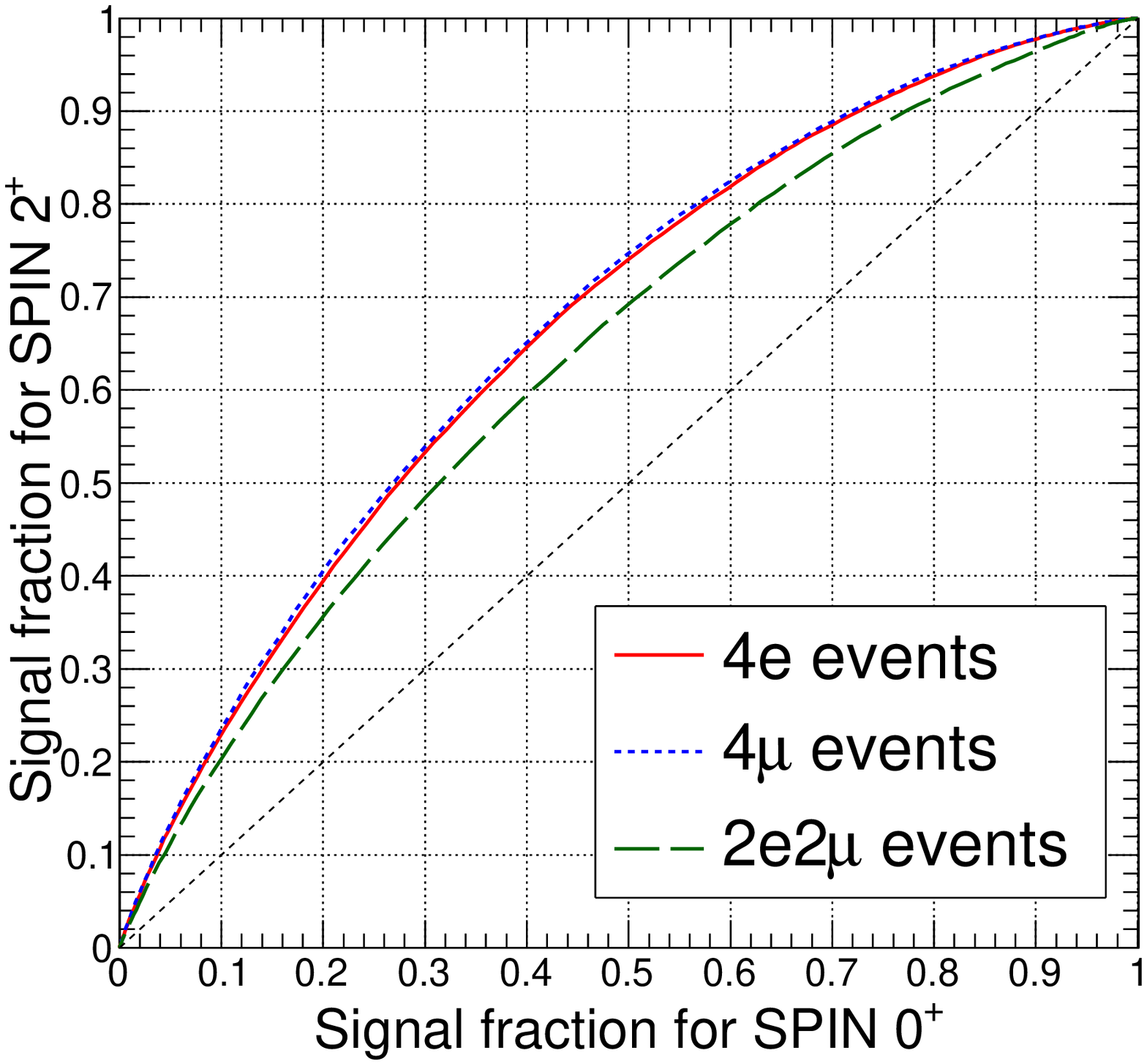}
\caption{
ROC curves, shown for all three final state flavors ($4e$, $4\mu$, $2e2\mu$), 
display the separation of alternative signal hypotheses:
(left) the pseudoscalar ($\mathrm{J^{CP}} = 0^-$) vs. the SM Higgs boson ($0^+$);
(right) the massive graviton ($\mathrm{J^{CP}} = 2^+_m$)  vs. the SM Higgs boson ($0^+$).
}
\label{fig:ROC_CPSpin_Diff_Flav}
\end{figure}

%=====================================================================
An important and well-studied use of the golden channel is for
determining the spin and parity of a putative Higgs boson. 
As we show below, the use of the exact matrix element for constructing
kinematic discriminants plays a crucial role in achieving the
best separation between alternative signal hypotheses. 
As an example, we consider two alternative signal models: 
a pseudoscalar boson ($\mathrm{J^{CP}}=0^-$) and
a massive graviton, which is, of course, a spin two boson, ($\mathrm{J^{CP}}=2^+_m$).
The couplings for the CP-odd scalar and the
massive graviton are defined in Sec.~\ref{sec:MG}.
To generate SM Higgs boson, pseudoscalar,
and spin massive gravitons events, we use {\sc MadGraph}.
The kinematic cuts on the leptons are the same as described in Sec.~\ref{sec:leptoncuts}.
We then use ROC curves for the kinematic discriminant 
$KD(\mathrm{J^{CP}}; \, \mathrm{H_{SM}})$ to quantify 
the differences in kinematics of the four lepton system
for the $\mathrm{J^{CP}}$ boson and the SM Higgs boson ($\mathrm{H_{SM}}$),
as shown in Fig.~\ref{fig:ROC_CPSpin_Diff_Flav}.

There are three important conclusions to be drawn from this figure.
First, one can see that the separation between the $0^-$ boson and $\mathrm{H_{SM}}$
is more pronounced in comparison with the $2^+_m$ boson case. This indicates
that one should be able to tell the difference between 
the $0^-$ boson and $\mathrm{H_{SM}}$ sooner than 
between the $2^+_m$ boson and $\mathrm{H_{SM}}$.
Second, there is a significant difference between 
the SF ($4e$ and $4\mu$) and DF ($2e2\mu$)
four lepton final states. This is a clear indicator that the 
allowed permutations of identical leptons and the associated interference
in the SF events play a significant role. The detriment to sensitivity 
in neglecting such effects in constructing $KD$ will be discussed below.
Third, for the $2^+_m$ boson, there is more expected separation
with the SM Higgs for the SF final state as opposed to the DF final
state, while for the $0^-$ boson case the opposite is true. 
Consequently, one should expect that the proper treatment of
permutations/interference in the construction of $KD$ should
lead to a greater ability to separate the $2^+_m$ boson from the SM
Higgs boson in comparison with
the $0^-$ boson case. 

In order to further elucidate the role of permutations and interference effects, 
we proceed with a more detailed analysis of the four muon final state,
$(\mu_1^- \mu_1^+) (\mu_2^-$ $\mu_2^+)$. 
Here, the first dimuon pair, $(\mu_1^- \mu_1^+)$, is formed from
the pair of opposite-charge muons whose invariant mass is closest to the $Z$
boson mass. We calculate three ``matrix elements'' for this final
state as follows:

\begin{enumerate}
\item \label{best} The complete leading order matrix element squared, 
$\left| {\mathcal{M}}_{4\mu} \right|^2$, 
with all permutations of identical leptons and associated interference included.
The choice of how one pairs muons in this case does not matter; the full matrix
element does not depend on how a human wants to group the final state leptons.
\item \label{sym} The ``symmetrized'' matrix element squared 
$\left| {\mathcal{M}}_{2e2\mu} \right|^2_{\mathrm{sym}}$ that takes into account
permutations of identical leptons, but ignores the associated interference effects:
\begin{equation}
\left| {\mathcal{M}}_{2e2\mu} \right|^2_{\mathrm{sym}} =
\left| \, {\mathcal{M}}_{2e2\mu}( \, (\mu_1^- \mu_1^+)_e \, (\mu_2^- \mu_2^+)_{\mu} \, ) \, \right|^2 + 
\left | \, {\mathcal{M}}_{2e2\mu}( \, (\mu_1^- \mu_2^+)_e \, (\mu_2^- \mu_1^+)_{\mu} \, )\, \right|^2,
\end{equation}
where the first pair, marked by a subscript $e$, is treated as if it were a pair of electrons.
\item \label{worst} The $2e2\mu$ matrix element squared without symmetrization, 
$\left| {\mathcal{M}}_{2e2\mu} \right|^2$,  that ignores both
permutations of identical leptons and the associated interference effects:
\begin{equation}
\left| {\mathcal{M}}_{2e2\mu} \right|^2 =
\left| \, {\mathcal{M}}_{2e2\mu}( \, (\mu_1^- \mu_1^+)_e \, (\mu_2^- \mu_2^+)_{\mu} \, ) \, \right|^2 ,
\end{equation}
where the first pair, marked by a subscript $e$, is treated as if it were a pair of electrons.
\end{enumerate}

\begin{figure}[t]
\centering
\includegraphics[width=\5]{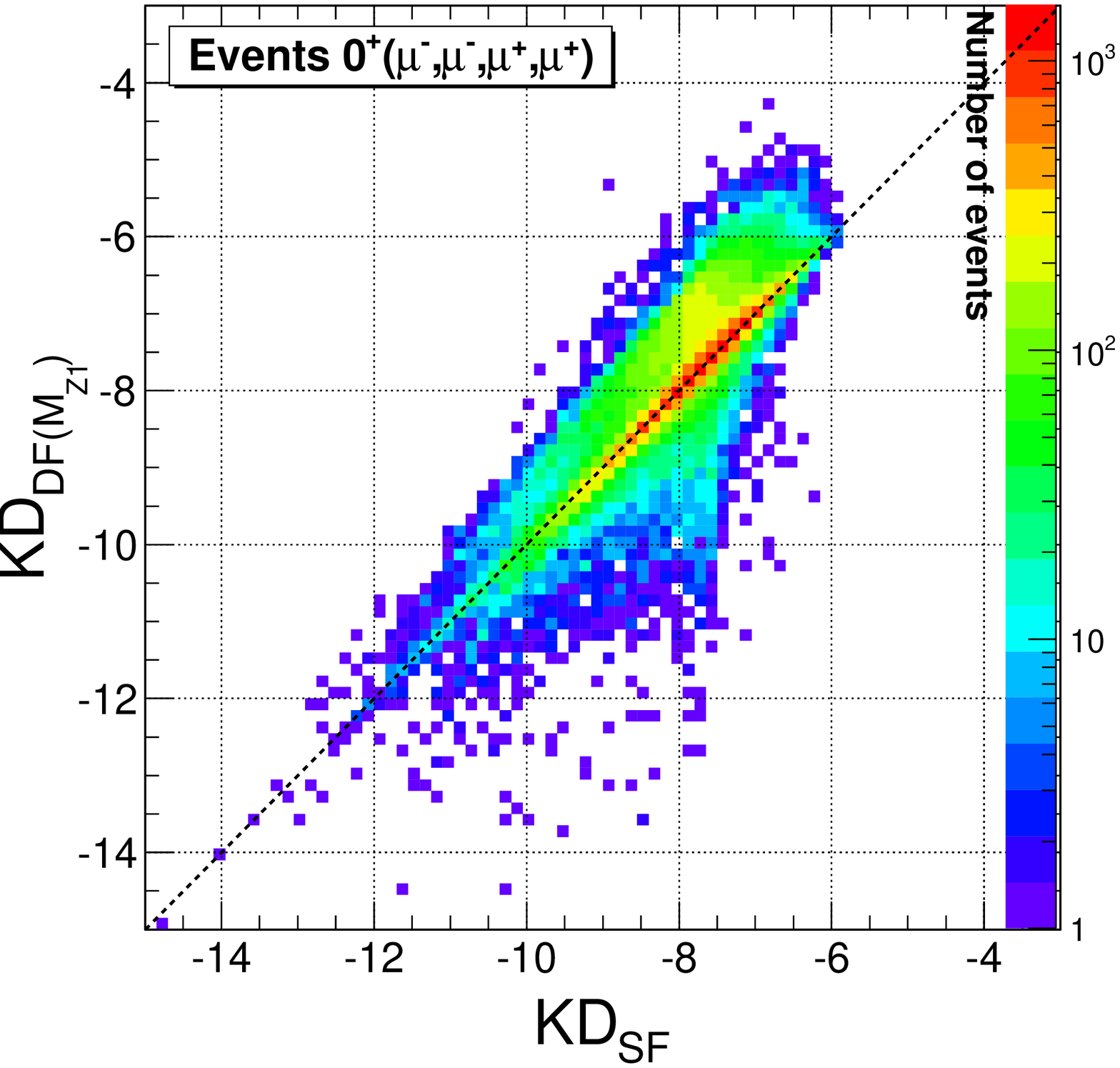}
\includegraphics[width=\5]{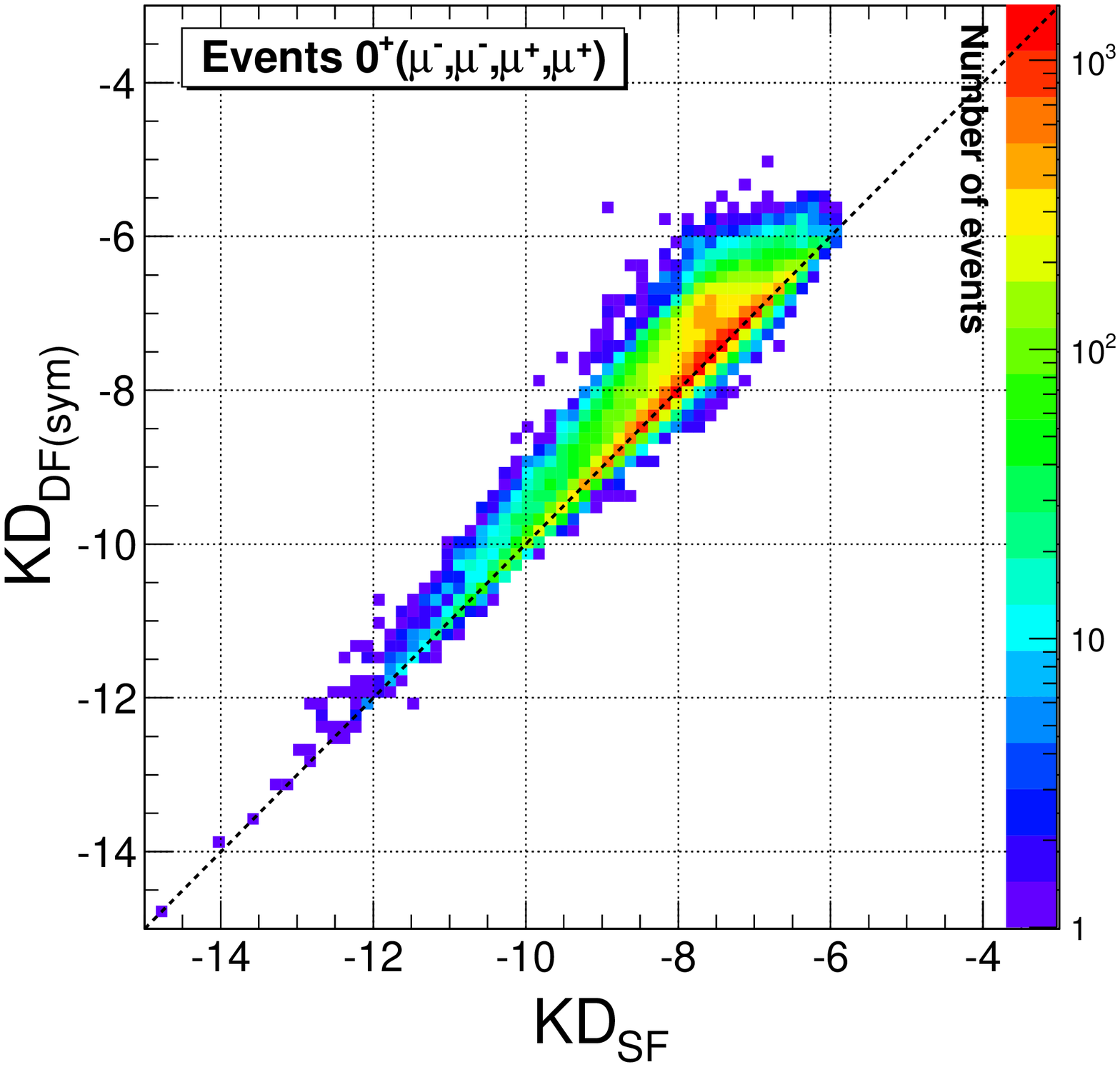}\\
\includegraphics[width=\5]{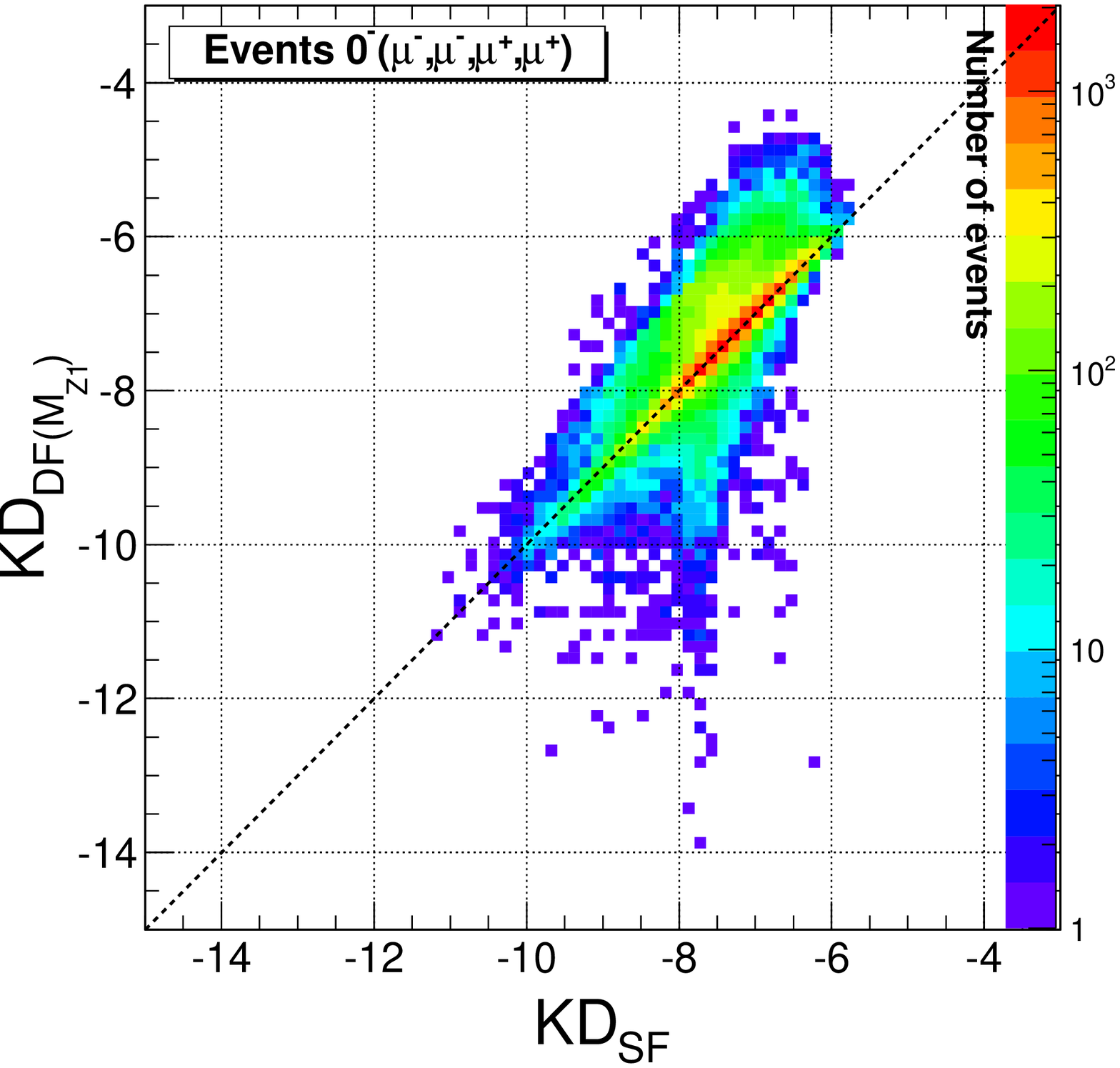}
\includegraphics[width=\5]{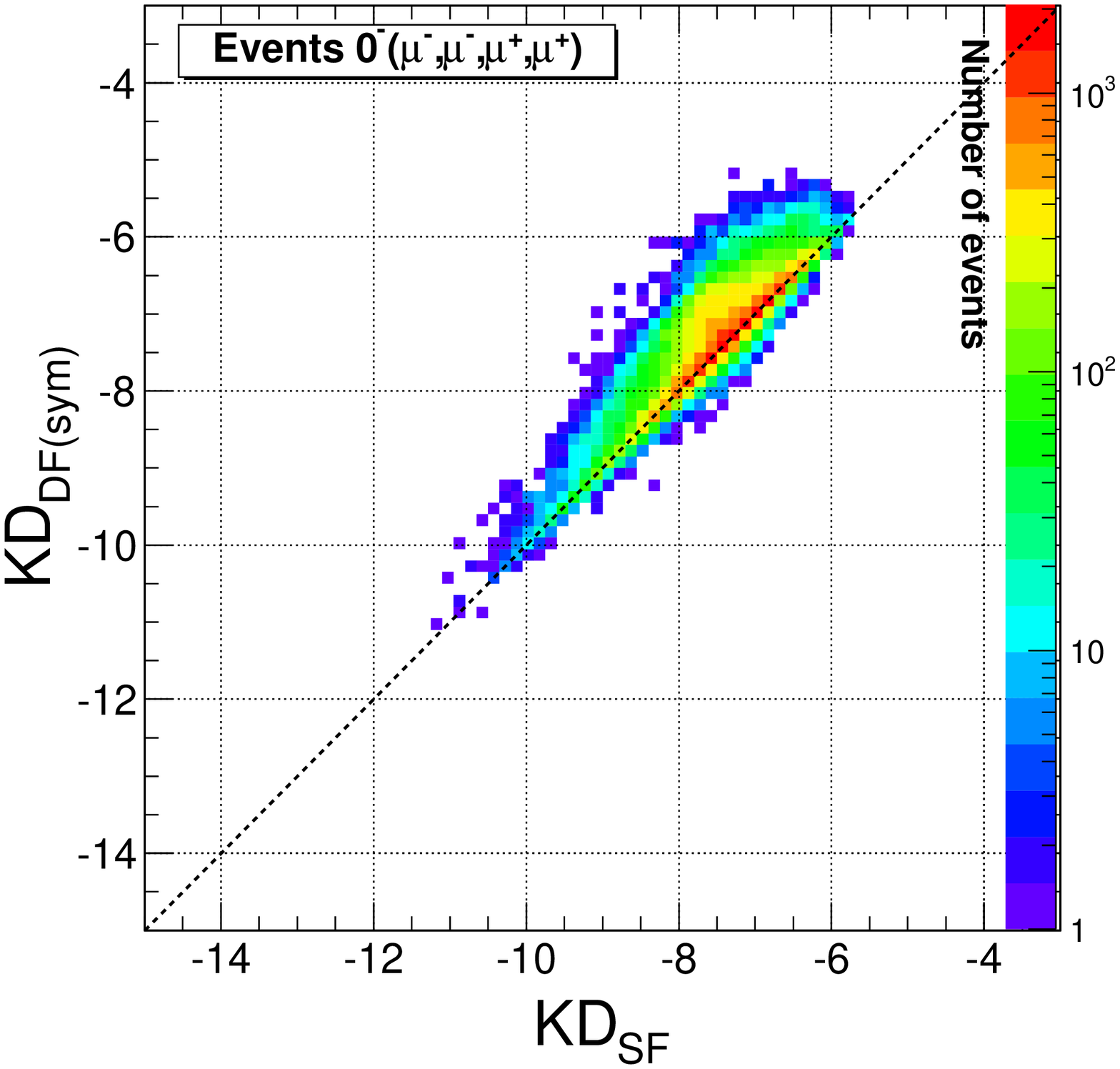}
\caption{
Comparison of kinematic discriminants $KD(0^-; \, \mathrm{H_{SM}})$ 
for the four muon final states calculated with different levels of approximations.  
In all four plots, the $x$-axis shows the $KD$ calculated using the complete matrix element squared, 
$\left| {\mathcal{M}}_{4\mu} \right|^2$.  
In the plots on the left, the $y$-axis gives the $KD$ calculated using
$\left| {\mathcal{M}}_{2e2\mu} \right|^2_{\mathrm{sym}}$, which takes into account
permutations of identical leptons, but ignores the associated interference effects.
In the plots on the right, the $y$-axis gives the $KD$ calculated using
$\left| {\mathcal{M}}_{2e2\mu} \right|^2$, which ignores both
permutations of identical leptons and the associated interference effects.
The top plots are for SM Higgs signal events, 
while the bottom plots are for a pseudoscalar.
}
\label{fig:KD_0+_0-_int}
\end{figure} 

\begin{figure}[t]
\centering
\includegraphics[width=\5]{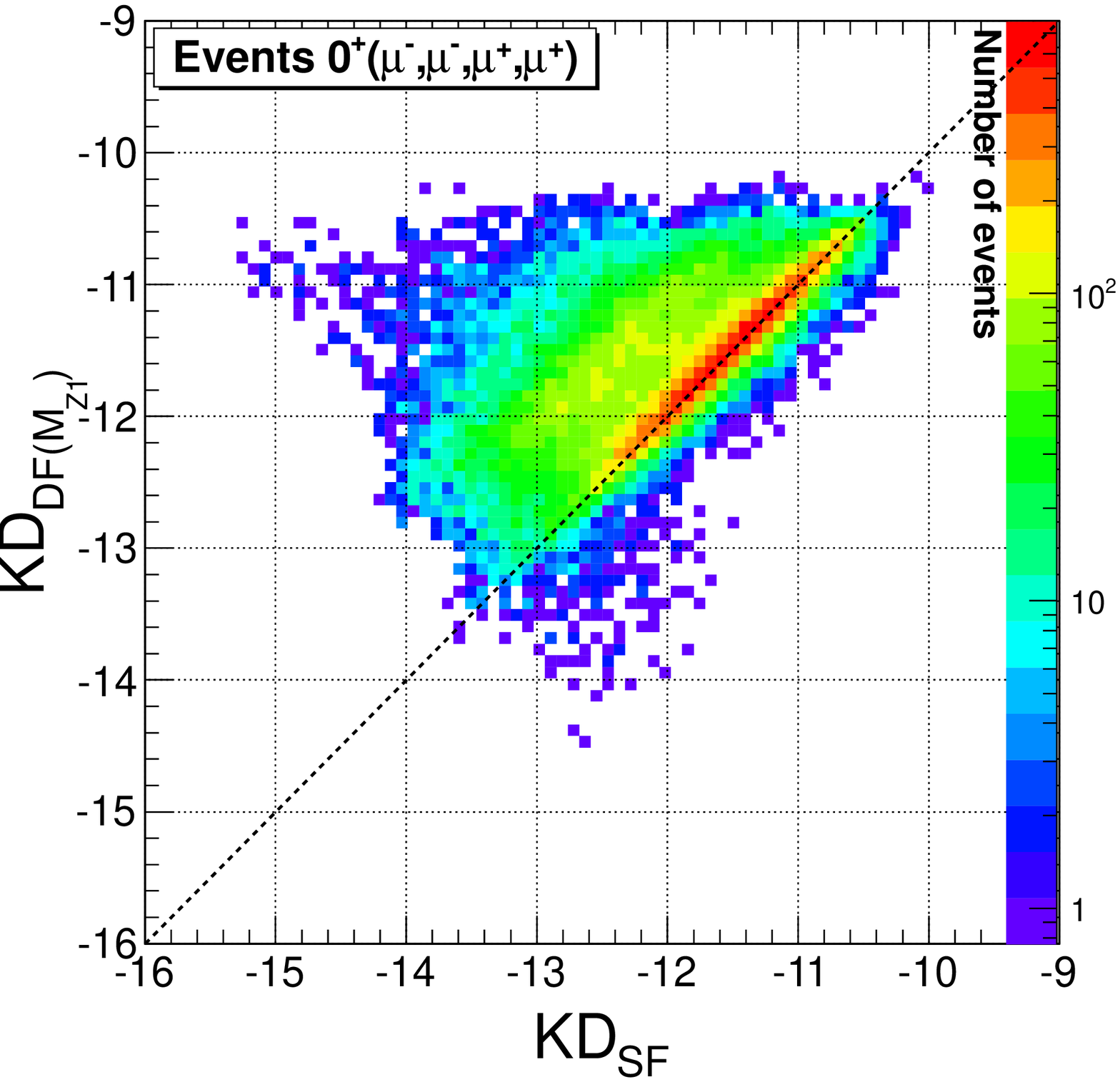}
\includegraphics[width=\5]{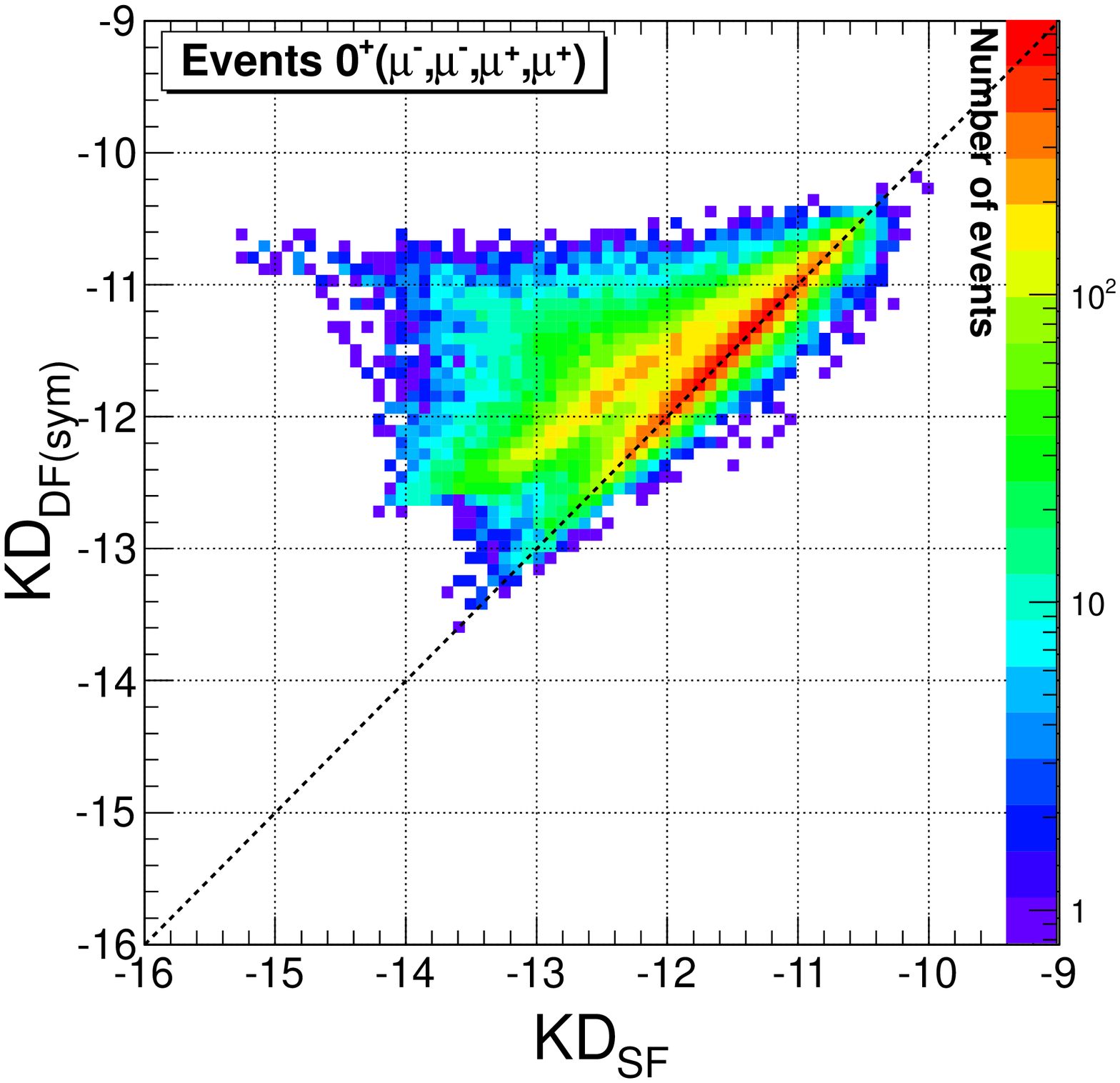}\\
\includegraphics[width=\5]{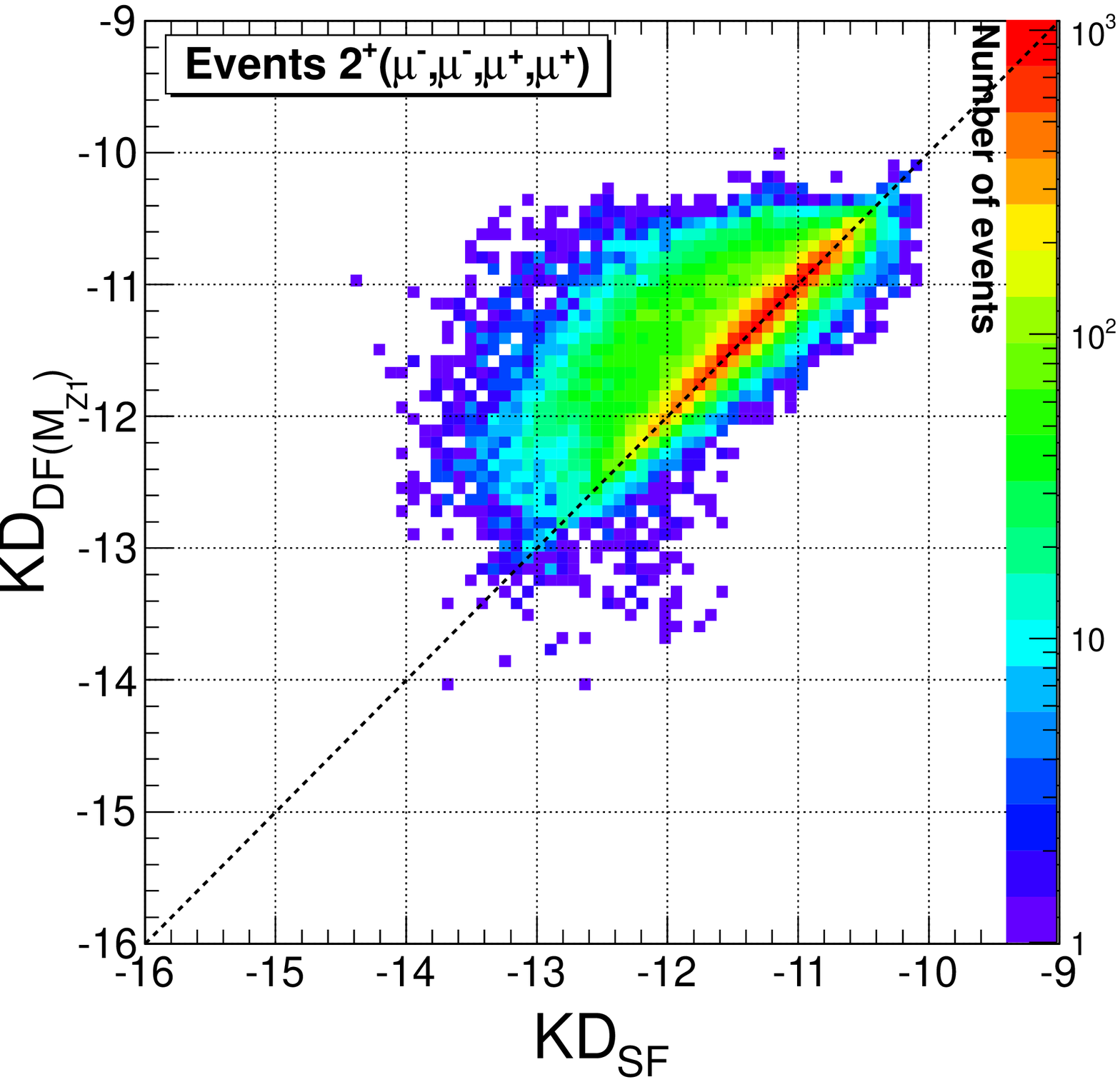}
\includegraphics[width=\5]{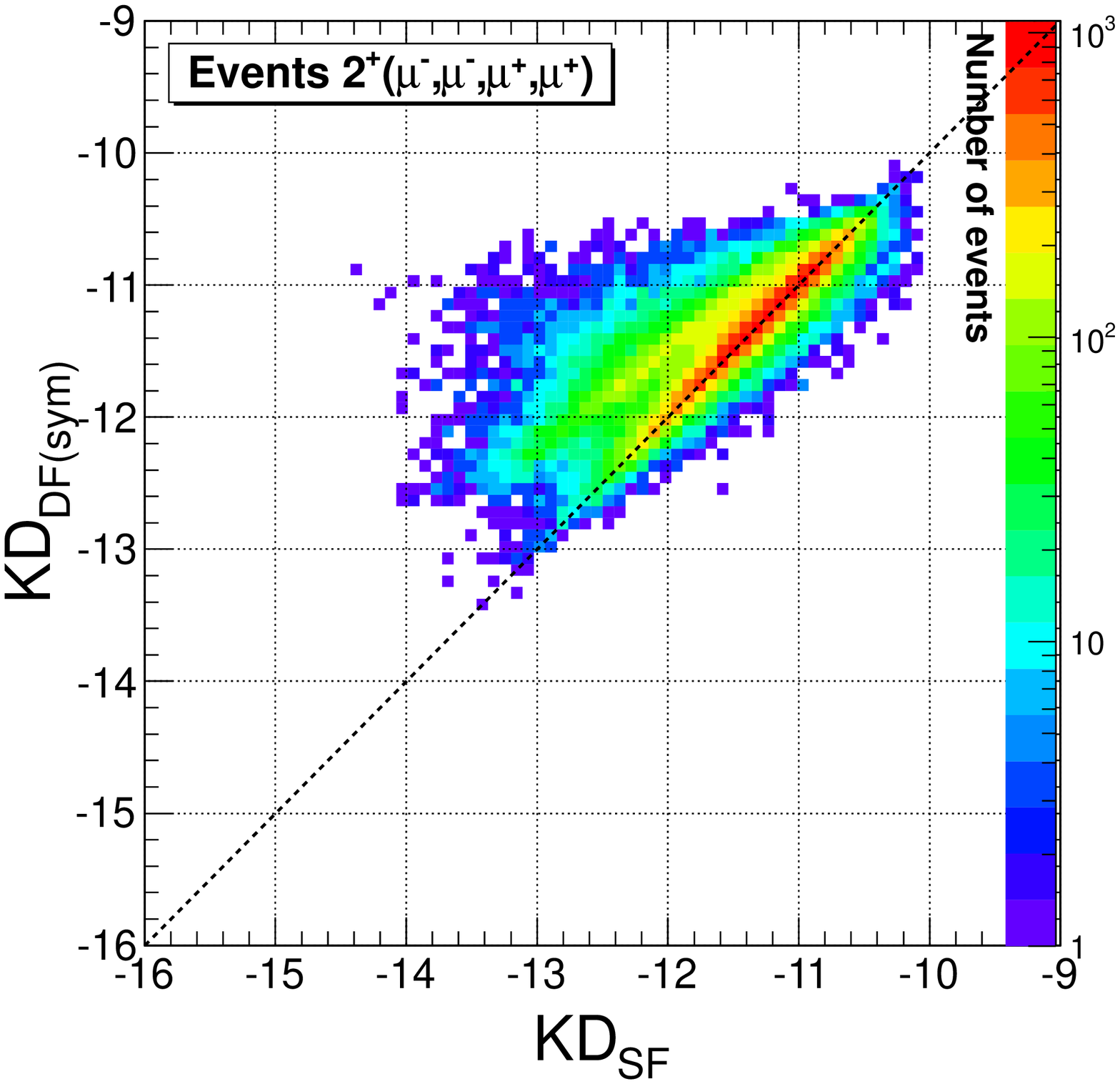}
\caption{
Comparison of kinematic discriminants $KD(2^+_m; \, \mathrm{H_{SM}})$ 
for the four muon final states calculated with different levels of approximations.  
In all four plots, the $x$-axis shows the $KD$ calculated using the complete matrix element squared, 
$\left| {\mathcal{M}}_{4\mu} \right|^2$.  
In the plots on the left, the $y$-axis gives the $KD$ calculated using
$\left| {\mathcal{M}}_{2e2\mu} \right|^2_{\mathrm{sym}}$, which accounts for
permutations of identical leptons, but ignores the associated interference effects.
In the plots on the right, the $y$-axis gives the $KD$ calculated using
$\left| {\mathcal{M}}_{2e2\mu} \right|^2$, which ignores both
permutations of identical leptons and the associated interference effects.
The top plots are for the SM Higgs signal events, 
while the bottom plots are for a massive graviton.
}
\label{fig:KD_0+_Grav_int}
\end{figure}

\begin{figure}[t]
\centering
\includegraphics[width=\2]{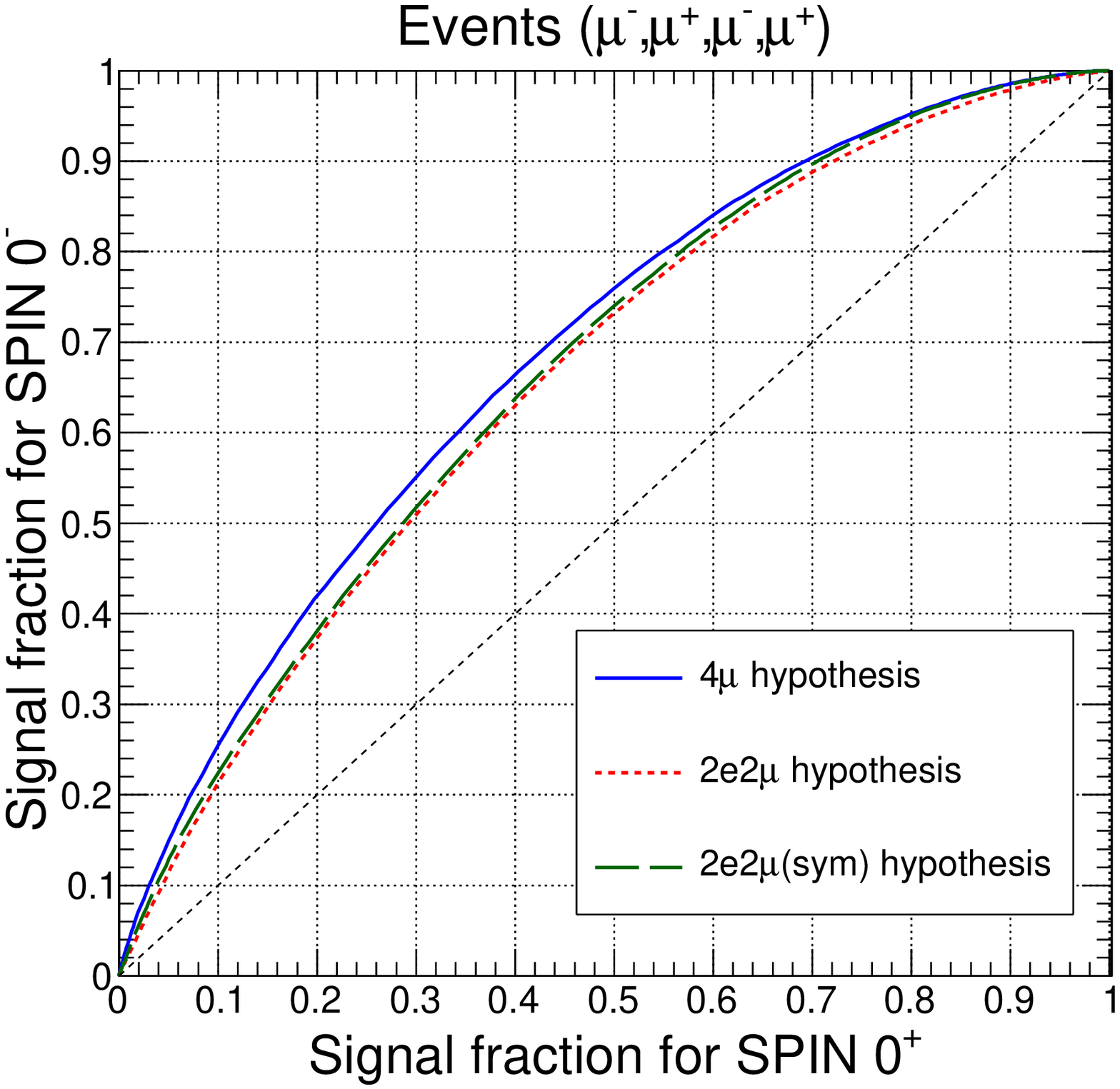}
\includegraphics[width=\2]{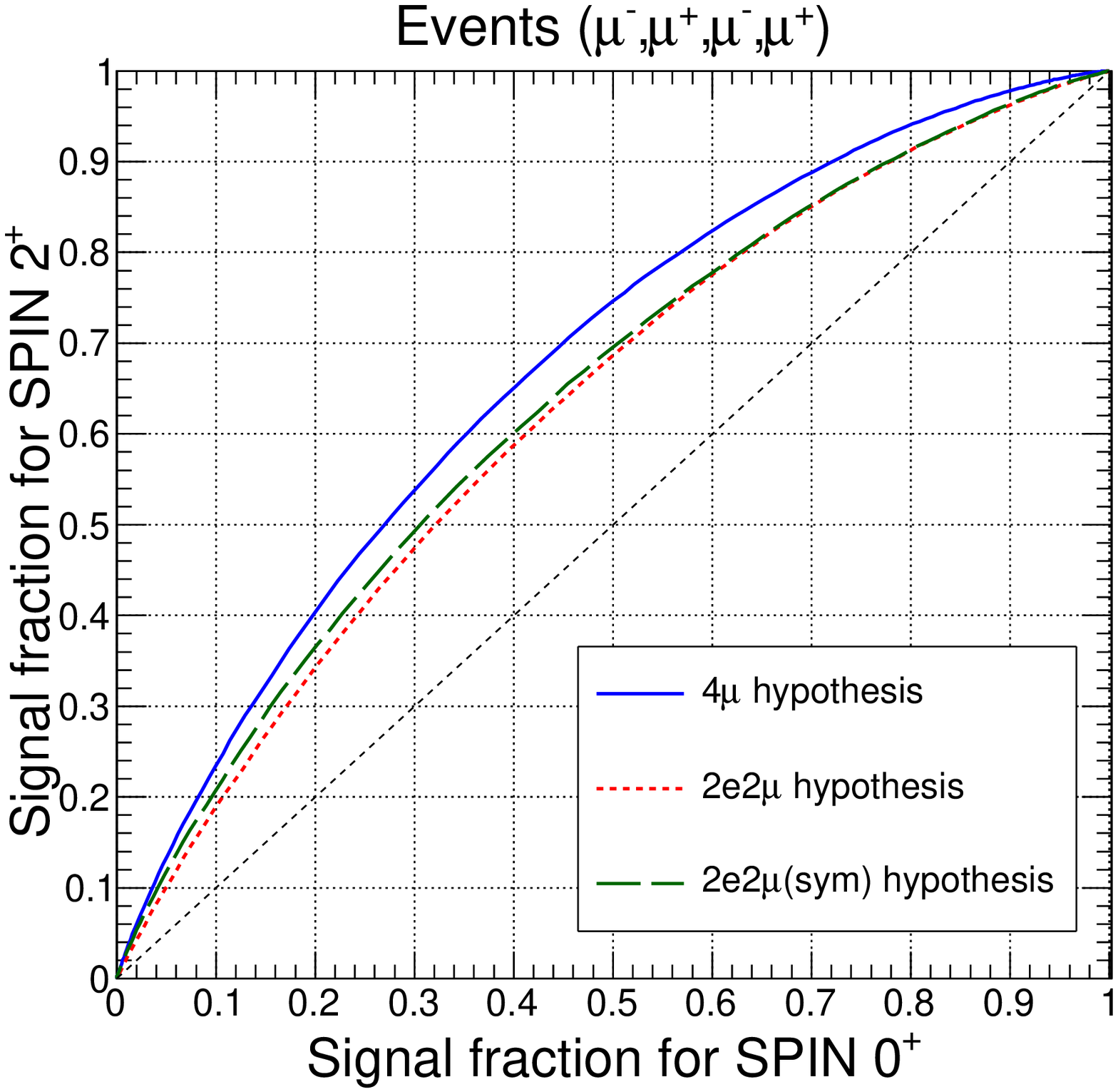}
\caption{ROC curves showing the separation of alternative signal hypotheses 
for the $\mu^+\mu^-\mu^+\mu^-$ final state: 
(left) the pseudoscalar ($\mathrm{J^{CP}} = 0^-$) vs. the SM Higgs boson ($0^+$);
(right) the massive graviton ($\mathrm{J^{CP}} = 2^+_m$)  vs. the SM Higgs boson ($0^+$).  
The solid blue curves are for $KD$s that use the
full expressions for the squared matrix element 
for the four muon final state.
The green dashed curves are for $KD$s that take into account
permutations of identical leptons, but ignore the associated interference effects.
The red dotted curves are for $KD$s that ignore both
permutations of identical leptons and the associated interference effects.
}\label{fig:ROC_CPSpin_Int}
\end{figure}

We compare the results obtained using each of these three treatments
of the SF final state in Figs.~\ref{fig:KD_0+_0-_int}, \ref{fig:KD_0+_Grav_int}, and \ref{fig:ROC_CPSpin_Int}.
Fig.~\ref{fig:KD_0+_0-_int} and Fig.~\ref{fig:KD_0+_Grav_int} 
compare the values of $KD(\mathrm{J^{CP}}; \, \mathrm{H_{SM}})$. 
In Fig.~\ref{fig:ROC_CPSpin_Int} the effect is shown
through ROC curves, corresponding to those shown in
Fig.~\ref{fig:ROC_CPSpin_Diff_Flav}. 
These figures clearly show that including or ignoring 
permutations and interference has a significant impact 
on the value of the kinematic discriminant calculated
and the separation power between alternative signal hypotheses.
Comparing Fig.~\ref{fig:KD_0+_0-_int} and Fig.~\ref{fig:KD_0+_Grav_int} 
(or the right and left plots in Fig.~\ref{fig:ROC_CPSpin_Int})
shows unambiguously that the impact on the expected separation power 
between the $2^+_m$ boson and $\mathrm{H_{SM}}$ is much larger 
than for the case of the $0^-$ boson versus $\mathrm{H_{SM}}$.

To translate the observed differences in the ROC curves into the
expected differences in our ability to tell apart different spin-parity hypotheses,
we perform a simplified statistical analysis using 
{\sc MEKD}-based discriminants. 
As a toy example, using {\sc CalcHEP}, we first calculate the expected number of events
(SM Higgs boson, $0^-$ pseudoscalar, $2^+_m$ massive graviton, and $ZZ$ background)
for an integrated luminosity of $25$ fb$^{-1}$ at $8$~TeV, with the following lepton
selection cuts applied: $p_T > 5$~GeV, $|\eta|<2.5$, $m_{\ell^+\ell^-}>12$~GeV,
$120< m_{4\ell} < 130 $~GeV. The bosons have mass $125$ GeV. 
The signal event count is scaled by K-factor $1.9$ to match the Higgs boson cross 
sections used by the LHC experiments.
The $ZZ$ background event count is scaled up by a factor of $1.5$ to approximately
account for the NLO K-factor and presence of reducible $4\ell$ backgrounds,
where one or more leptons are not prompt. 
This gives the following signal (background) event counts:
$4.0 (7.5)$ for the $4e$ final state,
$3.9 (7.0)$ for the $4\mu$ final state, and
$8.9 (15.5)$ for the $2e2\mu$ final state.
We assume that the alternative signal
hypotheses have the same cross sections as that of the SM Higgs boson
and thus cannot be distinguished from it by using the information 
on the event yields in the four lepton mass distributions alone.
We do not attempt to introduce any experimental event 
reconstruction/selection efficiencies. Thus the absolute numbers
do not reflect accurately the expected separation power
attainable by the LHC experiments; we are interested, rather, in
the relative results when one does or does not
include permutations/interference in the construction of
kinematic discriminants. 
The statistical analysis is based on 2D $pdf$s for alternative signal+background hypotheses:

\begin{itemize}

\item $pdf( \, x,y \, | \, \mathrm{H_{SM}} + \mathrm{bkg} \, )$ 
for the standard model Higgs boson with background;

\item $pdf( \, x,y \, | \, \mathrm{J^{CP}} + \mathrm{bkg} \, )$ 
for a signal of an alternative $\mathrm{J^{CP}}$ hypothesis with background,

\end{itemize}
where $x = KD( \mathrm{H_{SM}}; \, ZZ)$ and $y = KD( \mathrm{J^{CP}}; \, ZZ)$.
We build these 2D $pdf$s using events generated with {\sc MadGraph}. 
For background, we use the SM $q\bar q \to ZZ \to 4\ell$ process. 
Then we proceed with generation of pseudoexperiments, using the expected
event yields and the 2D $pdf$s. The pseudoexperiments are generated for two
different signal hypotheses, the standard model Higgs boson and 
a boson with alternative $\mathrm{J^{CP}}$, 
in both cases assuming the presence of the $ZZ$ background. For each pseudoexperiment,
we calculate as our test statistic, $q$, the negative log likelihood ratio:

\begin{equation}
q = -2 \, \ln 
\frac
{ \mathcal{L}( \, \text{pseudoexperiment} \, | \, \mathrm{J^{CP}} + \mathrm{bkg} \, ) }
{ \mathcal{L}( \, \text{pseudoexperiment} \, | \, \mathrm{H_{SM}} + \mathrm{bkg} \, ) }.
\end{equation}

\begin{figure}[t]
\centering
\includegraphics[width = \3]{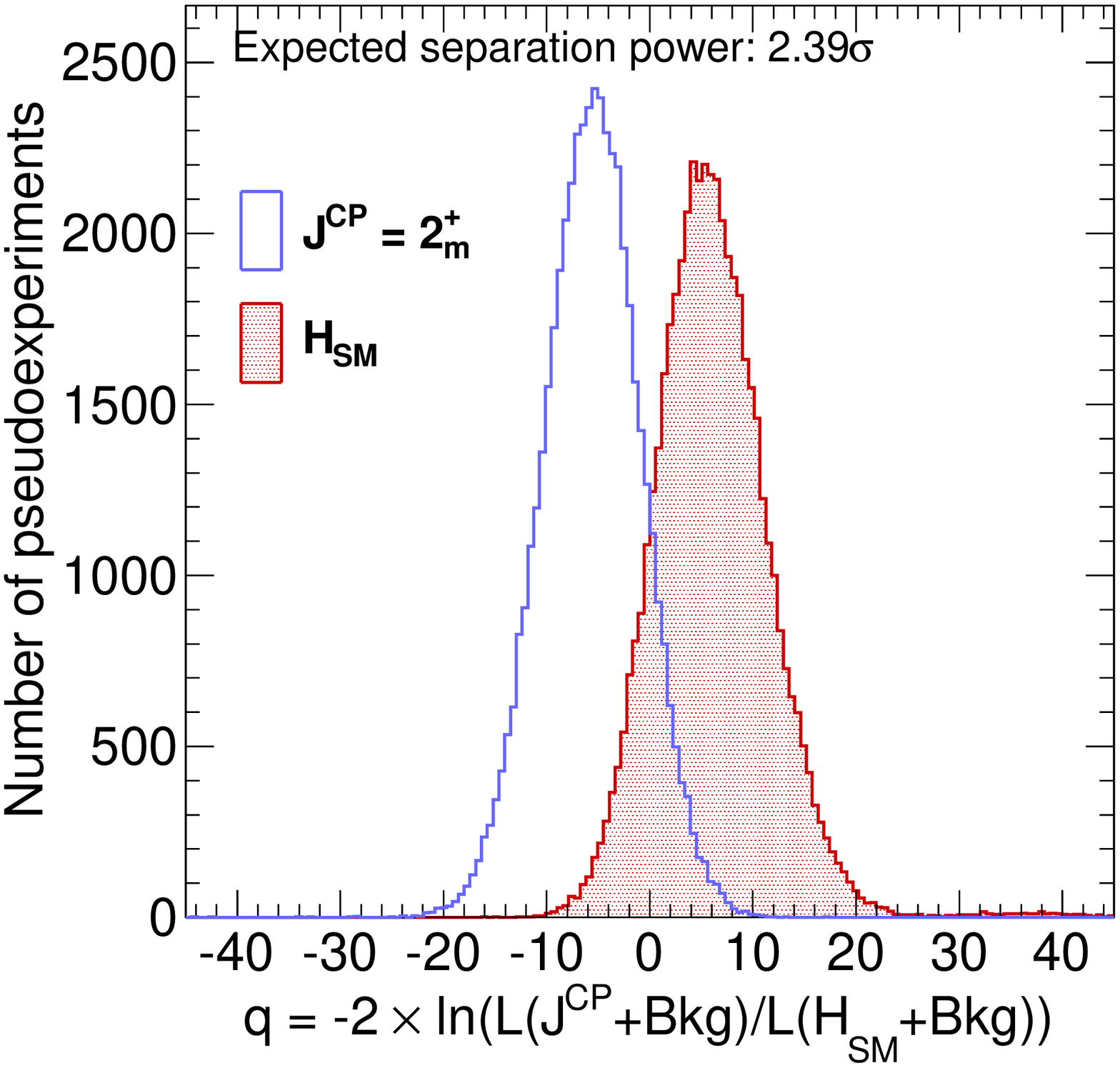}
\includegraphics[width = \3]{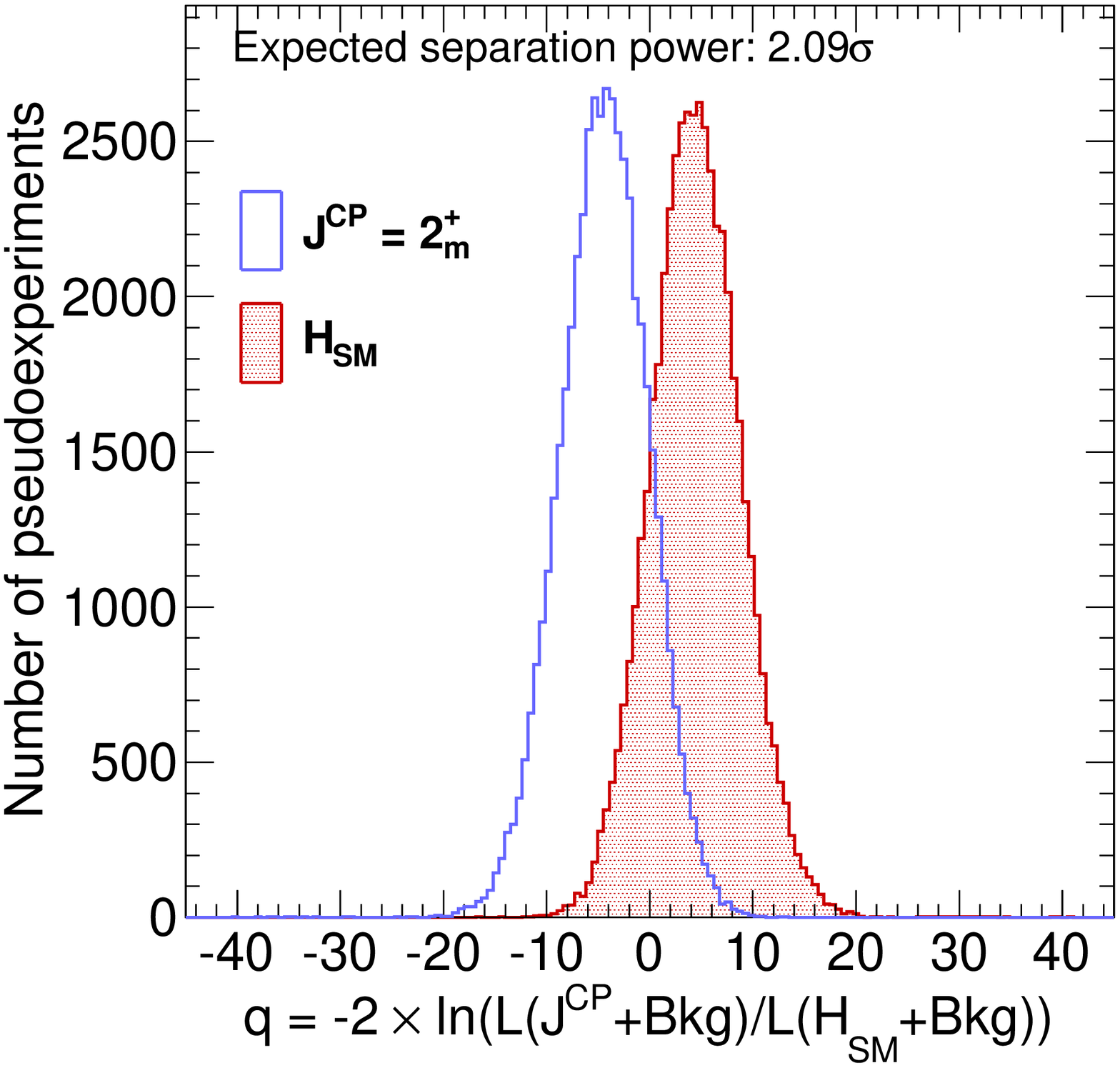}
\includegraphics[width = \3]{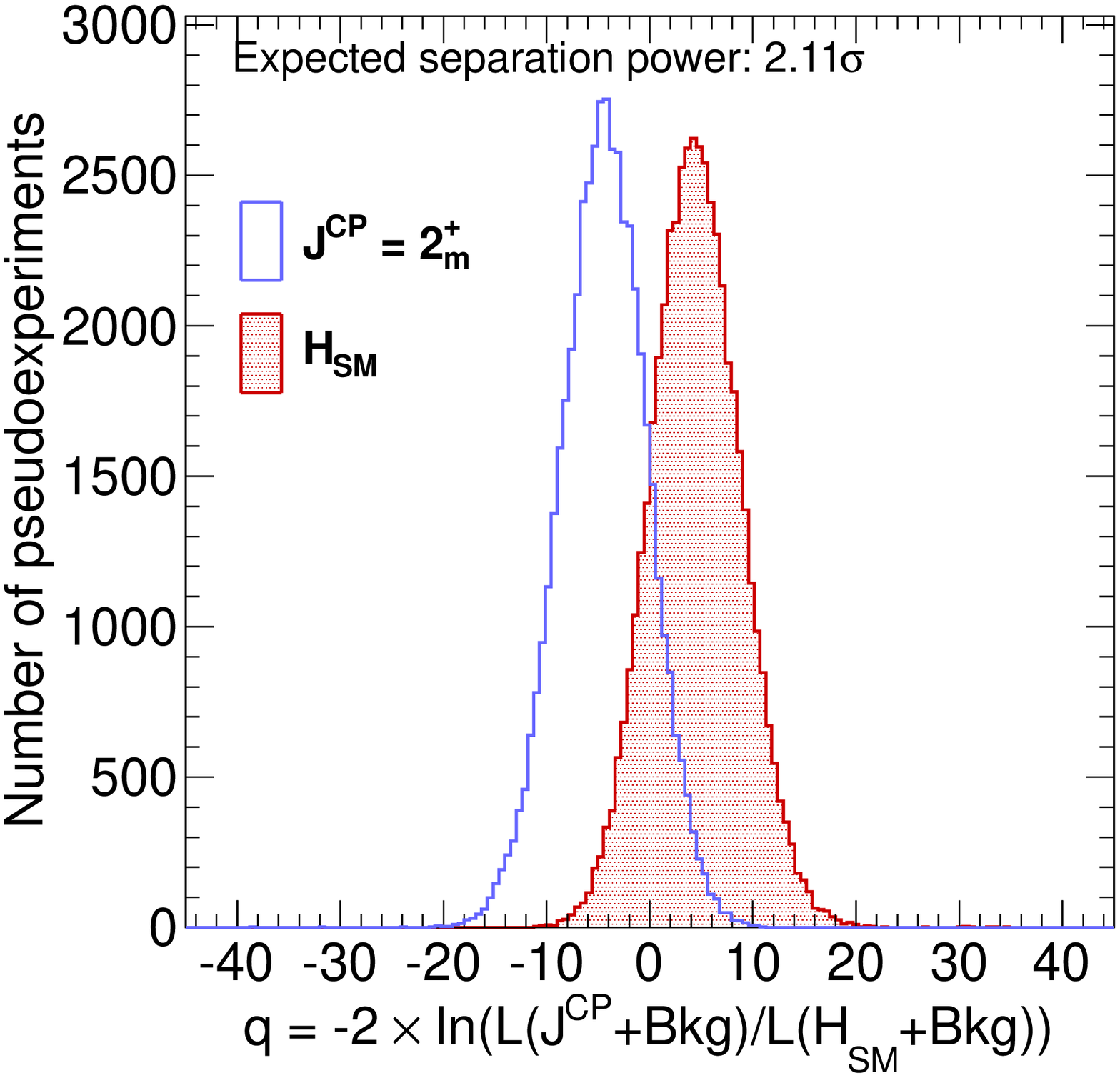}
\caption{
Test statistic distributions for the two alternative hypotheses: 
``massive graviton plus background'' and ``SM Higgs boson plus background'',
open and filled histograms, respectively. The three plots differ in
how the $4e$ and $4\mu$ final states are treated in calculations of
the matrix elements:
(left) complete leading order matrix element;
(middle) permutations of identical leptons in $4e$ and $4\mu$ final states are accounted for, but the associated interference between diagrams is ignored; 
(right) both permutations and interference in $4e$ and $4\mu$ final states are ignored.
The toy model used for generating pseudoexperiments is described in the text.
The projected hypotheses separations are stated on the plots.
}
\label{fig:grav}
\end{figure} 

\begin{figure}[t]
\centering
\includegraphics[width = \3]{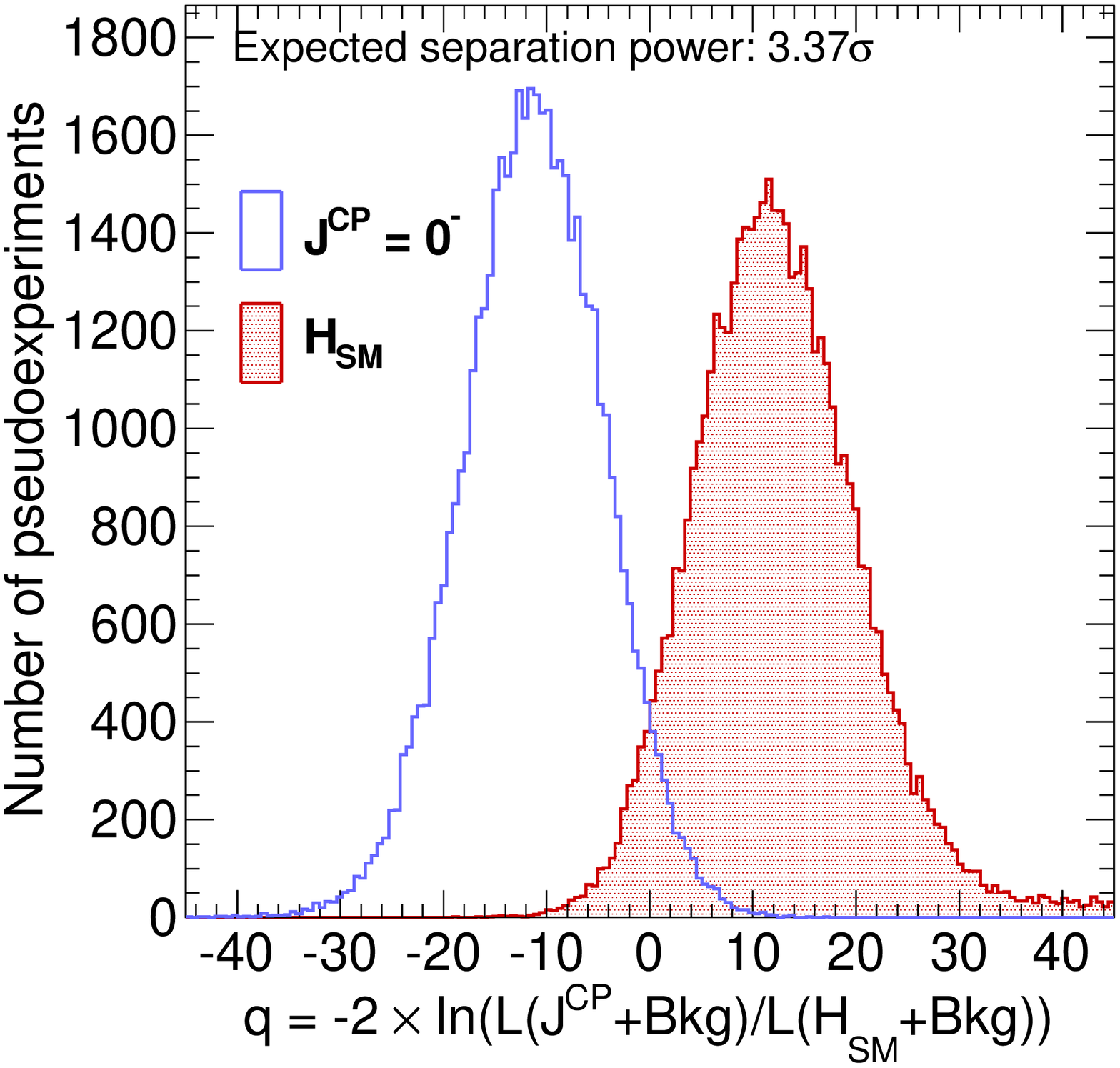}
\includegraphics[width = \3]{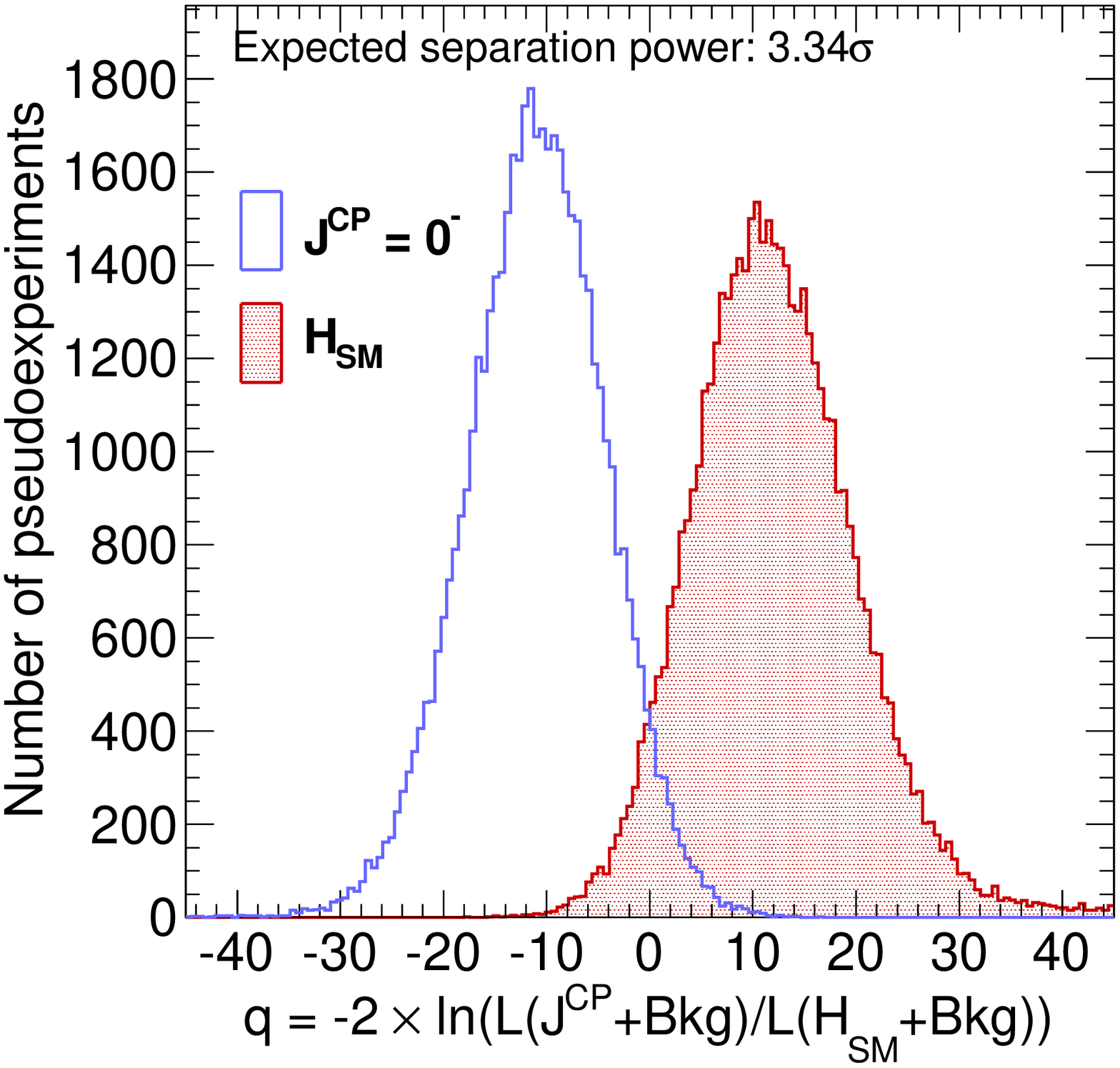}
\includegraphics[width = \3]{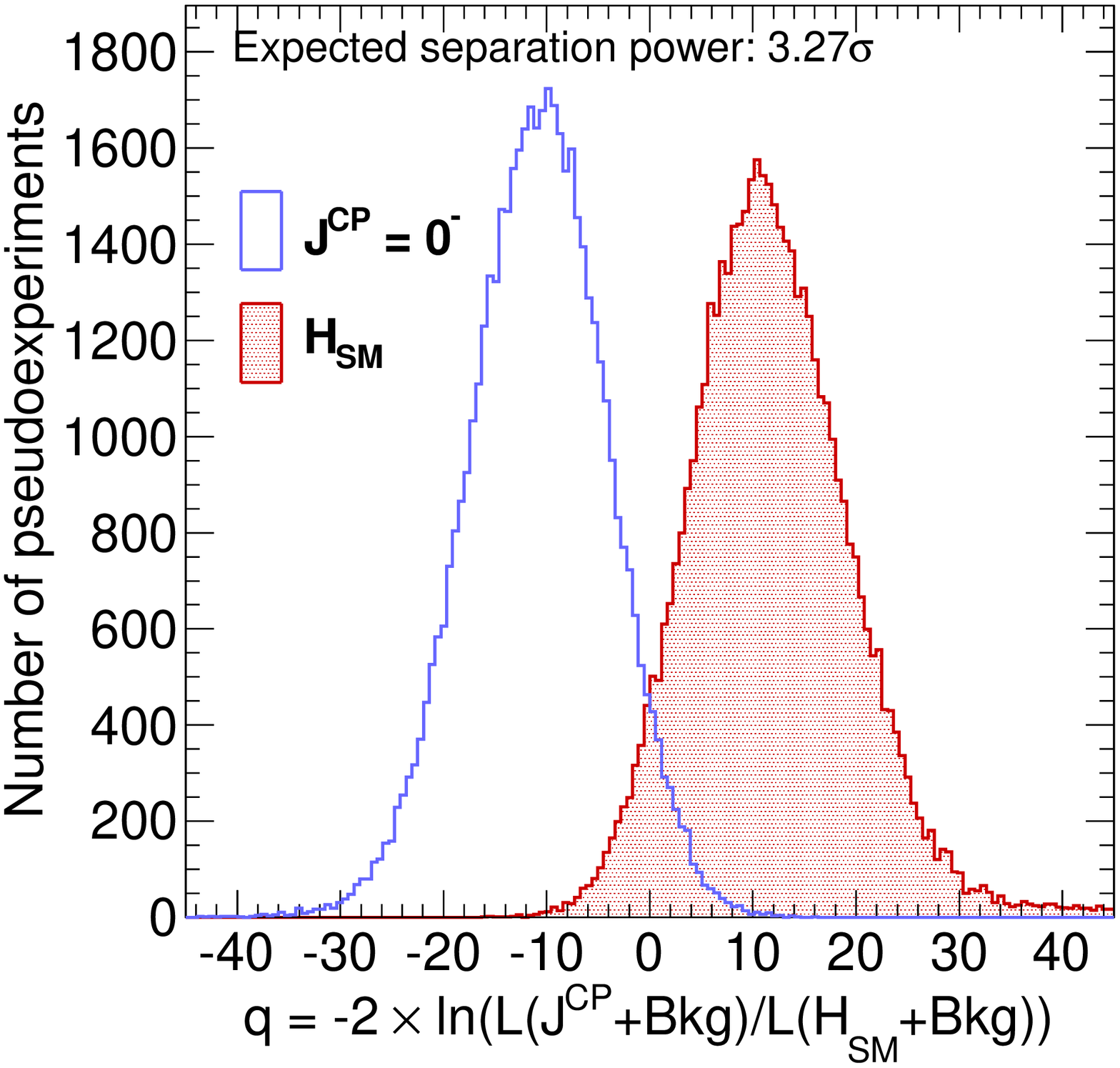}
\caption{
Test statistic distributions for the two alternative hypotheses: 
``pseudoscalar plus background'' and ``SM Higgs boson plus background'',
represented by open and filled histograms, respectively. The three plots differ in
how the $4e$ and $4\mu$ final states are treated in calculations of
the matrix elements:
(left) complete leading order matrix element;
(middle) permutations of identical leptons in $4e$ and $4\mu$ final states are accounted for, but the associated interference between diagrams is ignored; 
(right) both permutations and interference in $4e$ and $4\mu$ final states are ignored.
The toy model used for generating pseudoexperiments is described in the text.
The projected hypotheses separations are stated on the plots.
}
\label{fig:pseudo}
\end{figure}

Figure~\ref{fig:grav} shows the test statistic distributions for
two alternative hypotheses: 
``massive graviton plus background'' and ``SM Higgs plus background'', 
represented by open and filled histograms respectively. 
Figure~\ref{fig:pseudo} shows the test statistic distributions for
alternative signal hypotheses:
``pseudoscalar plus background'' and ``SM Higgs boson plus
background''. The three plots in each figure differ in
how the $4e$ and $4\mu$ final states are treated in calculations of
the matrix elements for signal, $\mathcal{M}_{\mathrm{H_{SM}}}$ and 
$\mathcal{M}_{\mathrm{J^{CP}}}$, and background,
$\mathcal{M}_{\mathrm{ZZ}}$, as follows:

\begin{itemize}
\item (left plot) complete leading order matrix element;
\item (middle plot) permutations of identical leptons in $4e$ and $4\mu$ final states are accounted for, but the associated interference between diagrams is ignored; 
\item (right plot) both permutations and interference in $4e$ and $4\mu$ final states are ignored.
\end{itemize} 

For the case of the massive graviton, 
there is about a $15$\% increase in sensitivity associated 
with the correct treatment of the SF four lepton final states.
The difference is large and may become a decisive factor
in the ability of ATLAS and CMS experiments to distinguish between
spin zero and spin two bosons with the data of the 2011+2012 LHC run.   
As expected for the pseudoscalar case, the gain from 
the correct treatment of the SF final states is smaller,
about $3$\%.

\section{Summary and outlook}
\label{sec:conclusions}
We have shown that matrix element analyses improve the extent to which
various Higgs-like signals can be discriminated from the background
and allow for the measurement of the spin and parity properties of
the Higgs-like boson in the $X\to ZZ^\ast\to 4\ell$ channel. 
We have demonstrated the importance of using the full matrix element
including permutations of identical leptons and the associated
interference terms for the SF four lepton final state.
The proper treatment of these effects enhances our ability to distinguish 
between different spin and parity signal hypotheses.
A gain in sensitivity as large as $15$\% can be expected for the case of 
SM Higgs vs. massive graviton hypothesis separation. 

As the ATLAS and CMS experiments move toward precision studies 
of the properties of the observed boson, the kinematic discriminants
will play larger and larger roles. Thus, the demand will grow for automated tools 
allowing for calculations of matrix element-based discriminants
that most accurately capture all features of the underlying physics
of the different signal and background processes involved.

The studies presented have been carried out with the {\sc MEKD} code,
which is now publicly available. The user of this code has the flexibility to 
construct signal matrix elements for an arbitrary set of allowed
couplings for different spin zero and spin two resonances;
the matrix element for the background process 
$q \bar q \to 4\ell$ is also a part of the package.
The general couplings of a spin one boson as well as the ability to
consider the final state with four leptons and a bremsstrahlung photon
will be added shortly \cite{newUFpaper}. 

\acknowledgments
We thank J.~Campbell, A.~Gritsan, I.~Low, J.~Lykken, R. Vega-Morales,
and C.~Williams for useful discussions.
J.~Gainer and K.~Matchev thank the Aspen Center for Physics (funded by
NSF Grant \#1066293) for hospitality during the completion of this work.
M.~Park is supported by the CERN-Korea fellowship through the National
Research Foundation of Korea.
Work supported in part by U.S.~Department of Energy Grant
DE-FG02-97ER41029 and NSF Grant 1007115.

\appendix

\section{Validations of {\sc MEKD} against {\sc CalcHEP} and {\sc NLOME}}
\label{sec:CH}
%=========================================================================

\subsection{Comparison with {\sc CalcHEP}}

Following the same procedure as in Sec.~\ref{sec:MG},
we also created an independent code to compute
$KD_{\textsc{CALC}}$, where the matrix element is calculated 
by {\sc CalcHEP}. As in Sec.~\ref{sec:MG}, the SM is implemented 
via {\sc FeynRules}, which ensures that our
$KD_{\textsc{MAD}}$ and $KD_{\textsc{CALC}}$
results are obtained with identical inputs for the SM parameters
(masses, couplings, etc.). Since there are no functionality differences between
{\sc MadGraph} and {\sc CalcHEP}, one would expect that
$KD_{\textsc{MAD}}$ and $KD_{\textsc{CALC}}$ should be the same; 
therefore this provides a useful cross-check.

\begin{figure}[ht]
\centering
\includegraphics[width=\3]{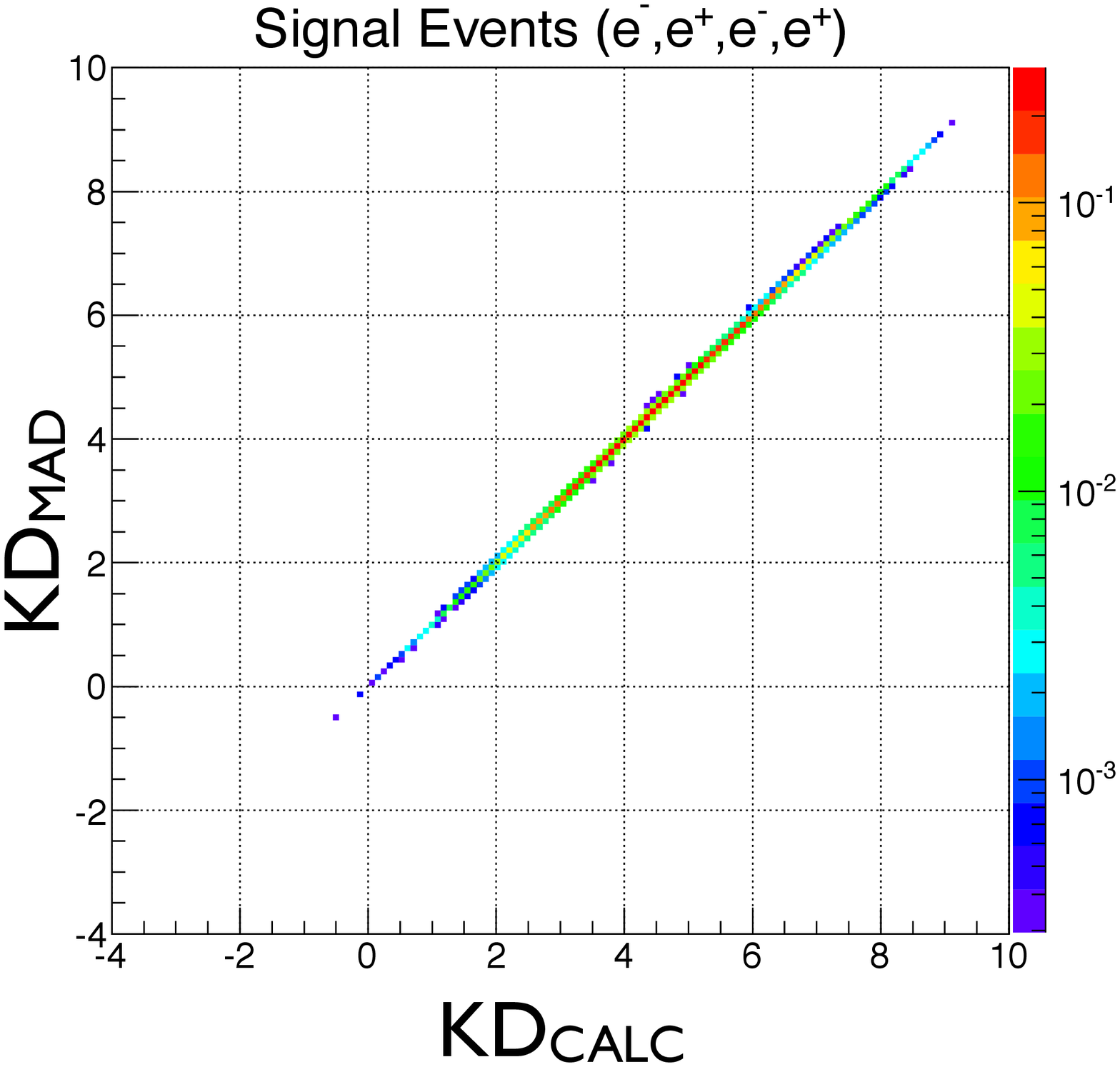}
\includegraphics[width=\3]{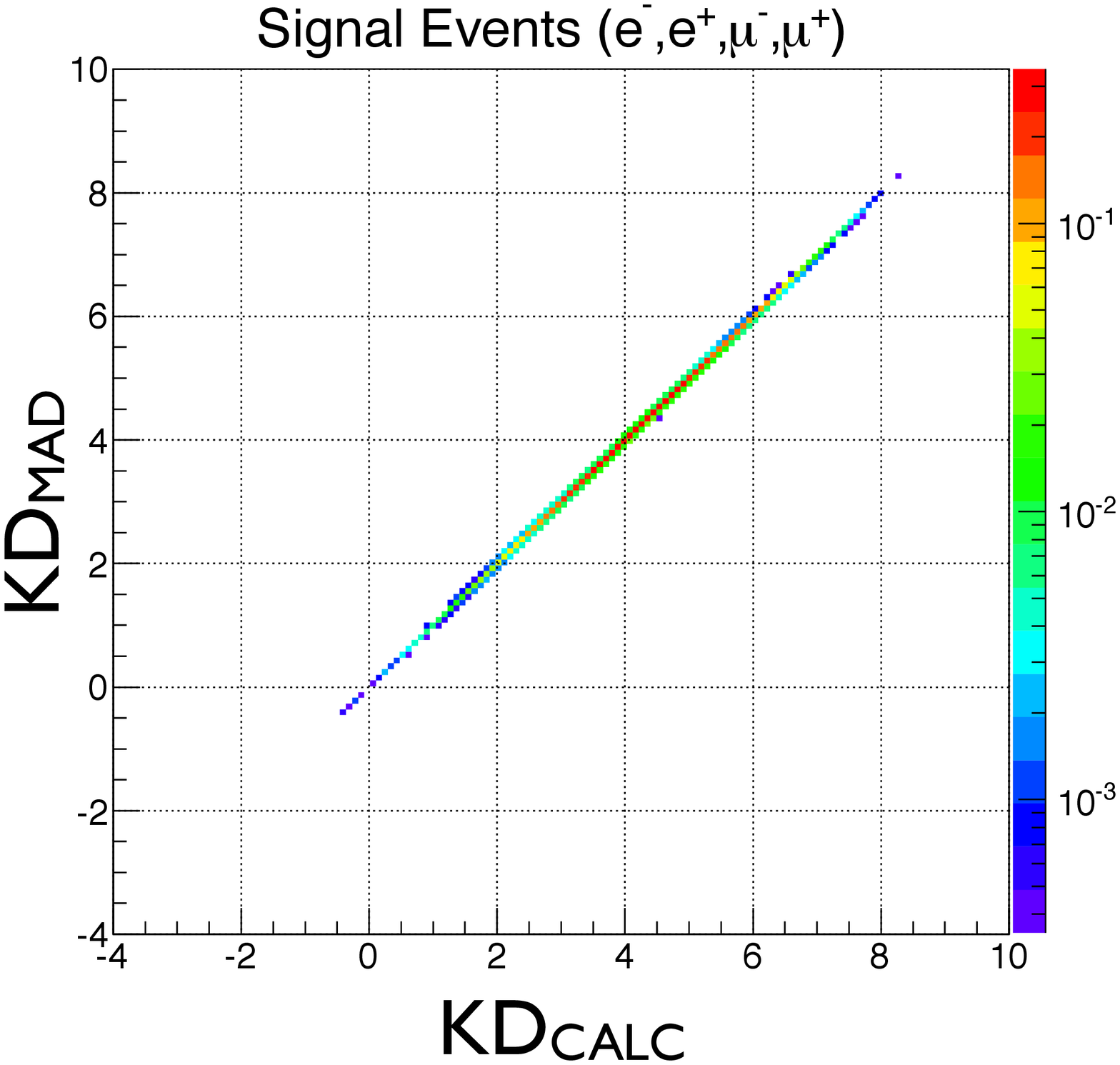}
\includegraphics[width=\3]{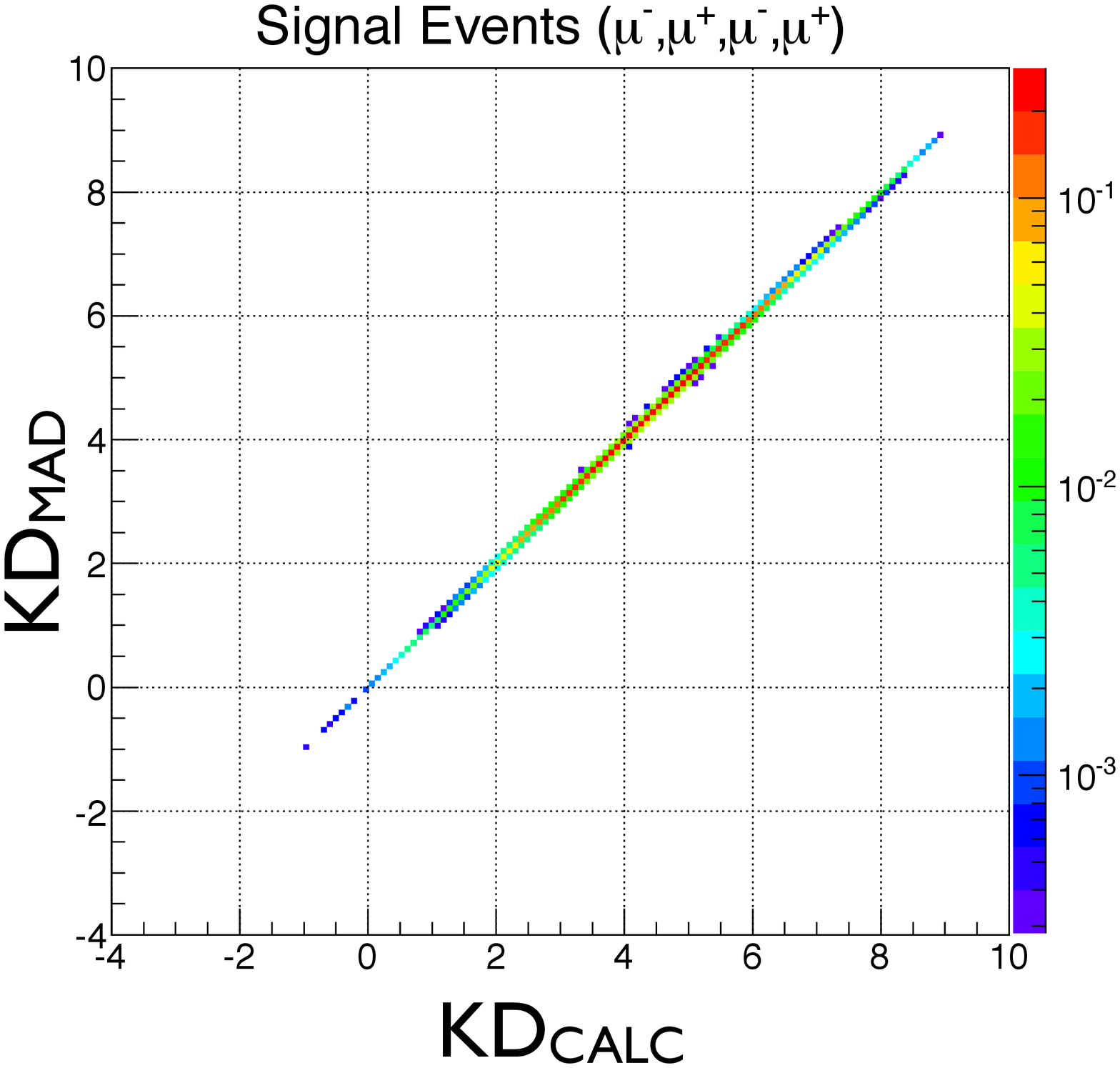}\\
\includegraphics[width=\3]{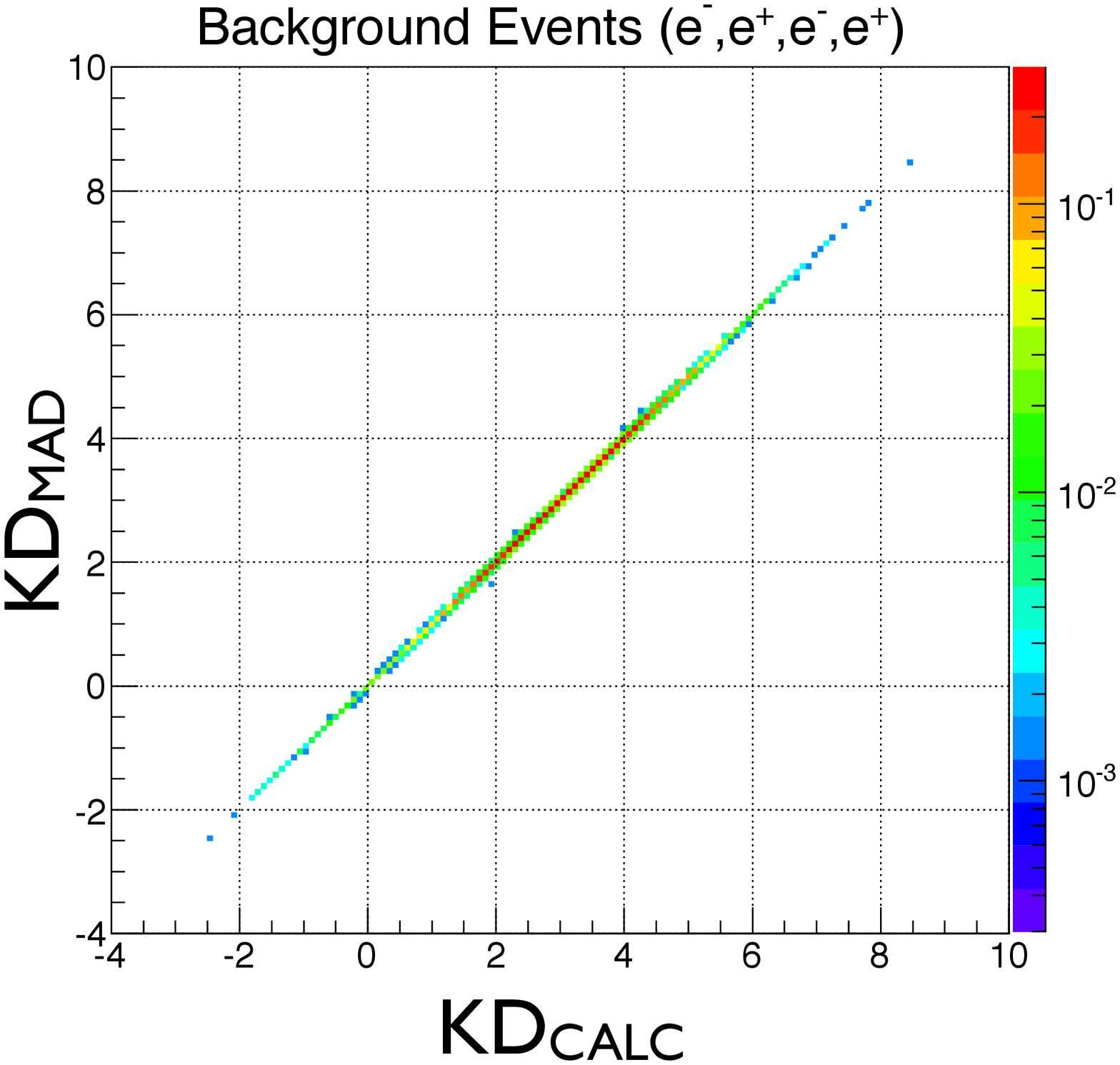}
\includegraphics[width=\3]{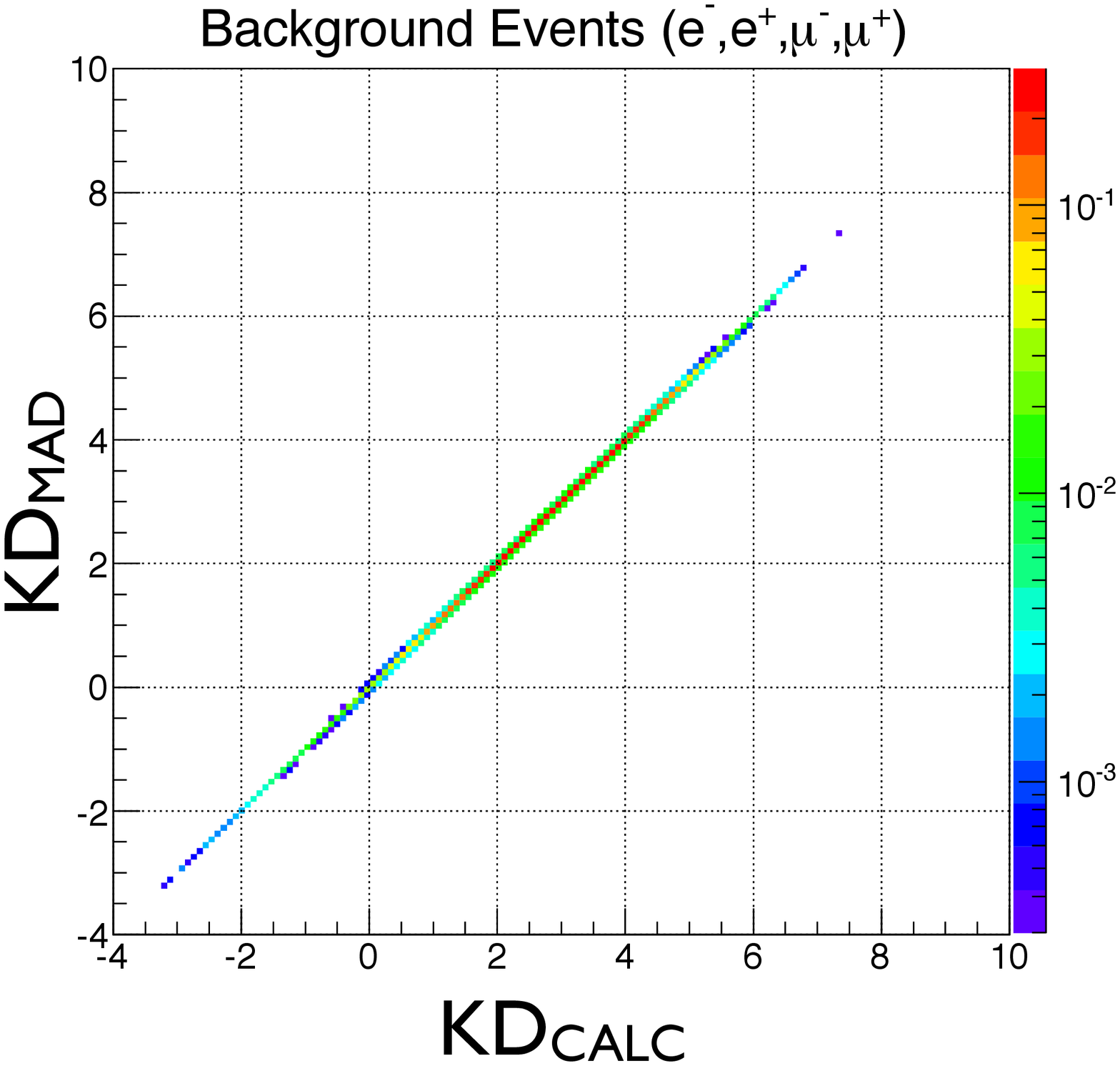}
\includegraphics[width=\3]{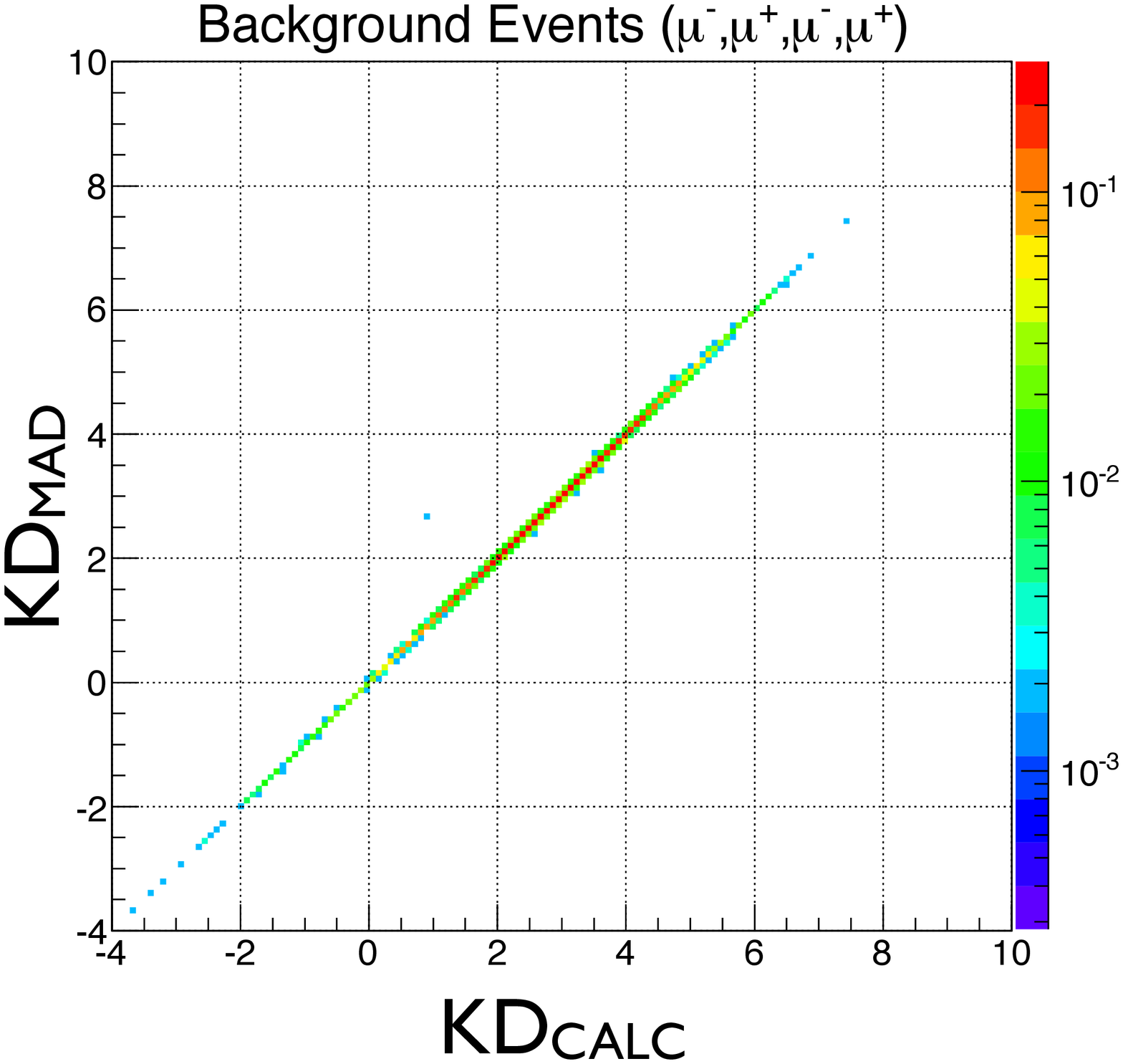}
\caption{\label{fig:CH_MG} Comparison of the kinematic discriminant,  $KD_{\textsc{CALC}}$, 
computed with {\sc CalcHEP}, and  the kinematic discriminant,
$KD_{\textsc{MAD}}$, computed by
{\sc MadGraph}, for SM Higgs signal events (top row) and background events (bottom row),
using $4e$ events (left column), $4\mu$ events (right column), and
$2e2\mu$ events (middle column).  }
\end{figure} 

We perform such consistency checks on both SM Higgs signal and 
background samples, for DF as well as SF final states.
The results are displayed in Fig.~\ref{fig:CH_MG}, which reveals that,
as expected, $KD_{\textsc{MAD}}$ and $KD_{\textsc{CALC}}$
are in excellent agreement. Perfect correlation is also
observed for the signal and background matrix
elements as calculated by the two tools. 
This synchronization exercise serves a dual purpose: first,
we are able to validate our code, and second, the level of agreement 
seen in Fig.~\ref{fig:CH_MG} provides a benchmark for the other
comparisons in this paper.

\subsection{Comparison with {\sc NLOME}}
%\label{sec:MCFM}
%==================================================================================
The {\sc MCFM} code \cite{Campbell:2010ff} can be inverted to compute 
the matrix element weight $|{\cal M}|^2$ from a given final state
kinematic configuration.
The corresponding code, {\sc NLOME} \cite{NLOME}, is still under development.
Here we use a beta version of {\sc NLOME} 
to cross-check against our results from the previous two sections. 
We should mention that as an NLO tool, {\sc NLOME} involves an integration over
the unknown longitudinal momentum of the additional jets, which are
recoiling against the $4\ell$ system. 
However, here we are using LO events, since we are interested in
comparing the {\em leading order} machinery implemented in the different tools. 
Therefore, we removed the additional longitudinal momentum integration in {\sc NLOME}
and we are running the {\sc NLOME} code in a purely LO mode.

\begin{figure}[ht]
\centering
\includegraphics[width=\2]{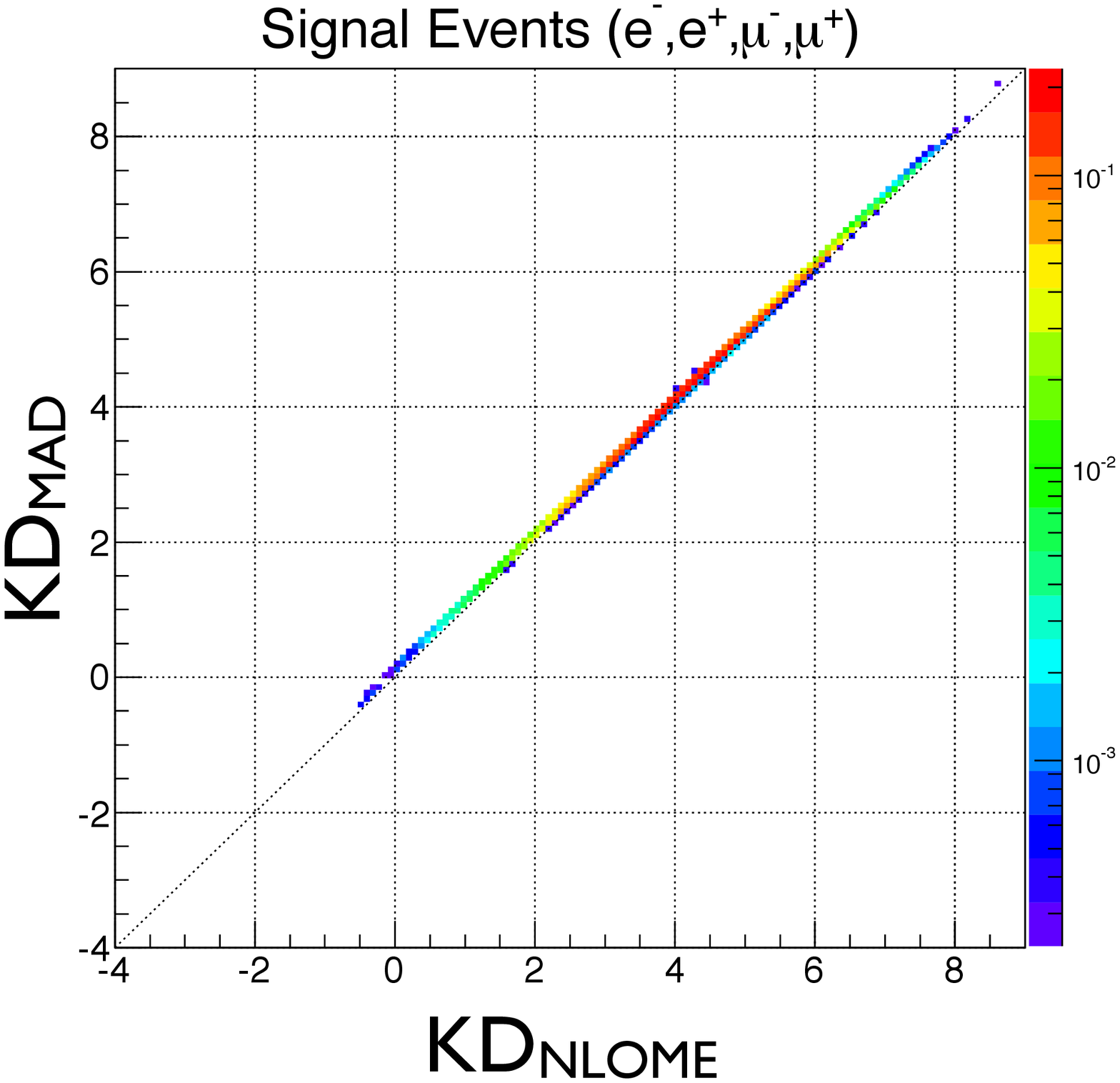}
\includegraphics[width=\2]{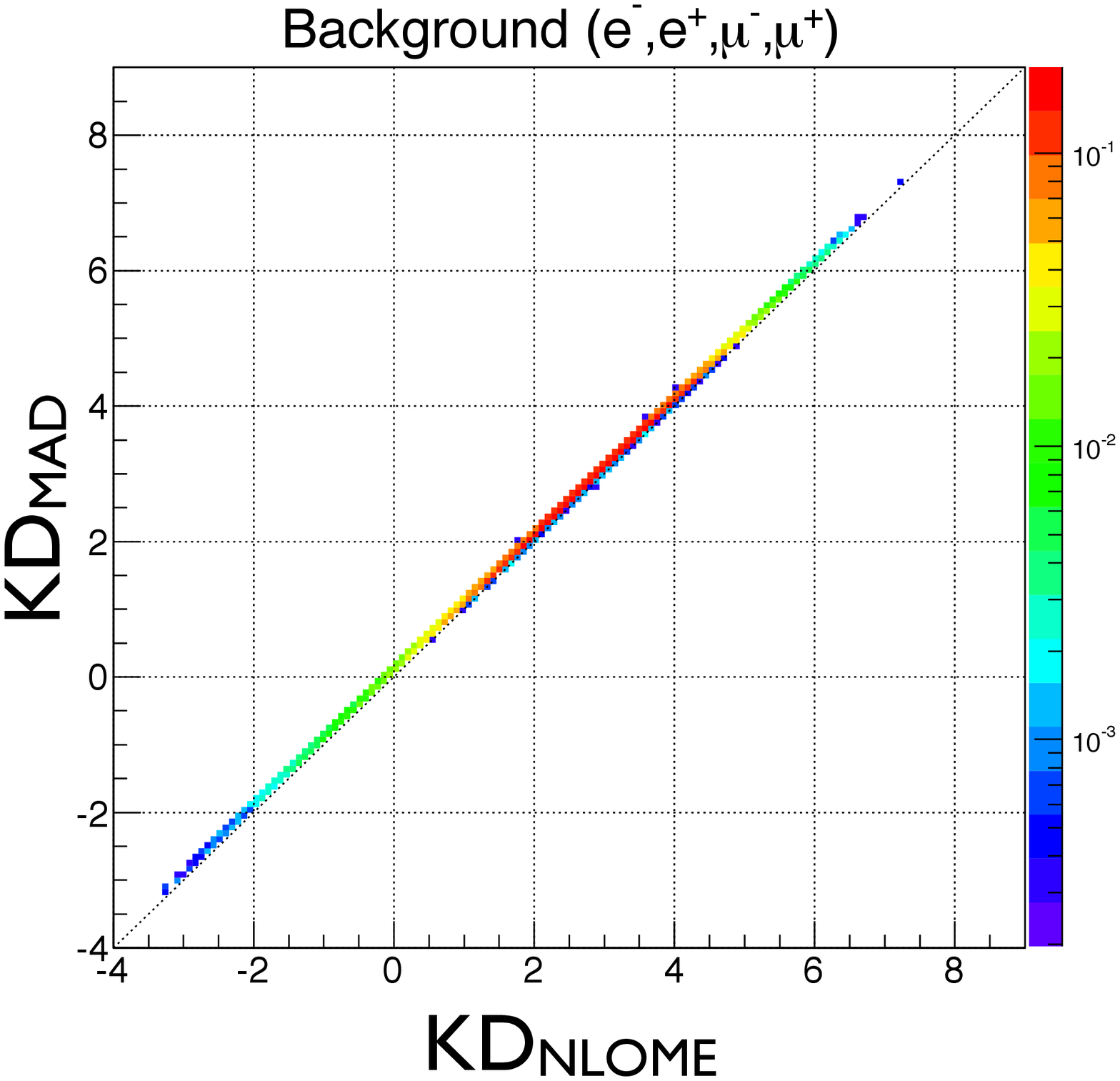}
\caption{\label{fig:CH_MCFM} Comparison of the 
kinematic discriminants $KD_{\textsc{NLOME}}$ and $KD_{\textsc{MAD}}$ 
for DF $2e2\mu$ events for SM Higgs signal (left plot) and background (right plot).}
\end{figure} 

The comparison between
$KD_{\textsc{MAD}}$ and $KD_{\textsc{NLOME}}$ for DF $4\ell$ events is shown in Fig.~\ref{fig:CH_MCFM}.
The two calculations for $2e2\mu$ final states are on equal footing 
and their results should agree.
Indeed, this is what Fig.~\ref{fig:CH_MCFM} shows: the level of agreement is 
excellent and comparable to what we observed earlier in Fig.~\ref{fig:CH_MG}.

\section{Notation and conventions}
\label{sec:notation}
\begin{figure}[ht]
\centering
\includegraphics[width=4.0in]{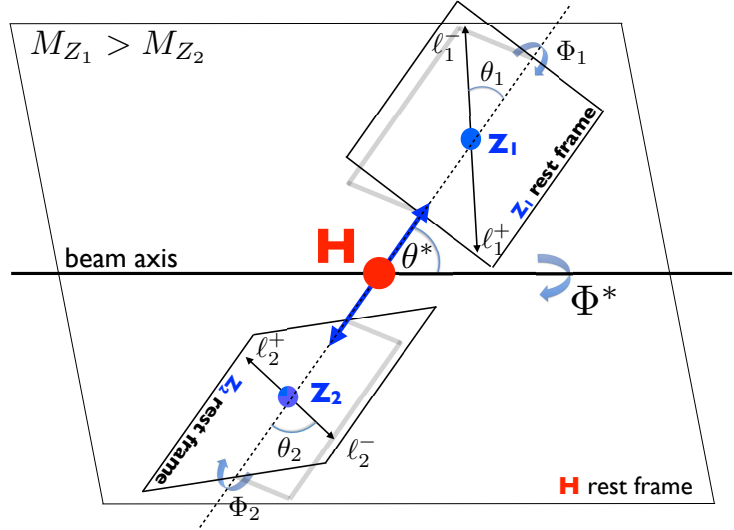}
\caption{\label{fig:angles} Definition of the angular variables relevant to the
$H\to ZZ^\ast \to 4\ell$ event topology.}
\end{figure}

At leading order, events in the Higgs golden channel are described by 
10 degrees of freedom\footnote{There are four particles in the final state, 
whose momenta give $4\times 3=12$ parameters, two of which are
removed by transverse momentum conservation.}. Following
\cite{Gao:2010qx,Bolognesi:2012mm}, we can take them to be as follows:
\begin{itemize}  
\item Three invariant mass parameters: $M_{4\ell}$, $M_{Z1}$ and $M_{Z2}$.
\item The rapidity $y_{ZZ^\ast}$ of the event in the LAB frame, see Sec.~\ref{sec:IS}.
\item Two angular variables defined in the CM frame of the whole event. 
These can be chosen to be the polar angle $\theta^\ast$ (measured from the beam axis)
and the azimuthal angle $\Phi^\ast$ of the $Z_1$ system, as shown in Fig.~\ref{fig:angles}. 
\item Two angular variables defined in the CM frame of the $Z_1$ system.
These can be taken to be the azimuthal angle $\Phi_1$ and the polar
angle $\theta_1$ 
(measured from the $Z_1$ direction in the Higgs CM frame) of the
lepton $\ell_1^-$ produced in the $Z_1$ decay (refer to Fig.~\ref{fig:angles}).
\item Two angular variables defined in the CM frame of the $Z_2$ system.
These can be taken to be the azimuthal angle $\Phi_2$ and the polar
angle $\theta_2$ (measured from the $Z_2$ direction in the Higgs CM
frame) of the lepton $\ell_2^-$ produced in the $Z_2$ decay. It is
often convenient to trade the angle $\Phi_2$ for $\Phi\equiv \Phi_2-\Phi_1$.
\end{itemize}

\section{The Matrix Element Kinematic Discriminant package, {\sc MEKD}}
\label{sec:MEKD}
%==============================================================================

The computer code for calculating $KD_{\textsc{MAD}}$ can be freely
downloaded from \cite{KDcode}. The website also includes instructions 
and examples for installing the package and running the producer, which for completeness
are also included here.

\subsection{Description of the code}

The Matrix Element Kinematic Discriminant, {\sc MEKD}, package provides tools to calculate
the leading order (LO) matrix elements for signal $gg\to X\to ZZ^\ast \to 4\ell$ and background
$q\bar{q}\to ZZ^\ast \to 4\ell$ processes, and to build kinematic discriminants $KD$ 
that can be used for separation between these processes. The supported signal processes 
include the production of a scalar and a spin-2 resonance $X$ through gluon-gluon fusion and its decay into four 
leptons via two $Z$ gauge bosons, where
the couplings of the scalar resonance to gauge bosons are kept general (parametrized).
The {\sc MEKD} package consists of two parts:
\begin{enumerate}
\item MEKD interface class (and the corresponding MadGraph libraries it depends on)
\item MEKD producer (macro defined in the \verb&runKD_MAD.cc&) 
\end{enumerate}
The MEKD class is declared in \verb&MEKD.h& and provides an interface to the methods to calculate the MEs and $KD$s for the chosen signal and background processes using the kinematic information about the four leptons in the final state. Detailed description of MEKD interface methods can be found at the dedicated MEKD class reference page at \cite{KDcode}. 

The macro \verb&runKD_MAD.cc& is an executable that can run over the input sample of four lepton events and as an output produces a file with MEs and $KD$s for each of the input events. Its functionality is based on the MEKD interface class. The {\sc MEKD} package provides the following functionalities:
\begin{enumerate}
\item Accepts as input the tabulated data file with the kinematic information of four leptons
   in the final state. The format of the input file is the following:
$$
   id_1\ id_2\ id_3\ id_4\ p_{1x}\ p_{1y}\ p_{1z}\ e_1\ p_{2x}\ p_{2y}\ p_{2z}\ e_2\ p_{3x}\ p_{3y}\ p_{3z}\ e_3\ p_{4x}\ p_{4y}\ p_{4z}\ e_4
$$
where $id_N$, $p_{Nx}$, $p_{Ny}$, $p_{Nz}$ and $e_N$ are the PDG id, 
spatial components and time component of the Nth lepton four momentum, respectively.
\item Allows the user to select the PDFs that should be taken into account in ME calculation.
\item Initializes the couplings of the resonance according to the user's signal selection.
\item Performs a purely transverse boost of each event to a reference frame where the transverse component of total momenta
   of the four lepton system is zero.
\item Feeds the boosted momenta of all four leptons to the ME calculator
  and computes MEs and the $KD$ for the given event.
\item Provides the output as the tabulated data file with the computed ME and $KD$ values.
   The format of the output file is the following:
$$
   |{\cal M}_{ZZ}|^2\ |{\cal M}_{XZZ}|^2\ KD 
$$
   where $|{\cal M}_{ZZ}|^2$, $|{\cal M}_{XZZ}|^2$ and $KD$ 
   are the squared ME for the background process, the squared ME for the selected signal
   process and the kinematic discriminant $KD$, respectively.
\item Prints out the status of the code execution and the summary message at completion.
 \end{enumerate}

\subsection{User instructions}

\subsubsection{Requirements}

The latest version of the code can be downloaded and installed
following the instructions at the {\sc MEKD} web site:
\begin{verbatim}
http://mekd.ihepa.ufl.edu/
\end{verbatim}
Compilation of the code requires an installed gcc compiler. 
GCC binary packages for multiple platform can be found here:
\begin{verbatim}
http://gcc.gnu.org
\end{verbatim}

\subsubsection{Setup of the {\sc MEKD} code}

The package can be installed from the downloaded tarball by typing in a terminal window 
\begin{verbatim}
tar xvf MEKD_Madgraph.tar
cd MEKD/macros
. setup.sh   
\end{verbatim}
The script will compile and link all necessary {\sc MEKD} libraries/macros.
One of the outputs is the main executable \verb&runKD_MAD&. If ROOT framework is installed and the environment properly set, the code will be compiled with ROOT support.

\subsubsection{Run the {\sc MEKD} producer}

In a terminal window, type 
\begin{verbatim}
./runKD_MAD [-f input_file] [-x x_resonance] [-p pdf_include] [-l log_file]
\end{verbatim}
   where the available options are described in Table~\ref{tab:options}.

\begin{table}[h!]
\centering
\begin{tabular}{|c|c|}
\hline
\verb&input_file&      & name of the input {\tt .dat} file (string, REQUIRED)                      \\ \hline
\verb&x_resonance& & choice of the signal resonance (string, DEFAULT = 'SMHiggs') \\ 
                                  & Available options: SMHiggs, Higgs0M, Graviton2PM and Custom.                   \\ \hline
\verb&pdf_include&  & name of PDFs, not used if \verb^pdf_include=''^ (string, DEFAULT=\verb^'CTEQ6L'^) \\ \hline  
\verb&log_file&          & name of log file, no logging if \verb&log_file=''& (string,DEFAULT=\verb&''&)\\ \hline
\end{tabular}
\caption{\label{tab:options} Options of the {\sc MEKD} producer.}
\end{table}

For example, to run the {\tt .dat} file provided in the package, execute command:
\begin{verbatim}   
./runKD_MAD -f DATA/Events/SIG_4l_30evt.dat
\end{verbatim}
The user can define a resonance with custom couplings by specifying as the type of resonance ''Custom'' and by specifying manually the desired couplings in the file:
\begin{verbatim}   
src/Cards/param_card.dat
\end{verbatim}
Examples of files with couplings parameters can be found in template .dat files located in the same directory.

Options and details on input parameters can always be printed out as:
\begin{verbatim}     
./runKD_MAD -h
\end{verbatim}

\subsubsection{Output from the {\sc MEKD} producer}

To run the code with a user-provided input file, \verb&yourInputFileName.dat&, evaluate
\begin{verbatim}   
./runKD_MAD -f yourInputFileName.dat
\end{verbatim}   
and the output will be stored in the {\tt .dat} file named
\begin{verbatim}   
yourInputFileName_withDiscriminator.dat
\end{verbatim}   
   It includes the MEs for selected signal and background, as well as the KD
   value. 
   
\subsubsection{Comparison of user \text{MEKD} results}

   The results of a custom user code which uses the \verb&MEKD.h& libraries
   can be compared to the results obtained by the \verb&runKD_MAD.cc& macro.
   User code should be run on the input
   {\tt .dat} files (SIG and BKG) which contain 30 events located here:
\begin{verbatim}
DATA/Events
\end{verbatim}
The output files with {\sc MEKD} results can be compared to the
reference output files of the "\verb^runKD_MAD^" macro located in the
same directory.

%====================================================================================

\end{document}